\documentclass{article}

\usepackage{arxiv}

\usepackage[utf8]{inputenc} 
\usepackage[T1]{fontenc}    
\usepackage{hyperref}       
\usepackage{url}            
\usepackage{booktabs}       
\usepackage{amsfonts}       
\usepackage{nicefrac}       
\usepackage{microtype}      
\usepackage{lipsum}
\usepackage{graphicx,subcaption}
\usepackage{amsmath}
\usepackage{amssymb,hyperref}

\usepackage[normalem]{ulem}
\usepackage{amssymb}
\usepackage{graphics}
\usepackage{amsmath,bm}
\usepackage{mdframed}
\usepackage{subcaption} 

\usepackage{multirow}
\usepackage{epstopdf} 
\usepackage{arydshln}
\usepackage{bbold}
\usepackage[bbgreekl]{mathbbol}
\usepackage{xcolor}
\usepackage[compatibility=false]{caption}
\usepackage{setspace}
\usepackage{xcolor}
\usepackage{amsmath,amsfonts,amssymb,amsthm}
\usepackage{mathtools}
\usepackage{latexsym}   
\usepackage{graphicx}
\usepackage{mathrsfs}
\usepackage{stmaryrd}
\usepackage{mathtools}
\usepackage{placeins}
\usepackage{tikz}
\usepackage{etoolbox}
\usepackage{anyfontsize}
\usepackage{multirow}
\usepackage{pifont}
\usepackage{moreverb}
\usepackage{colortbl}

\title{On the Condition Monitoring of Bolted Joints through Acoustic Emission and Deep Transfer Learning: 
Generalization, Ordinal Loss and Super-Convergence}

\author{
  Emmanuel Ramasso, Rafael de O. Teloli, Romain Marcel\\
  Department of Applied Mechanics\\
  Institut FEMTO-ST,  UFC/CNRS/SUPMICROTECH/UTBM, \\
  24 Rue Alain Savary, 25000 Besan\c con, France\\
  $^{*}$\texttt{emmanuel.ramasso@femto-st.fr}  
}

\begin{document}
\maketitle

\begin{abstract}
[This paper was accepted in \href{https://journals.sagepub.com/home/SHM}{SAGE/SHM journal}.]\\
 
This paper investigates the use of deep transfer learning based on convolutional neural networks (CNNs) to monitor the condition of bolted joints using acoustic emissions. Bolted structures are critical components in many mechanical systems, and the ability to monitor their condition status is crucial for effective structural health monitoring. We evaluated the performance of our methodology using the ORION-AE benchmark, a structure composed of two thin beams connected by three bolts, where highly noisy acoustic emission measurements were taken to detect changes in the applied tightening torque of the bolts. The data used from this structure is derived from the transformation of acoustic emission data streams into images using continuous wavelet transform, and leveraging pretrained CNNs for feature extraction and denoising. Our experiments compared single-sensor versus multiple-sensor fusion for estimating the tightening level (loosening) of bolts and evaluated the use of raw versus prefiltered data on the performance. We particularly focused on the generalization capabilities of CNN-based transfer learning across different measurement campaigns and we studied ordinal loss functions to penalize incorrect predictions less severely when close to the ground truth, thereby encouraging misclassification errors to be in adjacent classes. Network configurations as well as learning rate schedulers are also investigated, and super-convergence is obtained, i.e., high classification accuracy is achieved in a few number of iterations with different networks. Furthermore, results demonstrate the generalization capabilities of CNN-based transfer learning for monitoring bolted structures by acoustic emission with varying amounts of prior information required during training. 
\end{abstract}

\section{Introduction}
\label{Introduction}

Bolted structures play a key role in several mechanical systems used in industry to assemble parts together. Ensuring the healthy state of such jointed structures is essential {{in various industries, such as aerospace, automotive, and construction since failures in bolted joints can lead to significant safety risks, economic losses, affected performances and regulatory compliance}}. 

{{Detecting bolt loosening or degradation in its early stages enables prompt maintenance or repair, minimizing the risk of expensive downtime. Consequently, this facilitates the optimization of maintenance schedules and contributes to extending the longevity of equipment and structures.}} During in-service use, {{effective implementation of structural integrity monitoring systems}} is mandatory as the bolts can self-loosen leading to potentially catastrophic failures. 

{{Non-destructive techniques offer valuable tools for monitoring the condition of bolted joints. We can distinguish between active and passive techniques. In the recent review of Huang et al. \cite{huang2022comprehensive}, the authors described active techniques, summarized in three categories: sensor-based \cite{Li2022Sensors}, percussion-based~\cite{doi:10.1177/14759217231182305,YangNonDes2022} and vision-based~\cite{sun2022vision}. In the sensor-based category, ultrasonic testing is widely used. The fundamental principle of this active technique consists in an actuator, that generates an ultrasonic signal, and a sensor that receives the response, and in an algorithm that evaluates a change during transmission. The properties of the signal transmitted through the joint interface is modified as the properties of the latter evolve during in-service use. For example, the transmitted energy or spectral features of signals can be correlated with the tightening degree of the bolts. A similar principle was used in various applications such as the monitoring of corroding bolted joints \cite{shah2019ultrasonic}, and the detection of composite delamination \cite{peng2014integrated}.}}

{{One of the causes of self-loosening is the succession of microscale damage related to sticking and slipping between the contact surfaces during vibratory cycles \cite{rafik2019experimental}. This form of damage eludes detection through active techniques designed to assess the overall and global behavior of bolted structures. However, such microscale damages dissipate energy by itself, which can be recorded by a passive technique called Acoustic Emission (AE) \cite{ferrer2010discrete,geng2019using,feng2020review}}}. AE is defined as the detection of the subnanometric displacements of a material surface caused by the propagation of transient elastic waves generated by a sudden and permanent change in the material integrity. The evanescent nature of these waves requires continuous data collection using piezoelectric (PZTs) sensors, {{with a typical frequency range between dozens of kHz to 1 MHz, that convert displacements into voltage signals.

The main advantage of this technique lies on its \textit{passive} nature. Indeed, the detection of a defect depends on the energy emitted by the defect's source itself, and not by an external apparatus as in active techniques. Moreover, it is highly sensitive and can detect microscale damages. This accounts for its use in many applications, like material characterization and Structural Health Monitoring (SHM), and has high potential for condition monitoring of bolts. However, very few works have been published on {the application targeted by this work.} This technique was not considered in the review of Huang et al. \cite{huang2022comprehensive} because it has only recently been shown that a quantitative relationship can be established between AE signals features and bolt loosening \cite{zhang2019continuous}. Independently of the application, the significant challenge behind the use of AE is the \textit{extraction of representative and robust features}, which is of utmost importance for condition monitoring. Therefore, this paper aims at addressing the following questions: what kind of features are the most relevant for tightening level classification? How these features generalize through several measurement campaigns?}}

{{A beginning of answer to these questions was proposed by Zhang et al. \cite{zhang2019continuous}.
The AE technique was used to evaluate the failure of a bolted joint between composite materials. The energy released from {{AE}} signals was used to evaluate the residual torque of the joints within a limited range and the authors shown a shift of peak frequencies in the sprectal representation of AE signals. The study provided a proof-of-concept of the use of the AE technique for condition monitoring of bolted joints. However, the authors used handcrafted features based on empirical mode decomposition and Hilbert-Huang transform (HHT) extracted from {{AE}} signals. They also considered a sensor with a limited frequency range (20 kHz-160 kHz). Further research is thus needed to improve the feature extraction process as it can be time-consuming and both application/sensor dependent. Further work is also needed to evaluate if sensors with higher frequency range can be more valuable. Finally, it is necessary to study the performance of machine learning algorithms using the proposed features for SHM purposes.}}

Deep Convolutional Neural Networks (CNNs) are promising candidates for addressing the challenge behind \textit{automatic feature extraction}. This type of \textit{end-to-end} approach is of practical interest because there is no need of data preprocessing since it is already embedded through the numerous layers. {{It is still a relatively new research area in condition monitoring of bolted joints. The recent study from Fu et al. \cite{fu2023automatic} demonstrated the potential of CNNs for classifying discretized tightening {torque} levels of bolts from highly noisy {{AE}} data and using high-frequency sensors.}} They considered a recent benchmarking dataset called ORION-AE \cite{orionaedata} that allow performance comparison of detection methods. In their study, all tightening conditions were known \textit{in advance} during training and testing, which is not compatible with SHM. 

By considering all conditions in a training dataset, a data \textit{normalization} is implicitly applied, whereby the network is able to learn the testing conditions. This is because normalization is embedded in most of pretrained CNNs, like $\operatorname{Resnet18}$ or $\operatorname{GoogLeNet}$, through \textit{batch normalization (BN) layers} \cite{ioffe2015batch}. During testing, this type of layer keeps these statistics fixed, therefore the output of a BN layer in testing is normalized according to the mean and standard deviation of the batches estimated during training. If the training dataset comprises all conditions, the testing is thus biased. However, if all conditions are \textit{not} known in advance and different from the conditions represented in training data, then these statistics can be less representative and the generalization of the CNN more difficult. The normalization is an important issue for SHM \cite{FarrarBook} and is often understated. 
{{While there were many studies on the application of deep learning and transfer learning based on CNN to vibration data \cite{chen2021acoustic,xin2020fracture,pandiyan2022deep}, the study of generalization capabilities was not tackled so far for AE-based condition monitoring bolted joints.}}

{{A partial answer to the aforementioned questions was given in previous work. For the question ``What kind of features are the most relevant for tightening level classification?", it was shown in several works that the use of CNN prevents spending time on finding handcrafted features and enables automatic feature extraction through convolutional layers. However, the question "How these features generalize?" remains open - and contributions are needed on this point because generalization is crucial for SHM purposes. To fill this gap, this article contributes to this area by comparing deep learning methods and exploring parameter settings, particularly for the learning rate, to achieve high performance in generalization within a limited training time.

More specifically, our work is particularly innovative with the following contributions:
\begin{description}
    \item \textbf{Study of generalization for SHM purpose of bolted joints}: We provide the first study of generalization capabilities of four pretrained networks for condition monitoring of bolts, illustrated on the ORION-AE benchmark dataset. This comprises analyzing how well these networks perform in classifying different tightening levels of bolts based on AE data using distinct campaigns of measurements.
    By studying generalization, this work assesses the robustness and adaptability of these networks to various conditions encountered in SHM applications for bolted joints.
    
    \item \textbf{Super-convergence to reduce iterations:} This objective aims to explore the phenomenon of super-convergence in the context of SHM applications for bolted joints. Super-convergence refers to a rapid decrease in the number of iterations required to train a neural network while still achieving high performance. By observing and analyzing this phenomenon, this work provides insights into efficient training strategies that can significantly reduce the computational burden associated with training deep learning models for SHM purposes. 
    
    \item \textbf{Multiple sensors fusion:} This strategy is investigated for condition monitoring of bolted joints, as well as the \textbf{impact of pre-denoising} of {{AE}} data stream before CNN training and inference. By investigating these aspects, this study enhances the accuracy and reliability of AE-based SHM systems for bolted joints. It achieves this improvement by leveraging complementary information from multiple sensors and optimizing data preprocessing techniques.
    
    \item \textbf{Several ordinal loss functions are investigated} and compared to the standard cross-entropy loss. This objective focuses on investigating the effectiveness of ordinal loss functions in the context of SHM applications for bolted joints using CNNs. Ordinal loss functions are designed to take into account the order of classes, which is particularly relevant in SHM scenarios where the severity or level of damage may vary. By comparing ordinal loss functions with the standard cross-entropy loss, we aim to identify the most suitable loss function for accurately classifying different tightening levels of bolts based on AE data.
\end{description}
}}

Toward this background, this paper is structured as follows: Section ``\nameref{sec:Benchmark Description}" presents the ORION-AE benchmark, along with the experimental setup and the measurement protocol used to data acquisition. It also discusses the challenges encountered in performing SHM on bolted structures. Section ``\nameref{sec: Method}", in turn, describes the developed methodology, including the different modules employed in the workflow. Section ``\nameref{sec:NO-SHM case}" introduces the results obtained in this work for a so-called NO-SHM case, in which training, validation and test data come from the same experimental campaign. Therefore, section ``\nameref{sec:SHM case}" expands the results shown to a so-called SHM case, in which the generalization is of utmost importance, and the campaigns used to test the algorithm are not used at all during the training and validation stages.  In addition, the influence of adding \textit{a priori} knowledge about the testing campaign on the classification accuracy of the CNNs is also discussed. Next, ``\nameref{sec: Super convergence}" discusses, for the first time in the context of SHM on bolted joints, the super-convergence phenomenon. Finally, section ``\nameref{sec: Conclusions}" presents concluding remarks and outlines the next steps for further research.

\section{Benchmark Description: ORION-AE Dataset}
\label{sec:Benchmark Description}

\subsection{Experimental setup}
\begin{figure*}[htbp]
\centering
\includegraphics[width=0.7\linewidth]{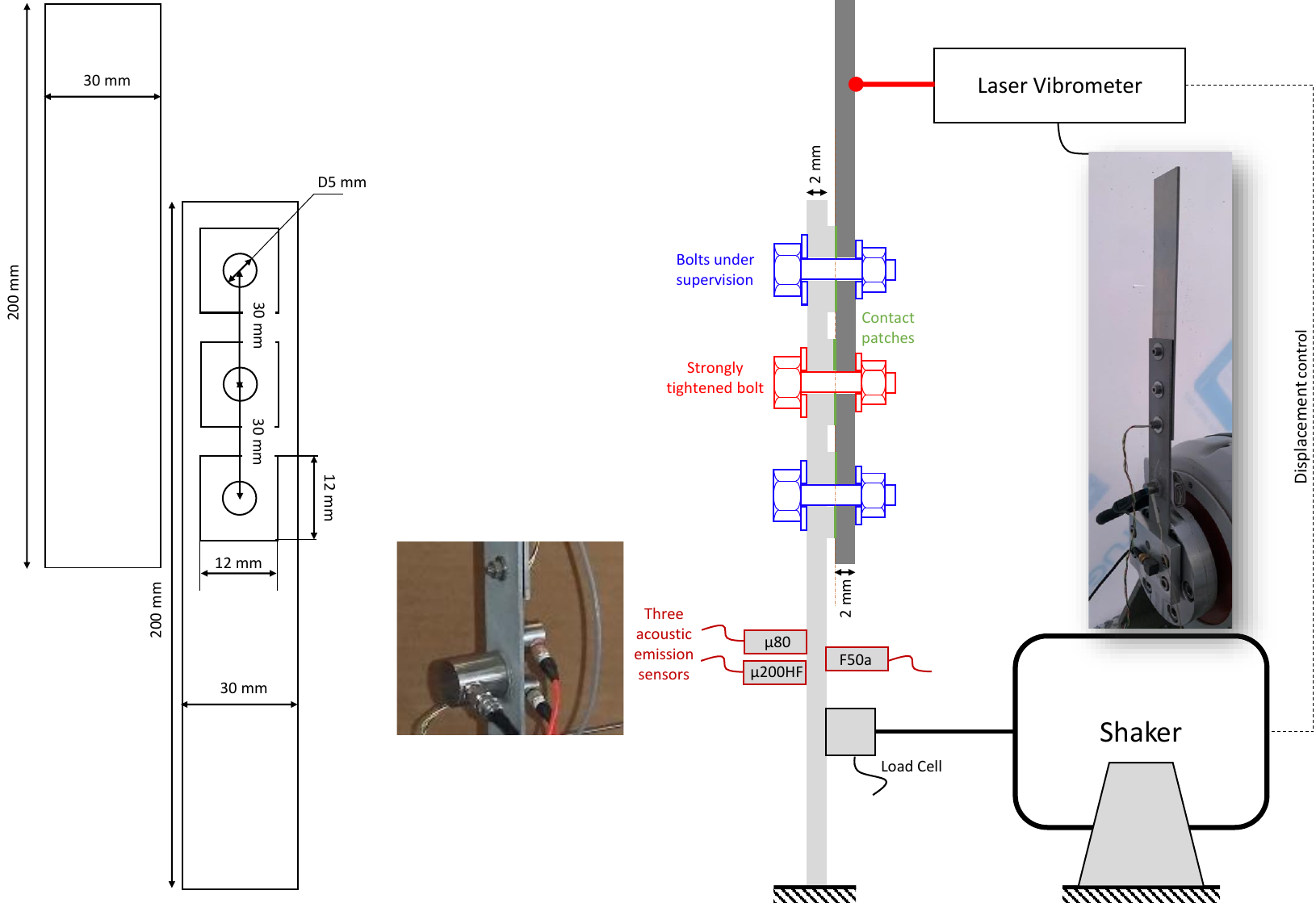}
\caption{ORION-AE benchmark and setup configuration \cite{ORIONdata}. \label{fig:setup}}
\end{figure*}

The experimental setup is illustrated in Figure \ref{fig:setup}. The structure used in the work, known as the Orion beam \cite{TELOLI2021107627, TELOLI2022108172}, comprises two duraluminium beams, each measuring $200 \times 30 \times 2$ mm. These beams are connected by three M4 bolts, spaced along a length of $30$ mm. There are contact patches at each bolt connection to retain the contact between both beams in a small area. These patches consist of a square of 12 $\times$ 12 mm$^2$ with an extra thickness of 1 mm. The ORION-AE benchmark dataset, available at Verdin et al.\cite{orionaedata} and described in Ramasso et al.\cite{ORIONdata}, encompasses data collected from a structural monitoring campaign using various sensors. The experimental setup includes a TIRA TV51120 permanent magnetic shaker, which applies a sinusoidal signal with a frequency of 120 Hz to excite the structure. A Polytec PSV500Xtra laser vibrometer is utilized to measure the velocity of the structure near its free end. Additionally, three {{AE}} sensors, namely $\operatorname{mu80}$, $\operatorname{F50A}$, and $\operatorname{mu200HF}$ manufactured by EuroPhysical Acoustics, are employed. These sensors are connected to a preamplifier set to $60$ dB (model 2/4/6 preamplifier made by EuroPhysical Acoustics) and possess frequency ranges of $200-900$ kHz, $200-800$ kHz and $500-4500$ kHz, respectively. The sensors are positioned on the lower beam, approximately $50$ mm above the clamped-end, using silicone grease. All data are sampled at a rate of $5$ MHz using a Picoscope 4824, which offers a bandwidth of $20$ MHz, low noise, 12-bit resolution, and a $256$ MS buffer memory. Note that the data collection is performed using a dedicated setup designed to replicate the loosening phenomenon observed in diverse industries, including aeronautics, automotive, and civil engineering, where bolted joints are commonly utilized for assembly. 

The dataset consists of five measurement campaigns ($\#B$, $\#C$, $\#D$, $\#E$ and $\#F$), each involving vibration tests conducted considering different tightening levels. It is important to stress that each campaign started after the structure had been completely disassembled and reassembled. By conducting multiple experimental campaigns, available data allow to assess the robustness of the detection and classification methodology in the presence of variations in the dynamics of the bolted joint. This variability provides valuable insights into the performance and generalization capacity of the proposed approach under different operating conditions and environmental factors. 

The upper bolt of the structure was subjected to a sequential decrease in tightening torque, while the other bolts remained fully tightened at a constant torque of $80$ cNm. The tightening levels are applied in the following order: $60$ cNm, $50$ cNm, $40$ cNm, $30$ cNm, $20$ cNm, $10$ cNm, and finally $5$ cNm. During the vibration tests, the torque modification process differed slightly for the tightening torque of $50$ cNm. In this case, the shaker was momentarily stopped before the torque modification and then resumed afterward. For the remaining tightening torque conditions, the modification was made sequentially. Each tightening level corresponds to a specific class, ranging from $1$ for $60$ cNm to $7$ for $5$ cNm. Notwithstanding, in each campaign, the displacement at the free-end of the beam was controlled through a feedback strategy from the laser vibrometer signal. The measure of the displacement was also recorded at 5 MHz. This displacement data is used to determine the vibration cycles applied during the experiments. 

In recent studies, the detection of torque loss in the bolts of the Orion beam has been investigated \cite{doi:10.1177/14759217211054150, 10.1115/1.4063794}. However, these studies primarily rely on modal parameters as features, which are extracted from vibration responses with low-frequency content. Although based on physical parameters, this approach has certain limitations, as the need for data preprocessing and subsequent modal analysis. Therefore, the quality of feature extraction is highly dependent on data and the specific processing techniques employed. Additionally, resonance frequencies, being global characteristics of dynamic systems, exhibit limited sensitivity to variations when compared to measurement uncertainties, particularly at higher levels of applied torque - the effects of microslip arising from relative motion between surfaces are not significant in this context. This work aims to overcome these limitations by utilizing {{AE}} data, which possess a rich high-frequency content and can be directly employed in deep neural networks without the need for extensive preprocessing or feature extraction.

\subsection{Challenges - Heterogeneous system}
\label{subsec: Challenges}
Gardner et al. \cite{gardner2021foundations} proposed a general framework for population-based SHM, discussing aspects that allow systems to be classified as homogeneous (e.g., similar geometry) or heterogeneous (e.g., significant difference in material properties). However, the challenge with generic frameworks lies in their inability to represent less typical cases. 

Joints are a source of epistemic uncertainties, which mainly appear from shape imperfections (non-flatness of surfaces), dynamic clearance, and pressure distribution over the contact area, they are directly related to mesoscale and microscale parameters in the joint design. Although after each assembly the structural condition is the same, i.e., the overall picture of the test-bench is maintained - which could be seen as a homogeneous system, pressure variations in the contact area caused by the tightening torque, or even by the order in which the bolts are tightened, can lead to significant changes in the vibratory behavior of the system \cite{BRAKE2019282, TELOLI2022108172}. One can also mention the occurrence of wear between the surfaces due to microslip, which is influenced by the surface roughness and, when accumulated, also results in global changes to the joint dynamics \cite{ZHANG201830, LI2020203411}. Note that all these microscale effects can directly influence {{AE}} measurements as these sensors, sensitive not only to noise but also to changes in material properties, generate an enormous amount of data.

\begin{figure*}[htbp]
\centering
\includegraphics[width=0.7\linewidth]{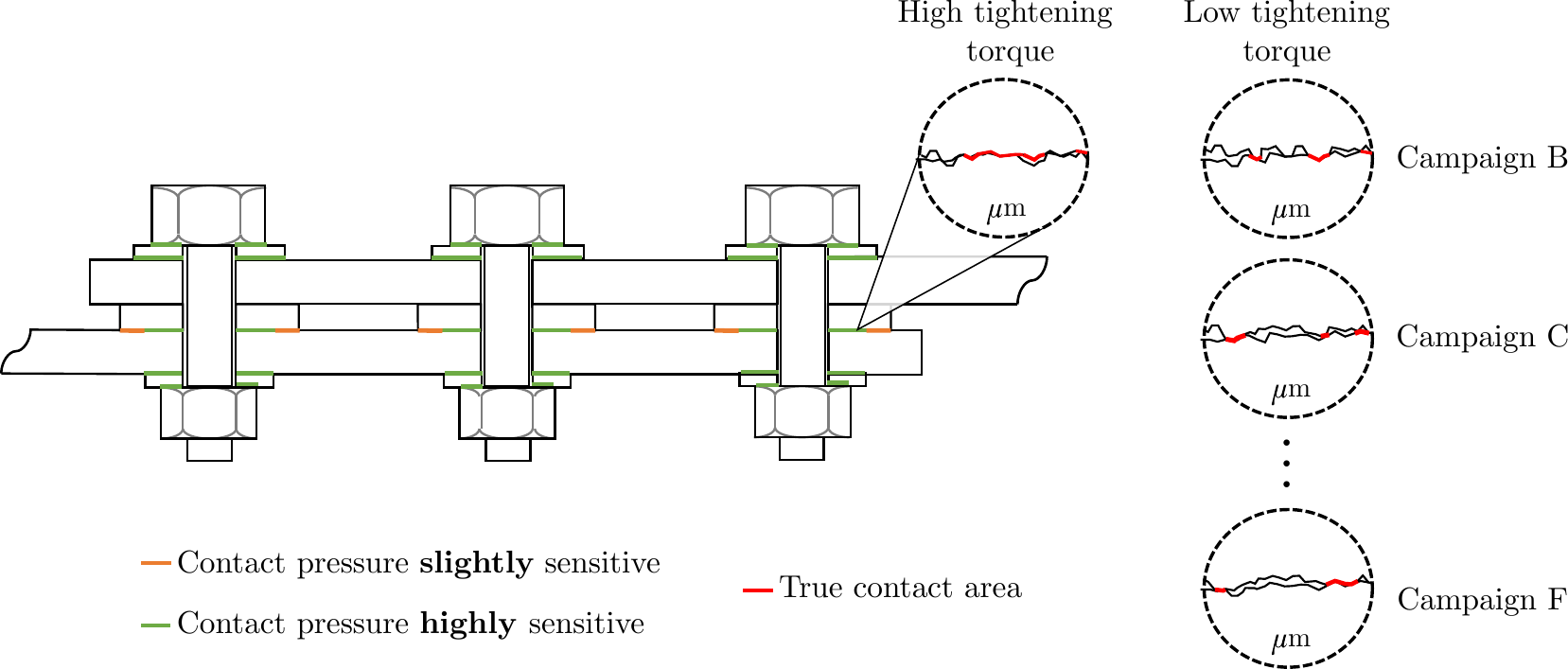}
\caption{Diagram depicting contact conditions of the Orion beam at interface, and a microscale contact area illustration which highlights the contact state at each campaign.  \label{fig:schematic}}
\end{figure*}

Figure \ref{fig:schematic} presents a schematic representation of the Orion beam illustrating, from the design perspective of the structure, zones that are more and less sensitive to contact pressure. Areas where the contact pressure is sensitive to shape imperfections, especially at edges far from the bolt hole, may exhibit wear as vibration tests are run and also show lower repeatability between tests. In general, normal contact areas are formed by micro-contacts randomly distributed over the surface, and thus, they are hardly arranged in the same configuration after complete disassembly \cite{stachowiak2013engineering}. This fact is illustrated by the schematic representation (see top right of Figure \ref{fig:schematic}) considering different campaigns, in such a way that, when there is a reduction of the tightening torque, repeatability of tests between experimental campaigns is difficult to guarantee, since the structure is disassembled and reassembled each time. 

The scale of analysis is thus important. From this perspective, micro effects have a non-negligible impact on {{AE}} patterns. Clusters of features have minimal intersection between campaigns on the ORION-AE benchmark \cite{ramasso2022clustering}. As mentioned by Bull et al. \cite{bull2021foundations}, systems in a population can be deemed heterogeneous if features differ while the systems are close. Therefore, it is argued that a new system is analyzed at each experimental campaign, hence the difficulty of developing SHM techniques for bolted joints, due to their heterogeneous character (although they are homogeneous systems from the geometrical perspective).

%


\section{Methodology}
\label{sec: Method}

\begin{figure*}[htbp]
\centering
\includegraphics[width=0.8\linewidth]{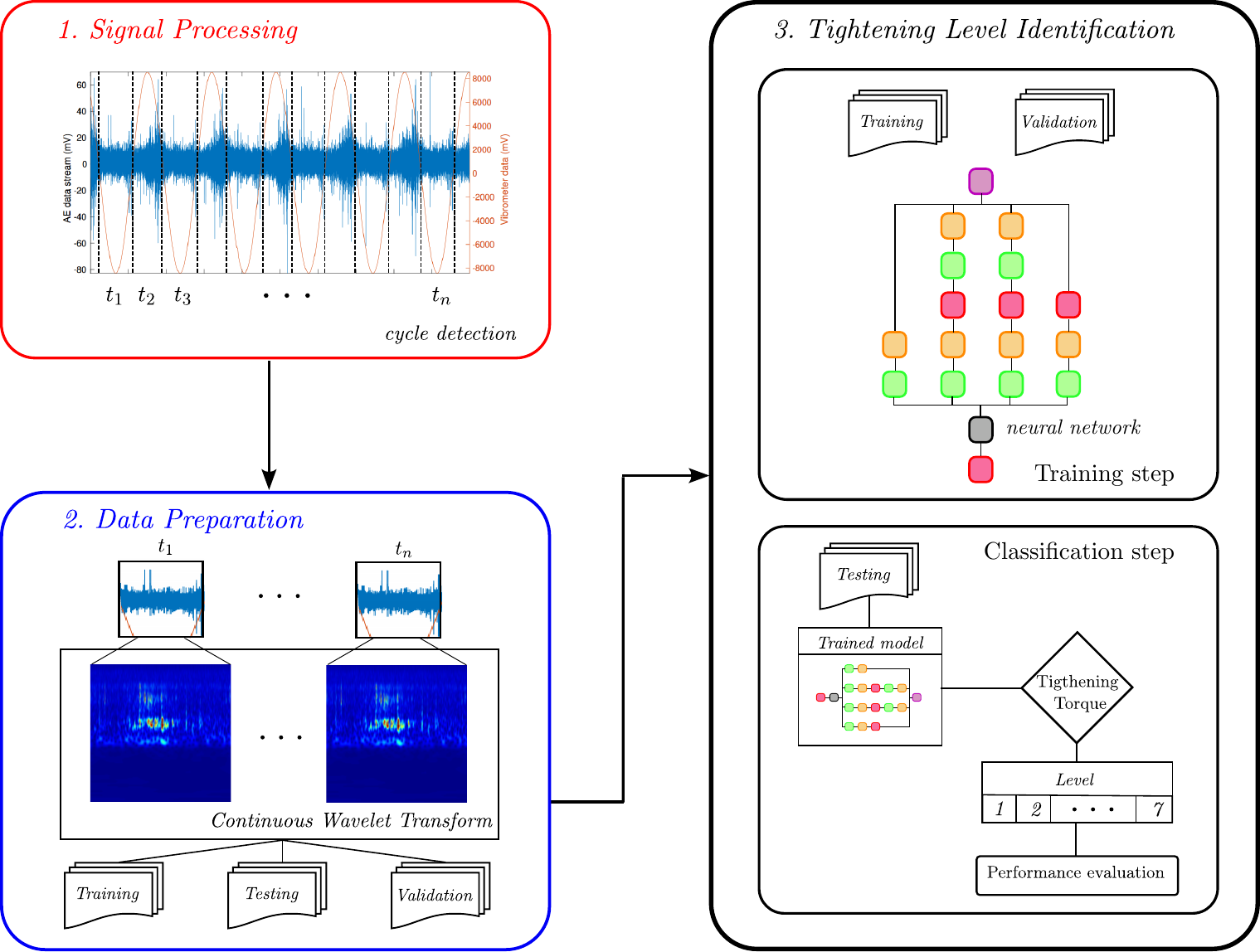}
\caption{Workflow containing three different modules: (1) Signal Processing, (2) Data Preparation and (3) Tightening Level Identification. \label{fig:workflow}}
\end{figure*}

The proposed methodology consists of three main modules, illustrated in Figure \ref{fig:workflow}, namely \textit{Signal processing}, \textit{Data preparation} and \textit{Tightening level identification}. 

\begin{figure*}[htbp]
\centering
\includegraphics[width=0.8\linewidth]{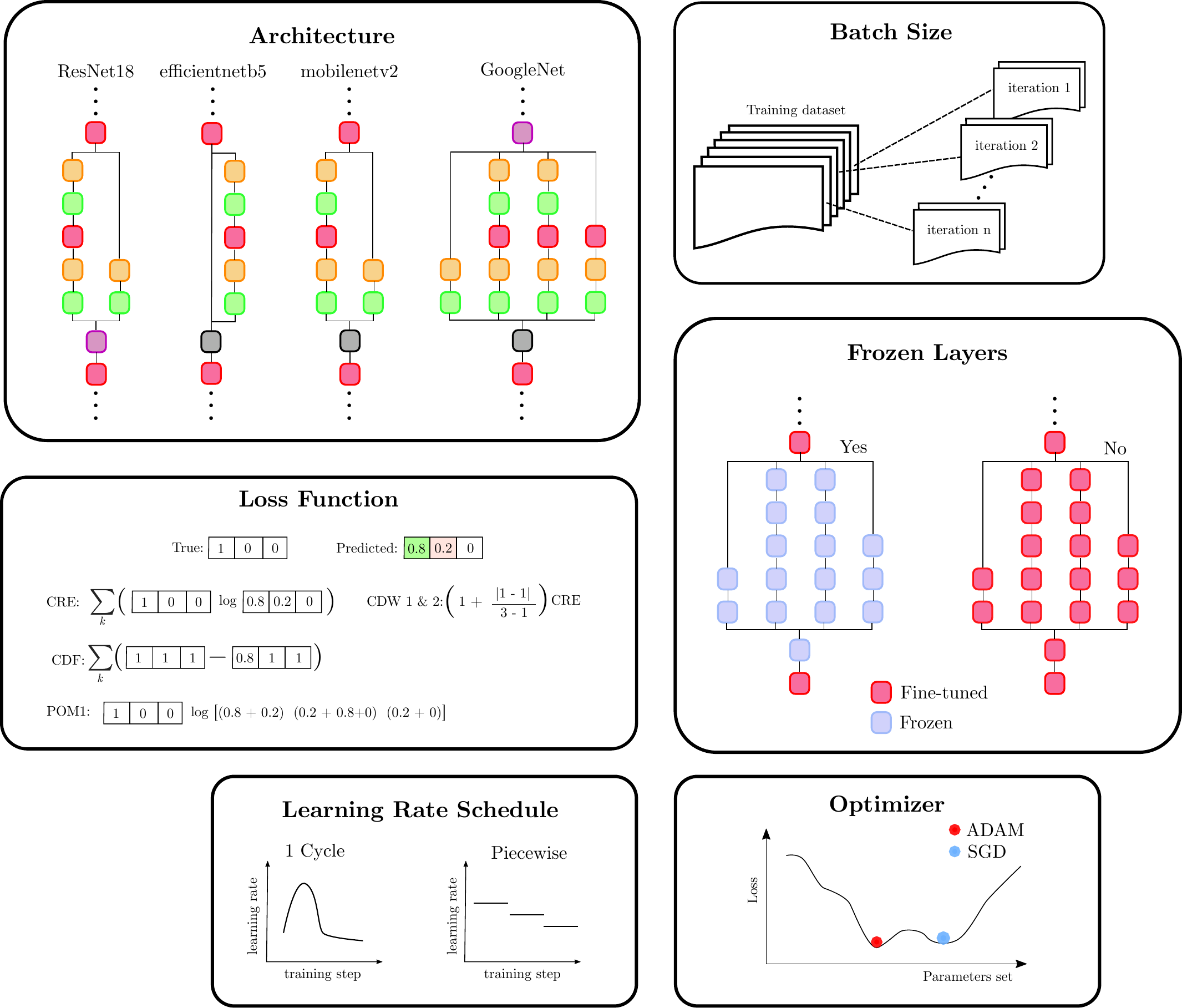}
\caption{Schematic representation of the studies carried out during the damage detection stage, considering different network architectures, batch sizes, loss functions, fine-tuned or frozen layers, learning rate scheduler and optimizer. In this figure, SGD stands for Stochastic Gradient Descent.\label{fig:schema}}
\end{figure*}

\subsection{Signal processing module} 
\label{sec:step1}

\begin{figure}
    \centering    
    \begin{subfigure}{0.39\textwidth}
        \includegraphics[width=1\linewidth]{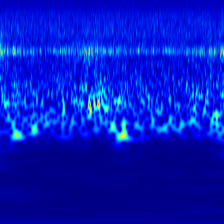}
        \caption{Without tapering.}
        \label{fig:05cnm}
    \end{subfigure}
    \hfill
    \begin{subfigure}{0.39\textwidth}
        \centering
        \includegraphics[width=1\linewidth]{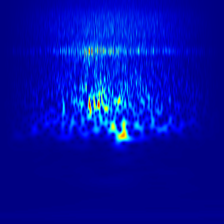}
        \caption{With tapering.}
        \label{fig:60cnm}    
    \end{subfigure}        
    \caption{Effect of apodization with a Hanning window on the CWT.}
    \label{fig:hanningcwt}
\end{figure}

Given an AE data stream, the objective is to estimate the level of tightening continuously. 
Classically, an AE data stream is decomposed into blocks before processing. Blocks can be obtained by different ways: hit detection procedure \cite{Pomponi2015110}, sliding window \cite{fu2023automatic}, or by exploiting additional sensors as in the present application. The vibration cycles obtained from the laser vibrometer signal are used to segment the AE data stream into blocks (top-left of Figure~\ref{fig:workflow}). More precisely, a zero-crossing detector is applied to detect the changes, from positive to negative, in the evolution of the vibrometer signal. This leads to $N$ raw AE signals, with about $1200$ number signals per tightening level (the number of cycles corresponds to the excitation frequency of $120$ Hz during $10$ seconds). Due to the cyclic nature of the experiment, the signals obtained are multiplied by a Hanning window before the next step. {{We can indeed observe that AE events in a given period generates side effects that can be retrieved in the next period as illustrated in Figure \ref{fig:hanningcwt}. Thus, the window acts as a weighting operation that diminishes the influence of data at the signal's edges. The benefit is to focus the analysis on the central portions of cycles.}}

\begin{figure*}
    \centering    
    \begin{subfigure}{0.49\textwidth}
        \includegraphics[width=1\linewidth]{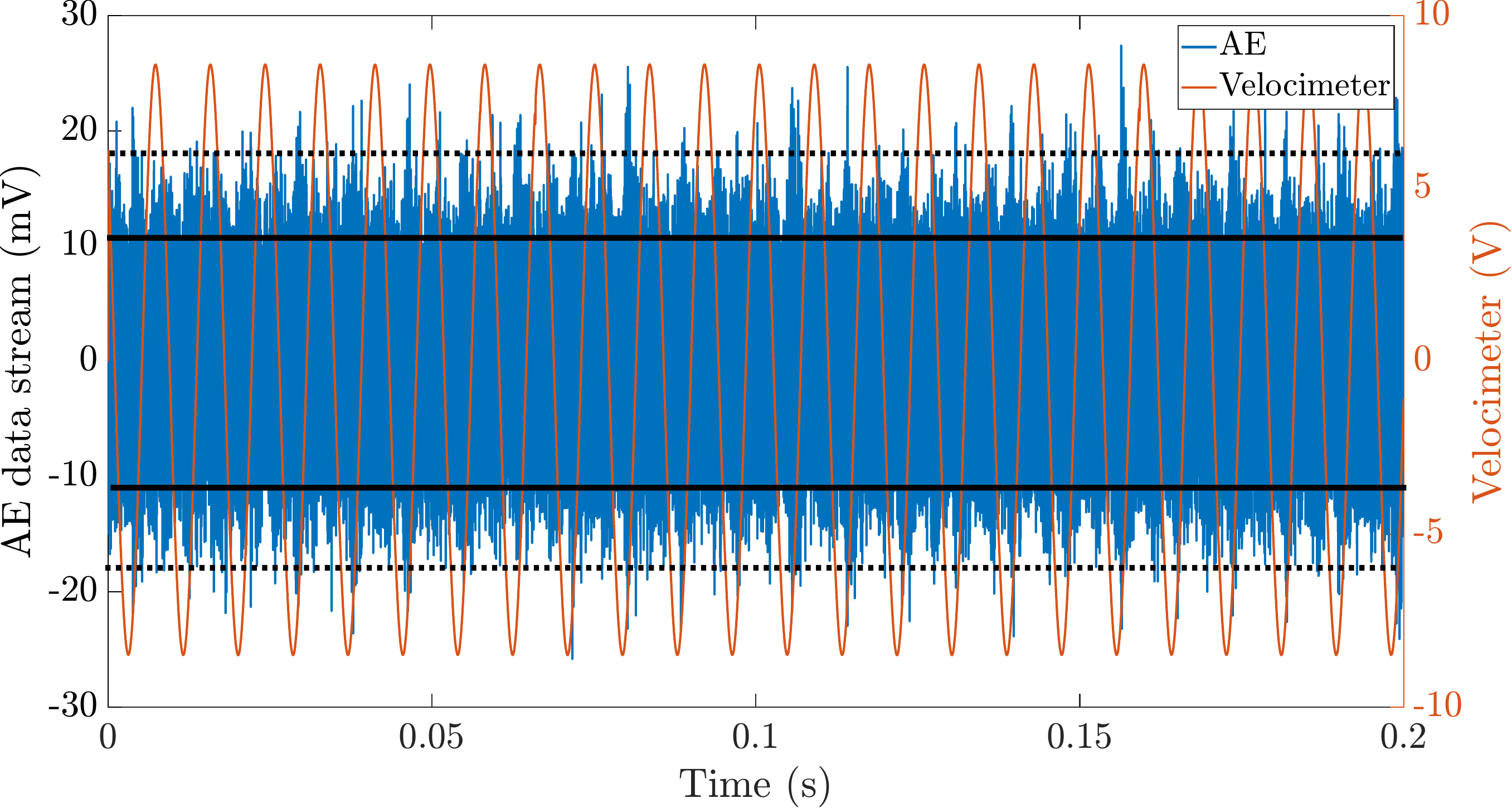}
        \caption{For 05cNm.}
        \label{fig:05cnm}
    \end{subfigure}
    \hfill
    \begin{subfigure}{0.49\textwidth}
        \centering
        \includegraphics[width=1\linewidth]{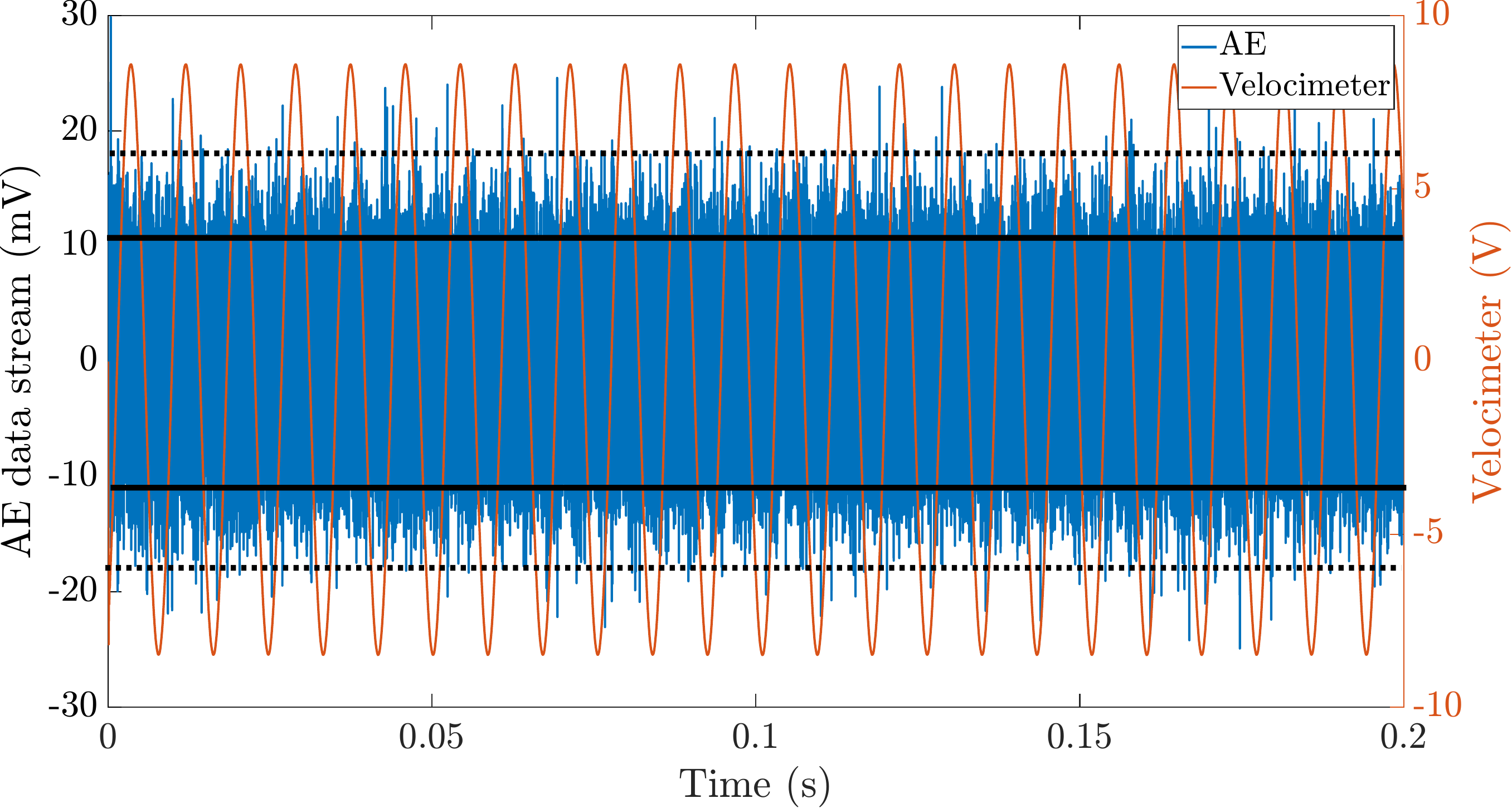}
        \caption{For 60cNm.}
        \label{fig:60cnm}
    \end{subfigure}
\caption{AE raw data stream to illustrate the unfavourable signal-to-noise ratio for two levels of tightening: a) for 05cNm; b) for 60 cNm. Continue and dashed horizontal lines represent the approximate range of noise and signal, respectively.}
    \label{fig:snr}
\end{figure*}

{{Samples of AE data stream are shown in Figure \ref{fig:snr}, which illustrates the amount of noise and the complexity associated. 
Acoustic emission data are particularly noisy for bolts looseness detection because the damages occur at the microscale (see Figure \ref{fig:schematic}), whereas the structure is under vibration with a relatively large amplitude. An illustration of the amount of noise is given in Figure \ref{fig:snr}, where bold and dotted horizontal lines represent the approximate range of noise and signal, respectively. By considering peak-to-peak values for the noise and signals, the signal-to-noise ratio (SNR) is approximately 
$$\text{SNR} \approx 10\log_{10}\frac{36}{21}\approx 2.3\text{dB}.$$ The SNR is highly unfavorable, where only peaks can be distinguished while signal onsets cannot be accurately estimated from the raw data. For comparison, typical AE applications utilize a detection threshold of 40 dB or higher, which significantly exceeds the threshold in our case.}}

\subsection{Data preparation module}
\label{sec:step2}

Data preparation (bottom-left of Figure~\ref{fig:workflow}) follows with a transformation of the AE signals into images to feed the CNN. This module considers the Continuous Wavelet Transform (CWT) using a filter bank, based on the analytic Morse $(3,60)$ wavelet (localized in frequency) with scales discretized by $12$ filters. It is possible to impose frequency limits in the filter bank, for example by using the bandwidth of the sensors, but this work uses directly the raw signals (which makes the procedure simpler). The result of this step is a set of $N$ RGB images with size $224\times 224\times 3$. For each campaign of measurements, about $\sim 8400$ images are generated ($1200$ per level), leading to a total of about $42000$ images per sensor and for all campaigns.

\begin{figure*}[htbp]
    \centering
    \begin{subfigure}[b]{0.33\textwidth}
        \includegraphics[width=\textwidth]{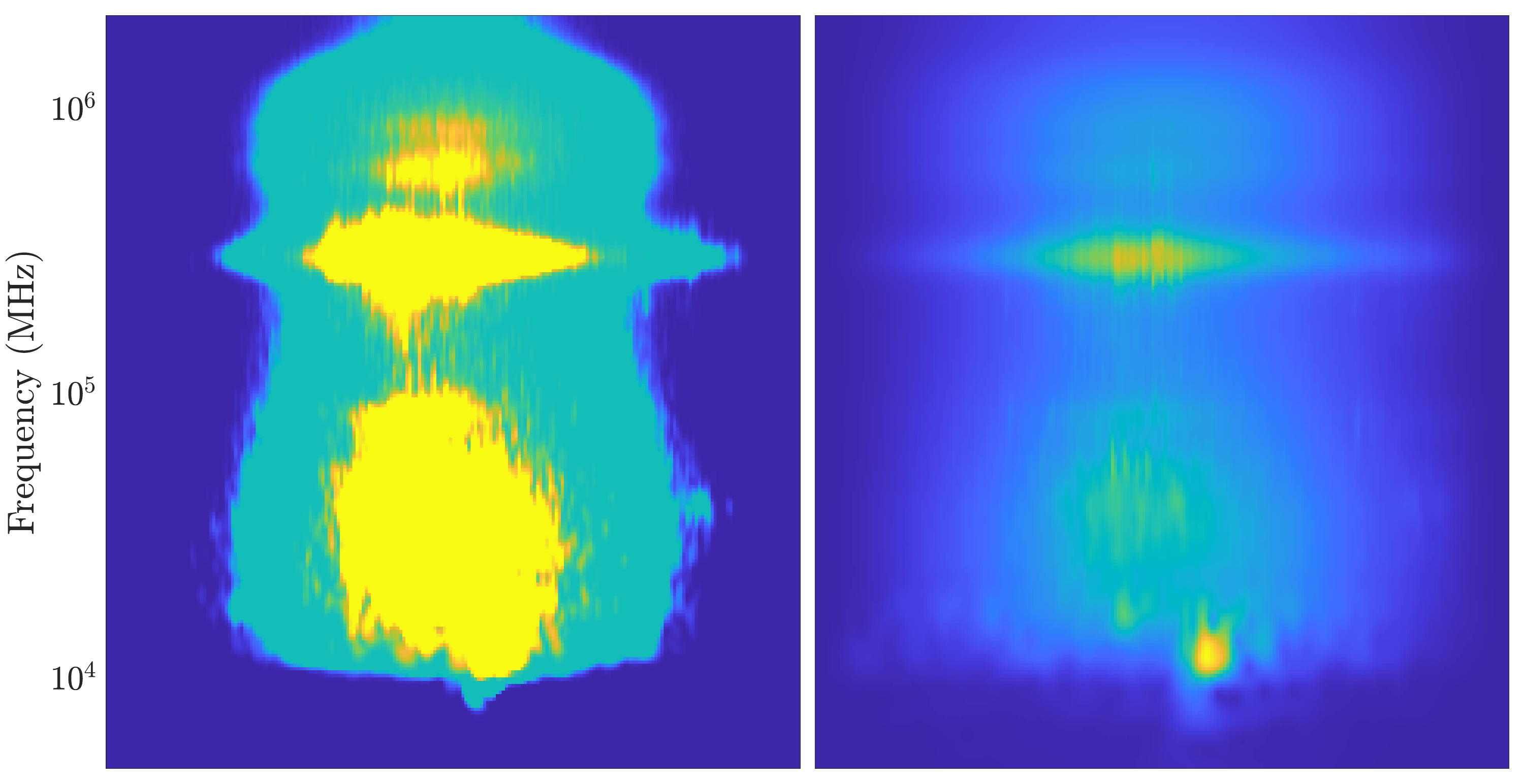}
        \caption{Sensor $\operatorname{mu80}$, $5$cNm}
    \end{subfigure}
    \begin{subfigure}[b]{0.33\textwidth}
        \includegraphics[width=\textwidth]{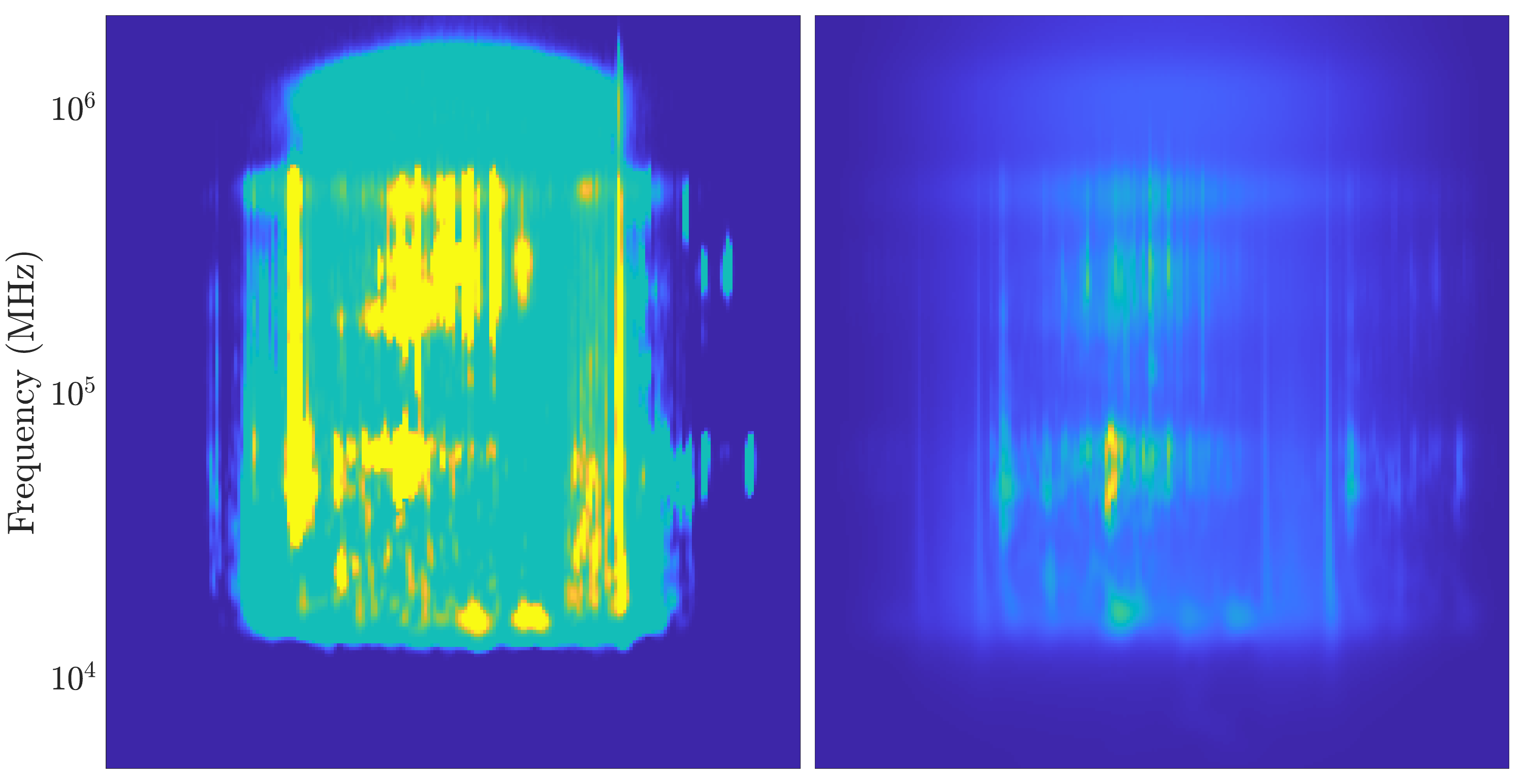}
        \caption{Sensor $\operatorname{F50A}$, $5$cNm}
    \end{subfigure}
    \begin{subfigure}[b]{0.33\textwidth}
        \includegraphics[width=\textwidth]{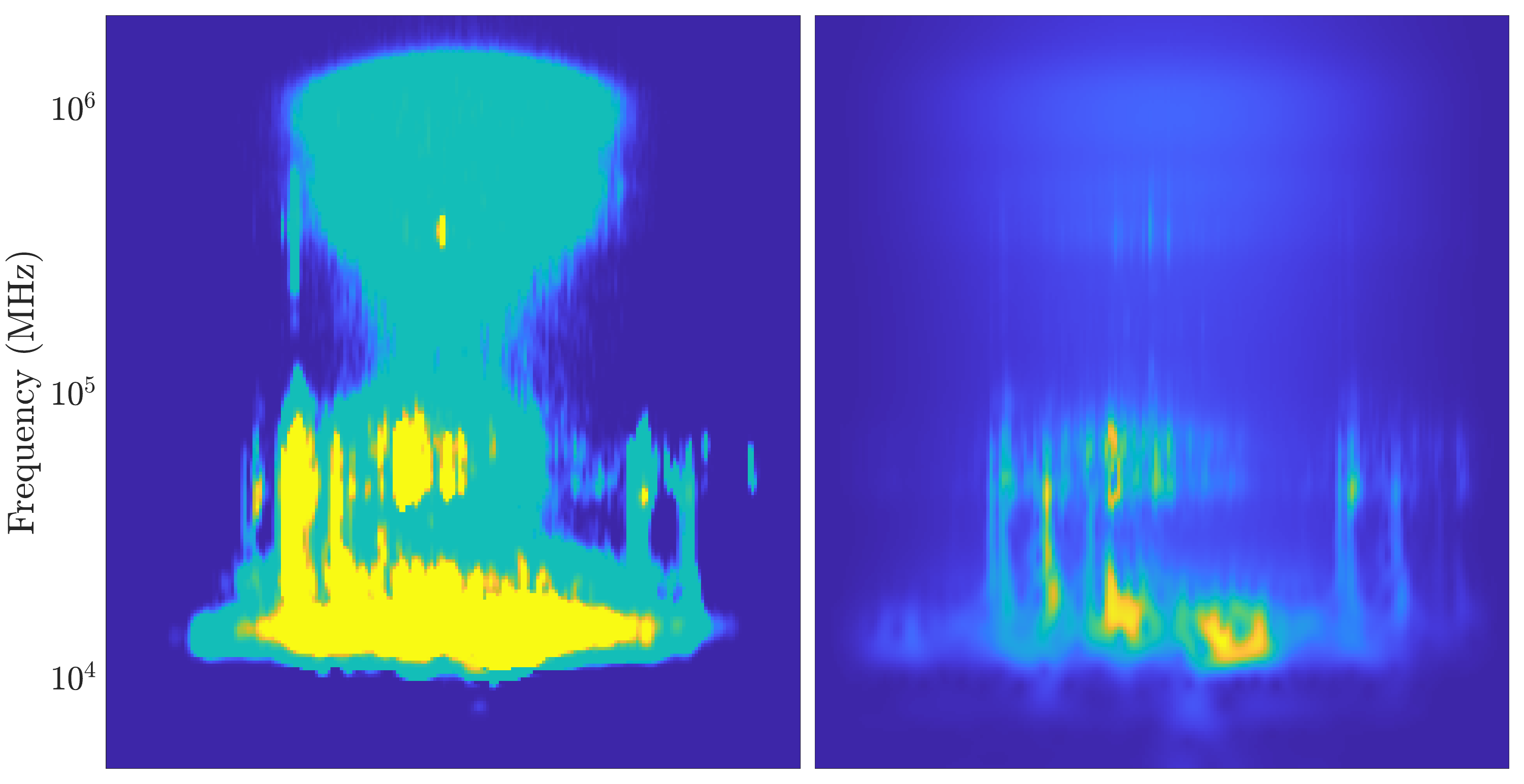}
        \caption{Sensor $\operatorname{mu200HF}$, $5$cNm}
    \end{subfigure}    
    \begin{subfigure}[b]{0.33\textwidth}
        \includegraphics[width=\textwidth]{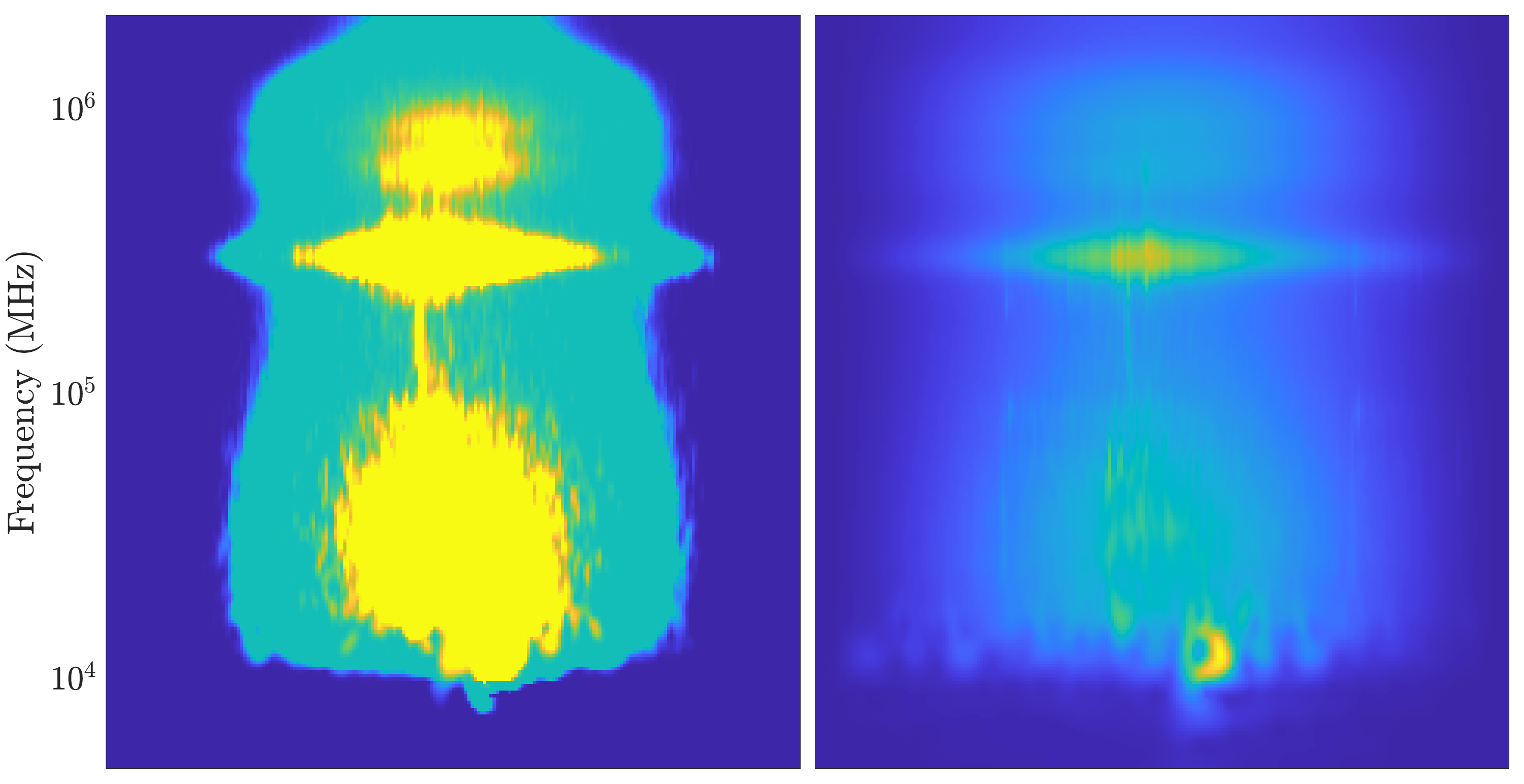}
        \caption{Sensor $\operatorname{mu80}$, $10$cNm}
    \end{subfigure}
    \begin{subfigure}[b]{0.33\textwidth}
        \includegraphics[width=\textwidth]{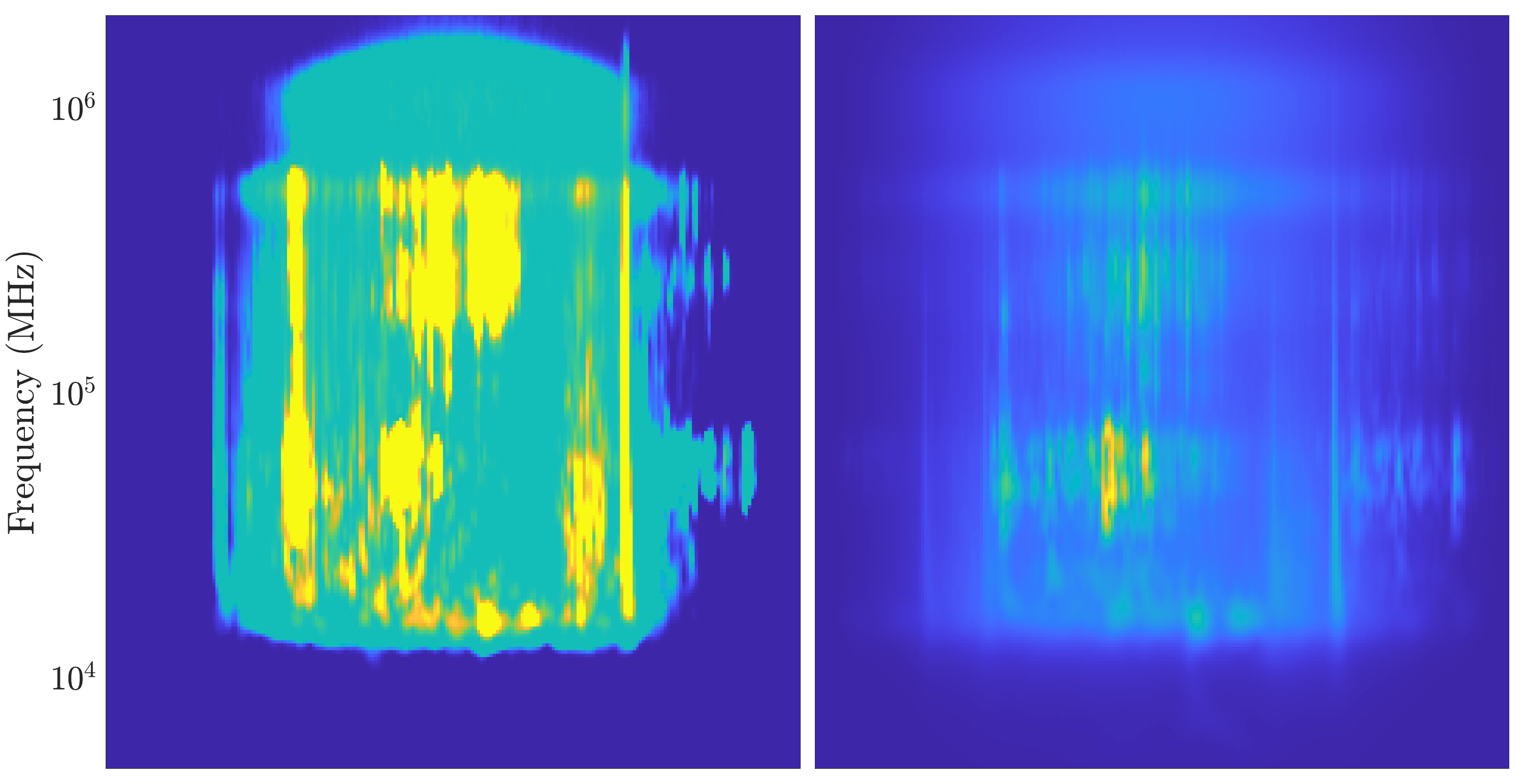}
        \caption{Sensor $\operatorname{F50A}$, $10$cNm}
    \end{subfigure}
    \begin{subfigure}[b]{0.33\textwidth}
        \includegraphics[width=\textwidth]{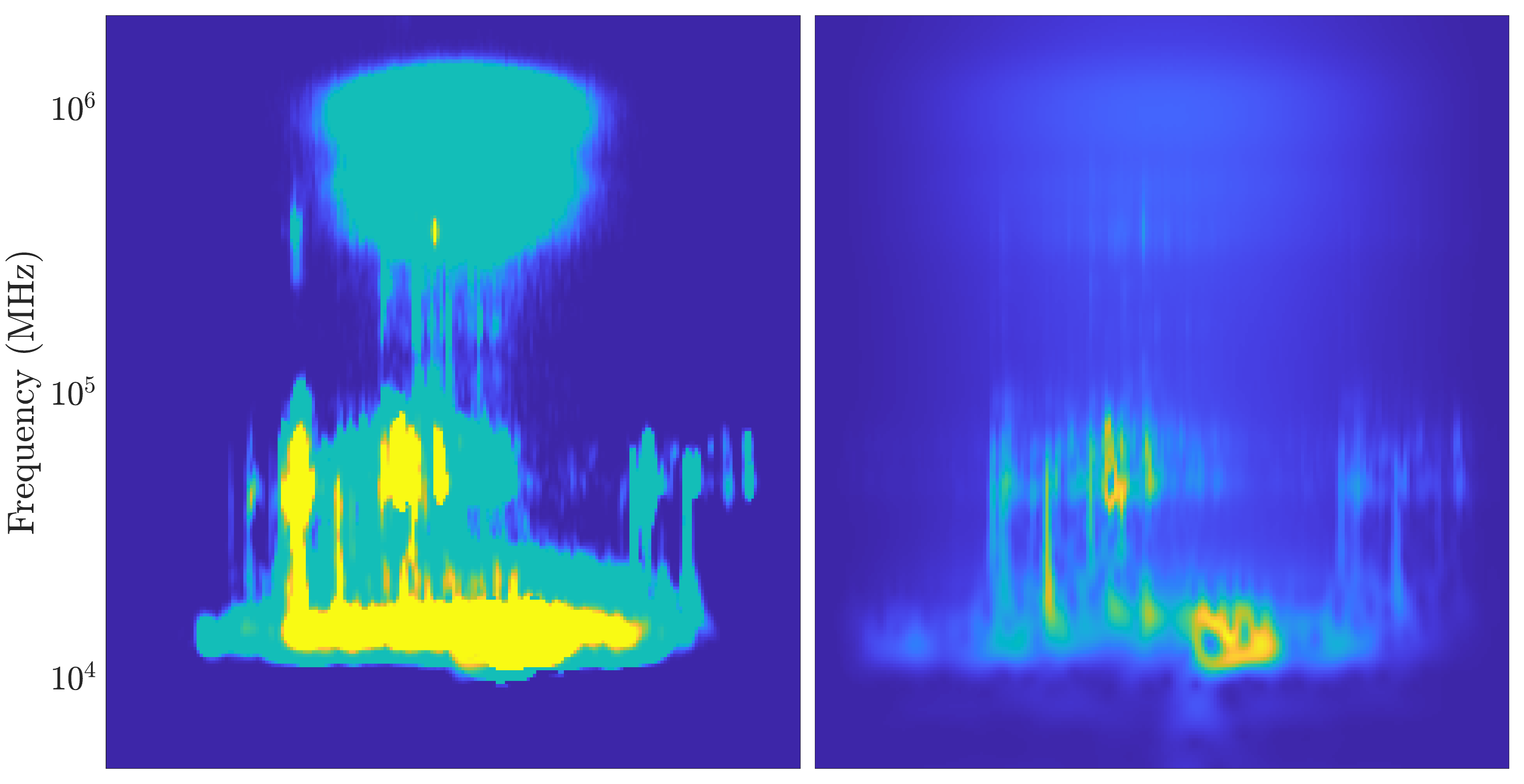}
        \caption{Sensor $\operatorname{mu200HF}$, $10$cNm}
    \end{subfigure}
    \begin{subfigure}[b]{0.33\textwidth}
        \includegraphics[width=\textwidth]{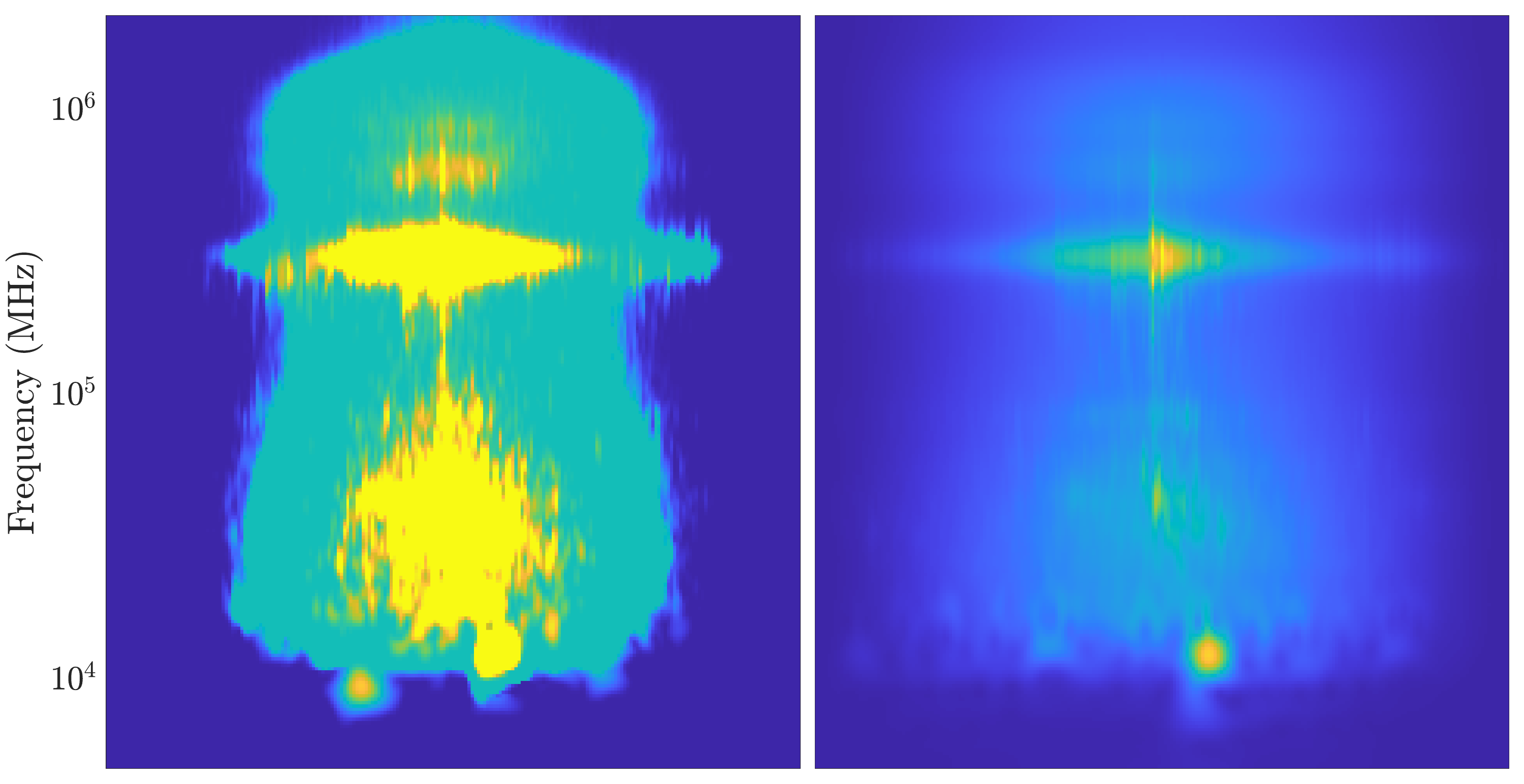}
        \caption{Sensor $\operatorname{mu80}$, $20$cNm}
    \end{subfigure}
    \begin{subfigure}[b]{0.33\textwidth}
        \includegraphics[width=\textwidth]{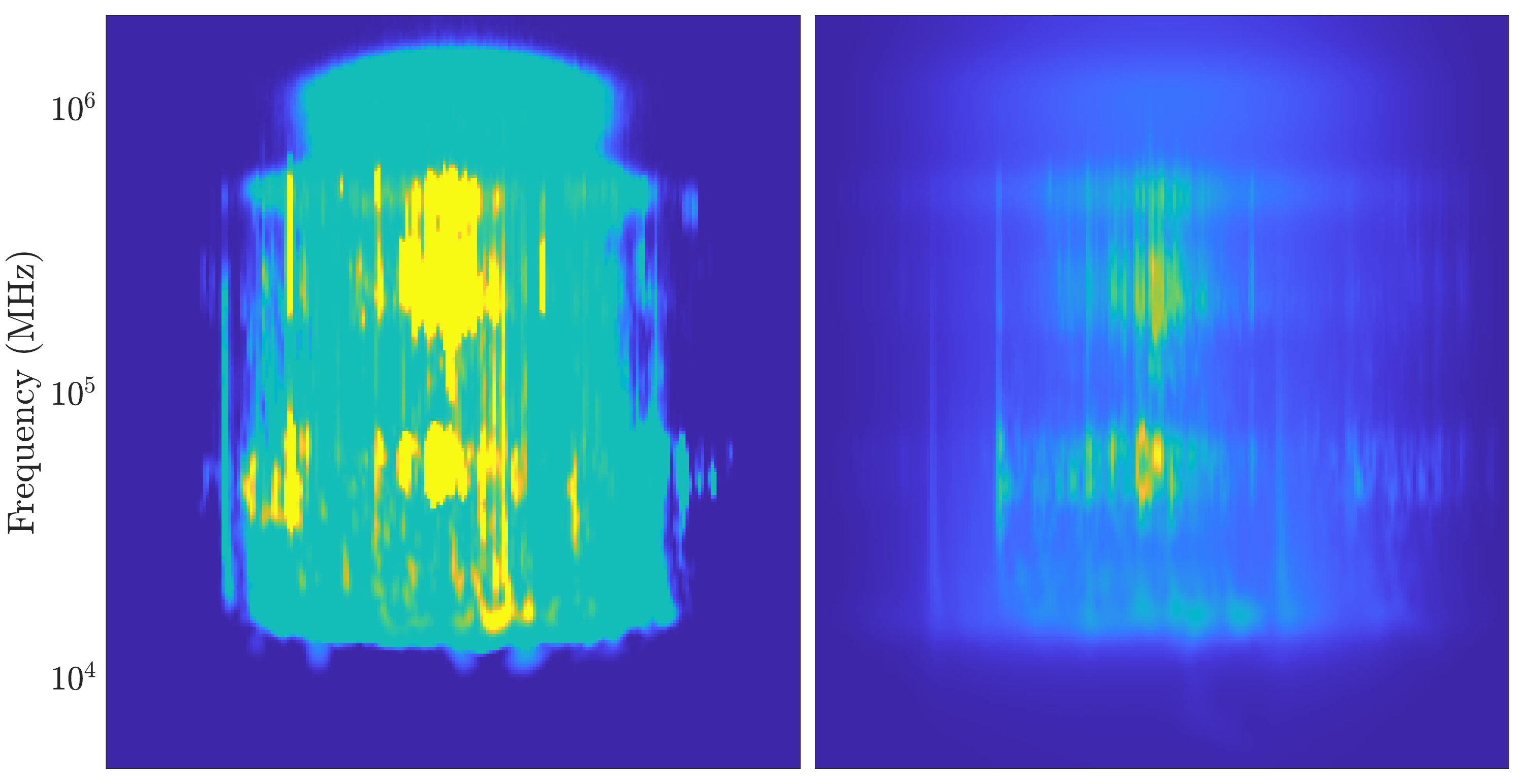}
        \caption{Sensor $\operatorname{F50A}$, $20$cNm}
    \end{subfigure}
    \begin{subfigure}[b]{0.33\textwidth}
        \includegraphics[width=\textwidth]{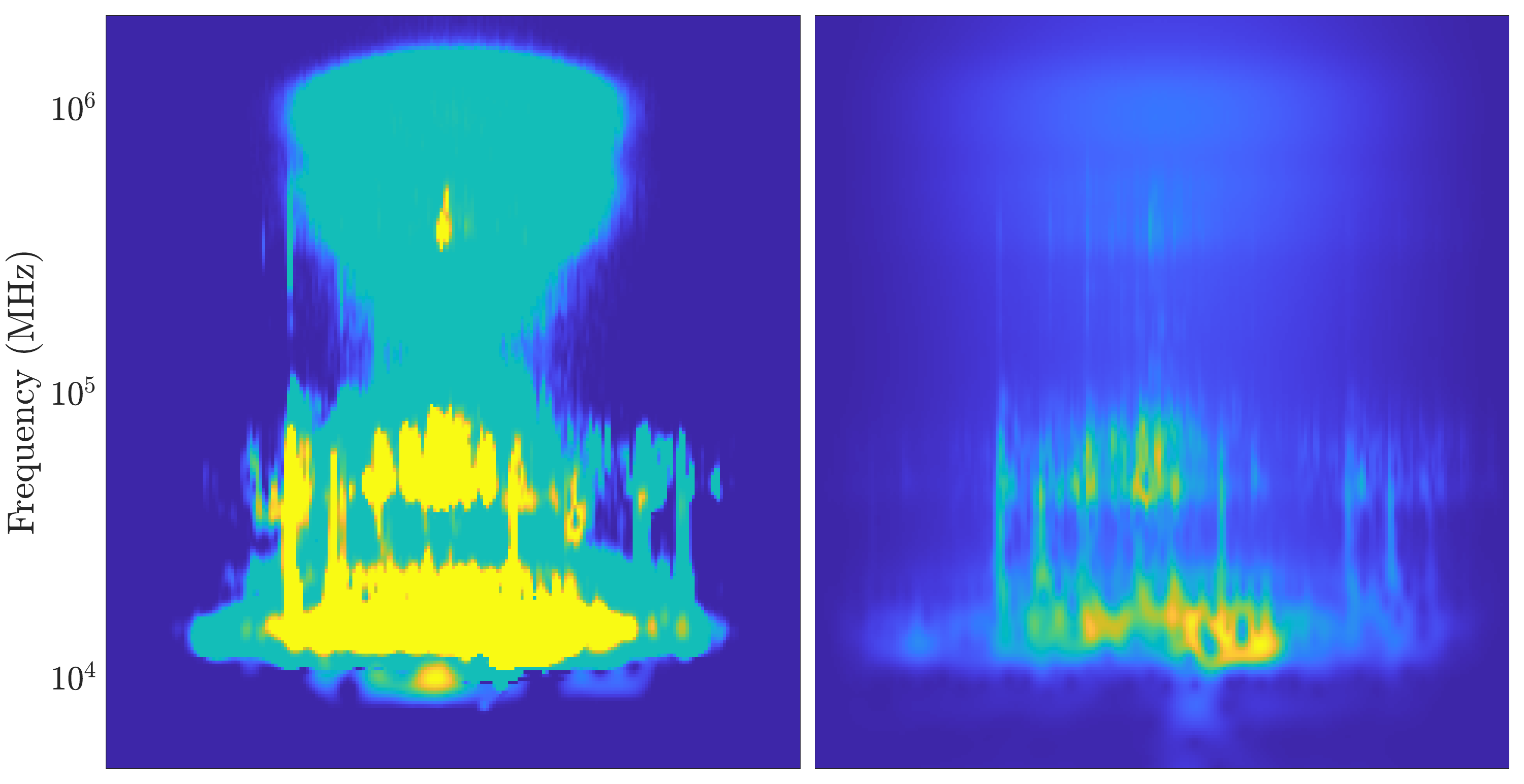}
        \caption{Sensor $\operatorname{mu200HF}$, $20$cNm}
    \end{subfigure}
    \begin{subfigure}[b]{0.33\textwidth}
        \includegraphics[width=\textwidth]{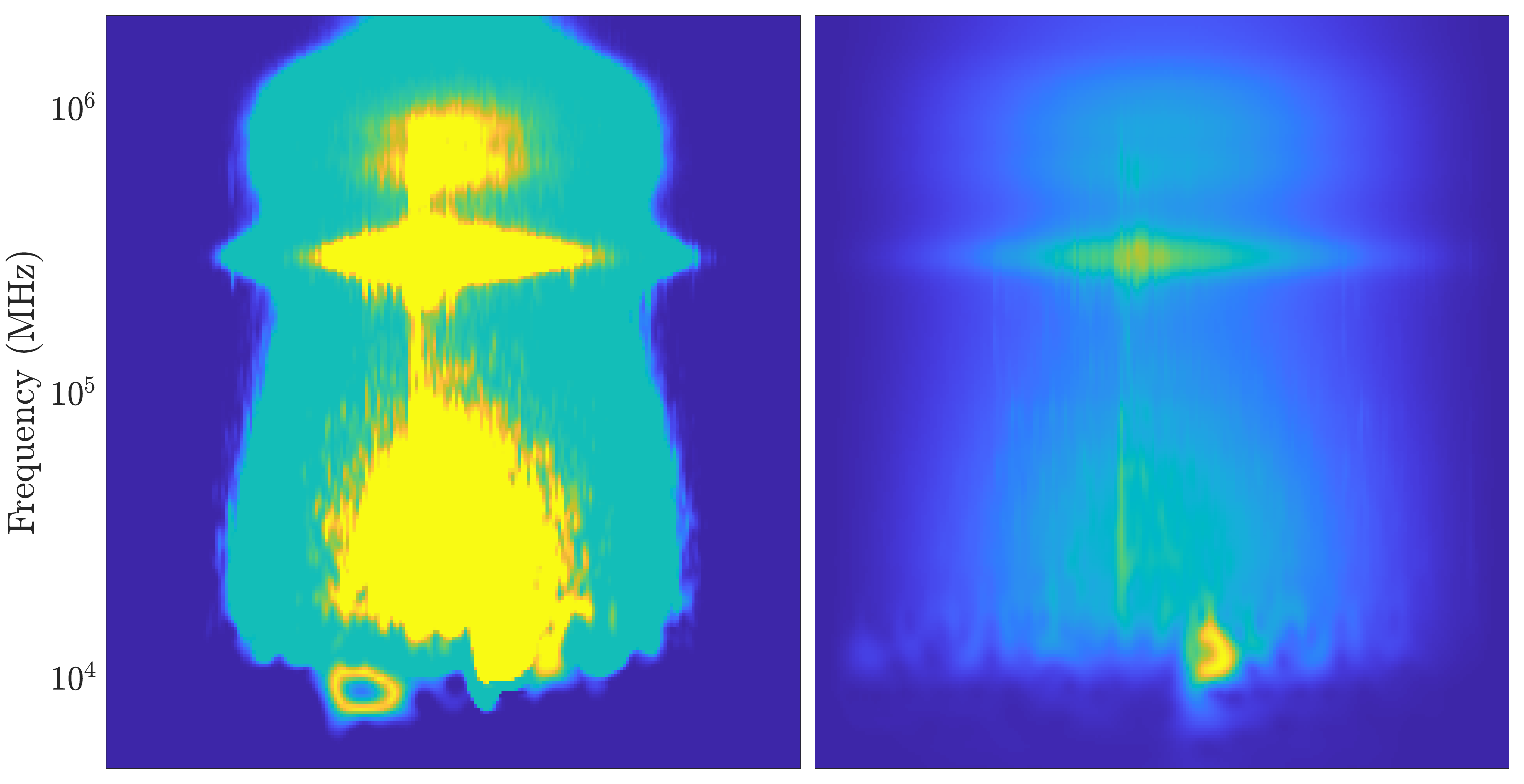}
        \caption{Sensor $\operatorname{mu80}$, $30$cNm}
    \end{subfigure}
    \begin{subfigure}[b]{0.33\textwidth}
        \includegraphics[width=\textwidth]{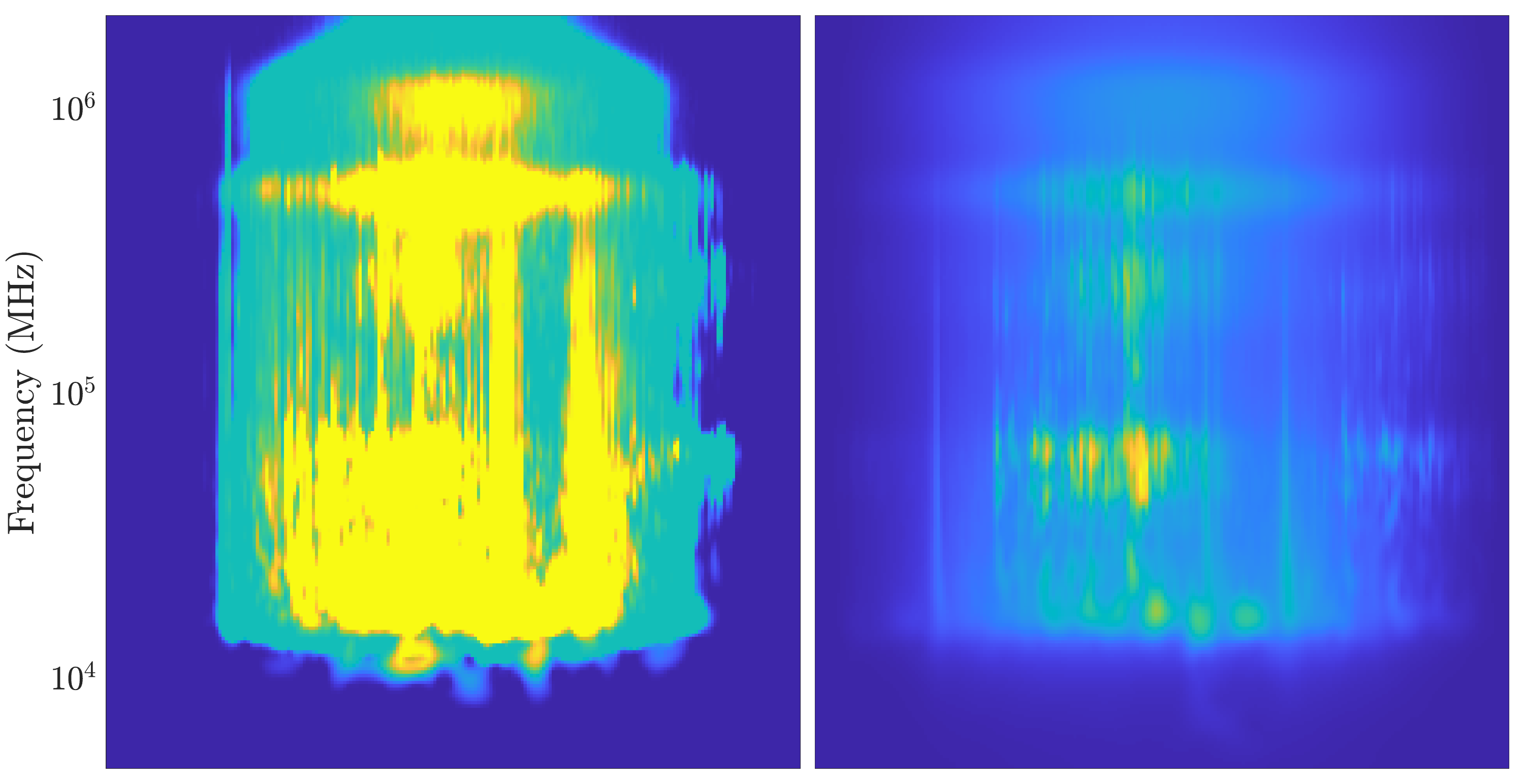}
        \caption{Sensor $\operatorname{F50A}$, $30$cNm}
    \end{subfigure}
    \begin{subfigure}[b]{0.33\textwidth}
        \includegraphics[width=\textwidth]{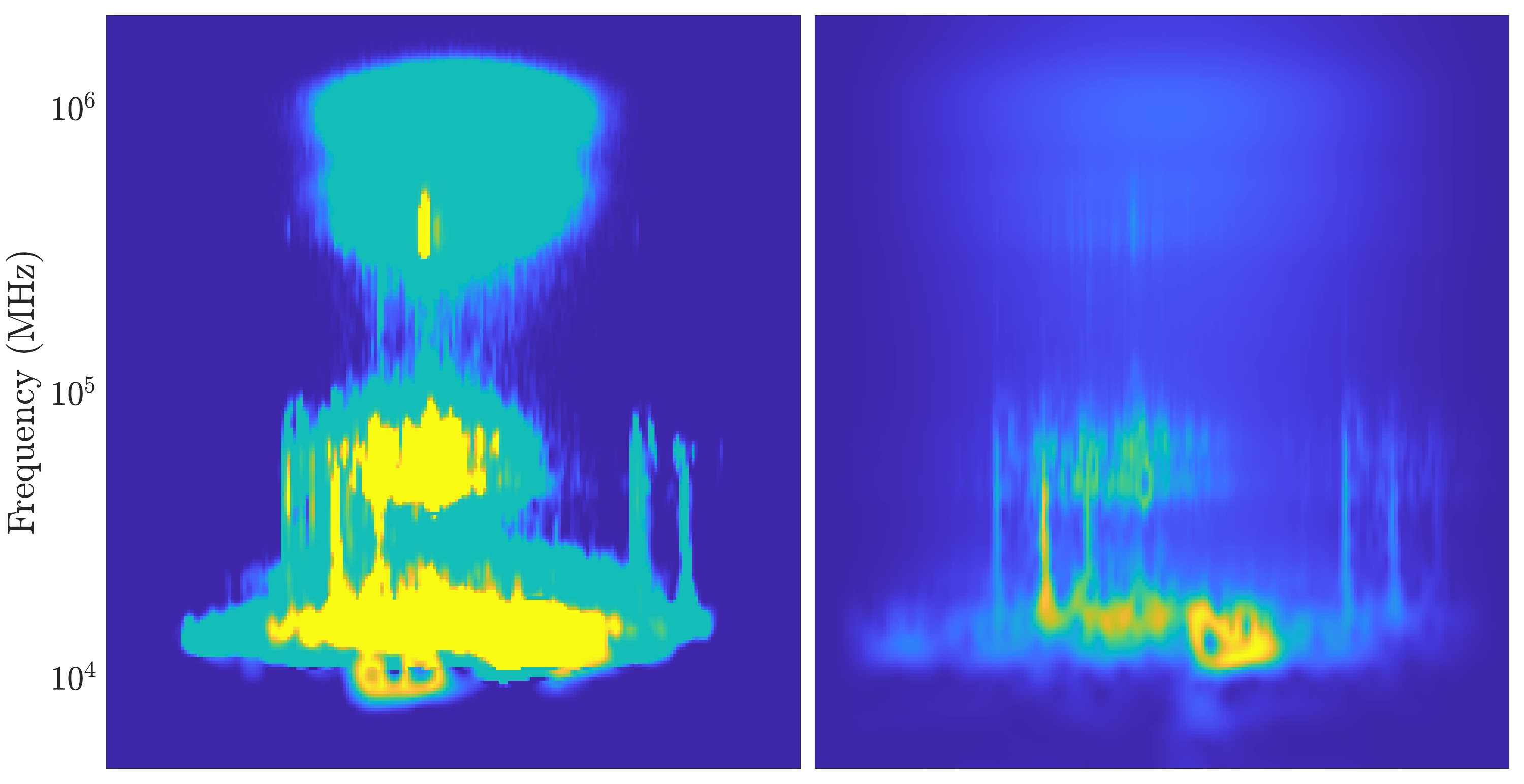}
        \caption{Sensor $\operatorname{mu200HF}$, $30$cNm}
    \end{subfigure}
    \begin{subfigure}[b]{0.33\textwidth}
        \includegraphics[width=\textwidth]{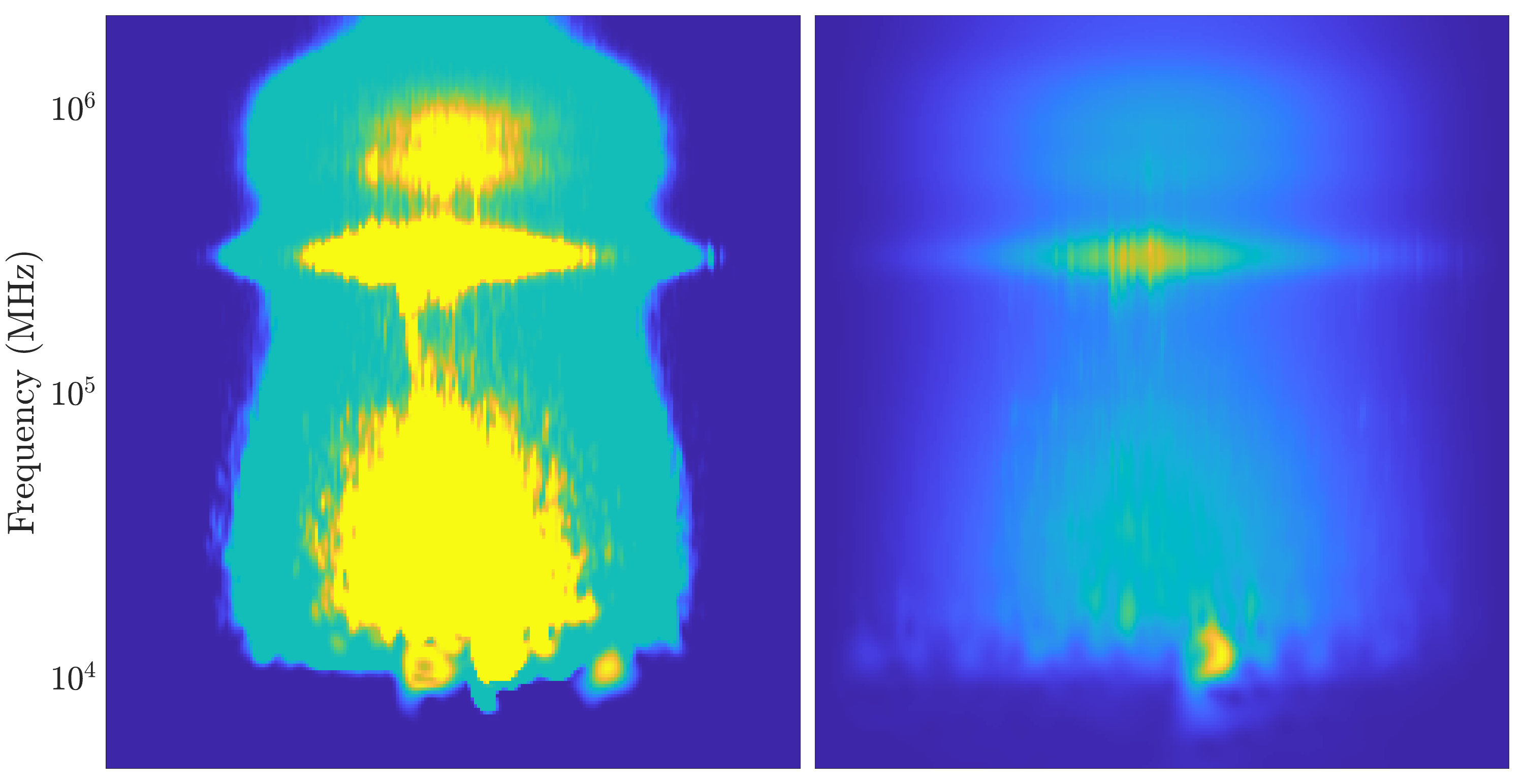}
        \caption{Sensor $\operatorname{mu80}$, $40$cNm}
    \end{subfigure}
    \begin{subfigure}[b]{0.33\textwidth}
        \includegraphics[width=\textwidth]{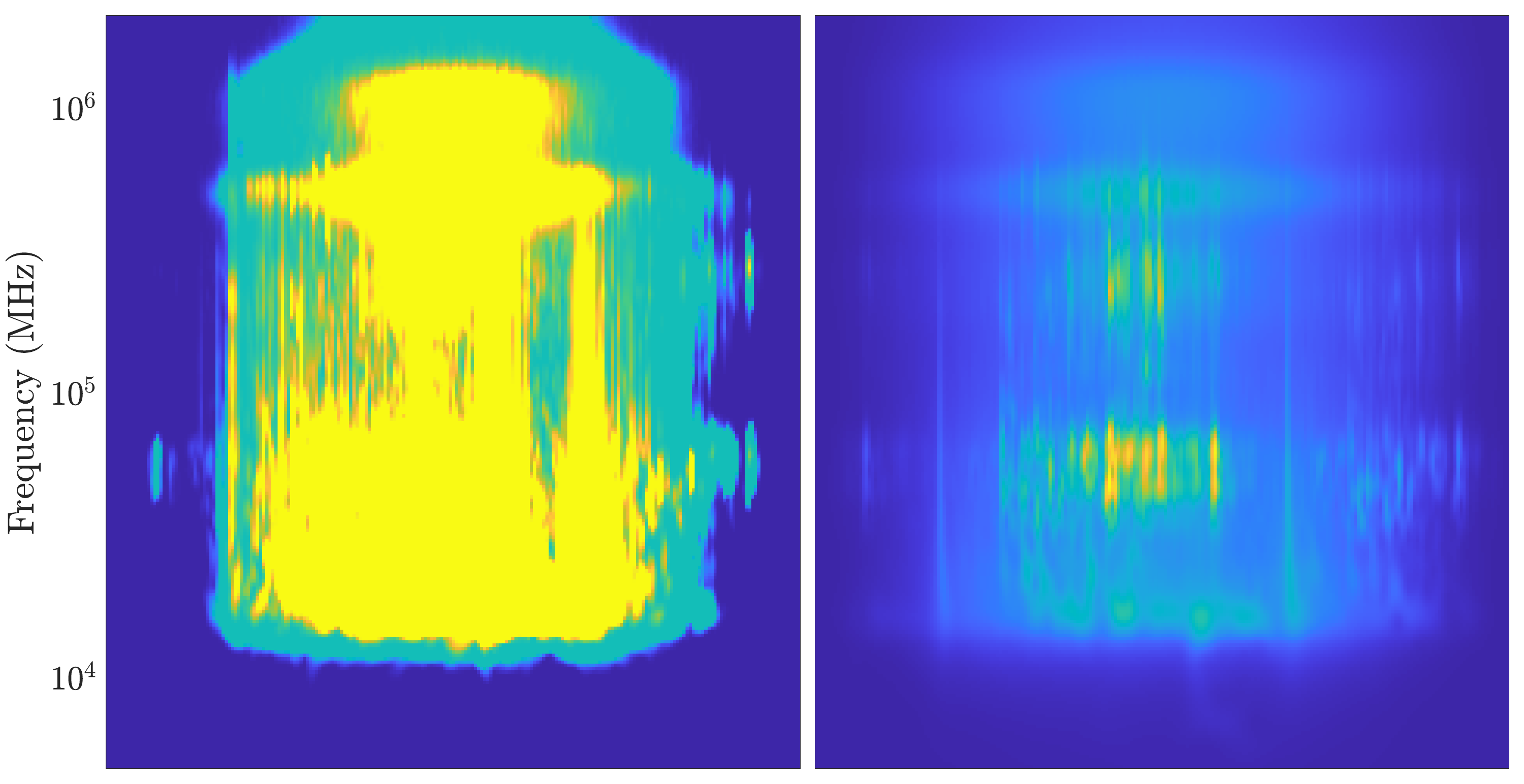}
        \caption{Sensor $\operatorname{F50A}$, $40$cNm}
    \end{subfigure}
    \begin{subfigure}[b]{0.33\textwidth}
        \includegraphics[width=\textwidth]{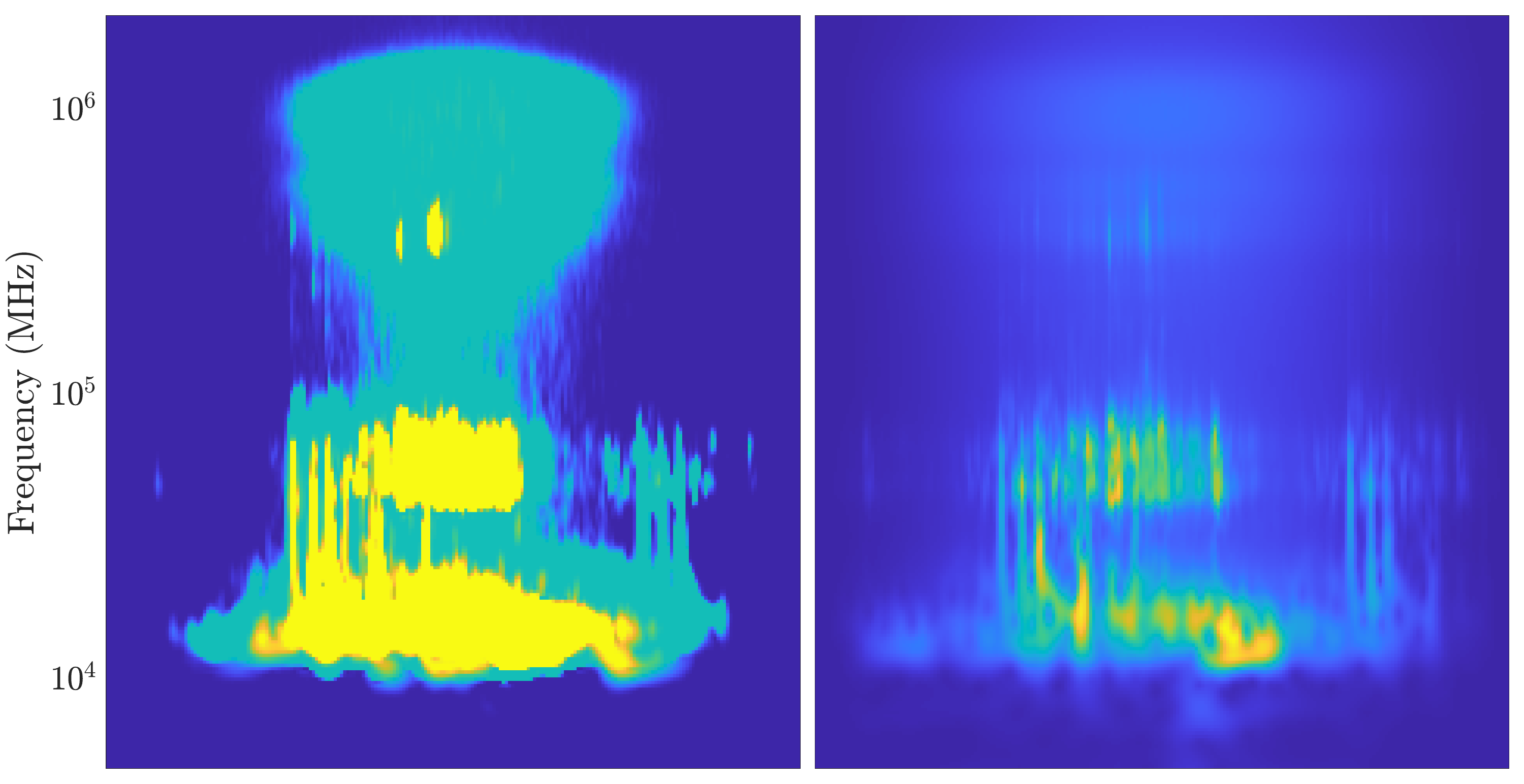}
        \caption{Sensor $\operatorname{mu200HF}$, $40$cNm}
    \end{subfigure}
    \begin{subfigure}[b]{0.33\textwidth}
        \includegraphics[width=\textwidth]{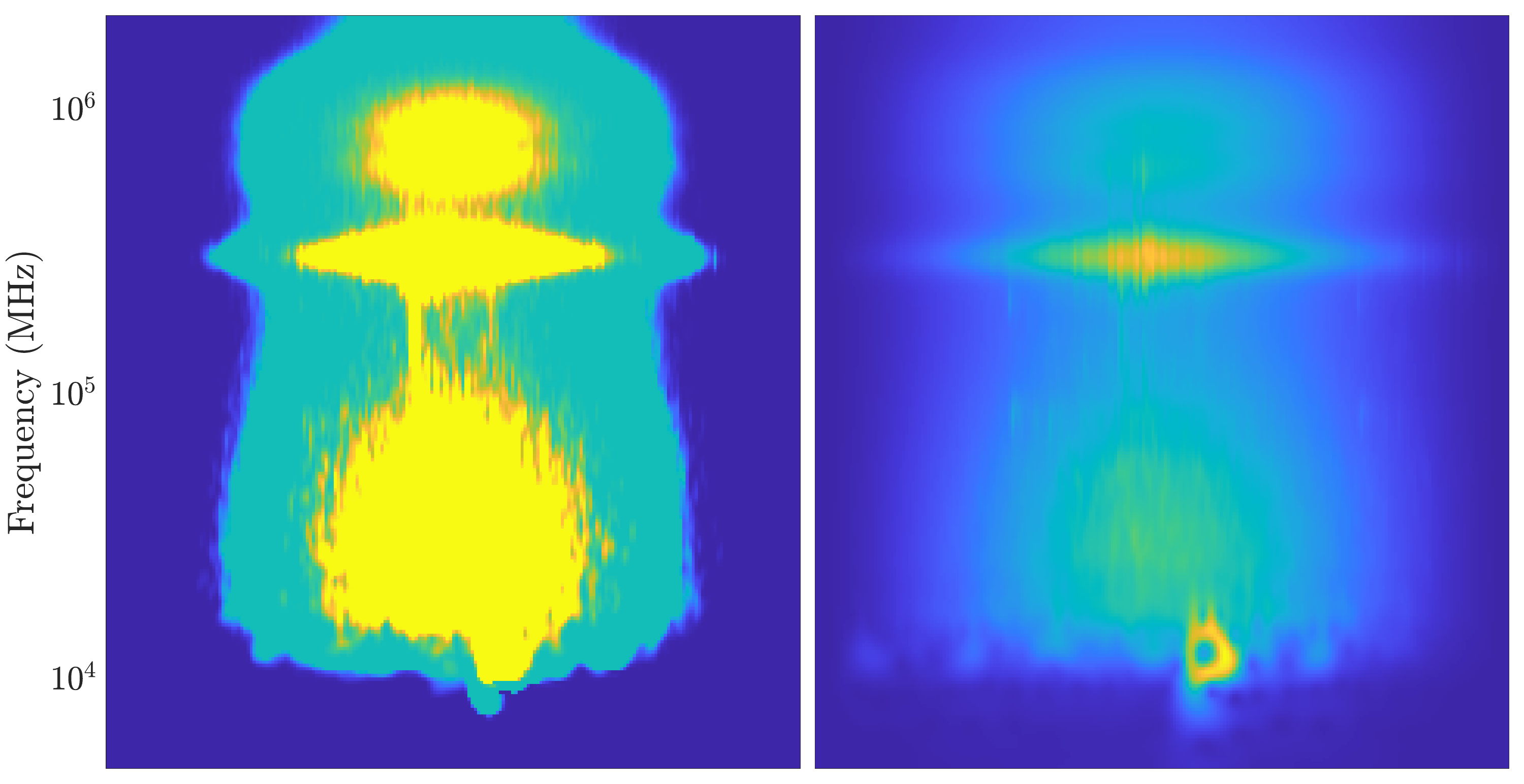}
        \caption{Sensor $\operatorname{mu80}$, $50$cNm}
    \end{subfigure}
    \begin{subfigure}[b]{0.33\textwidth}
        \includegraphics[width=\textwidth]{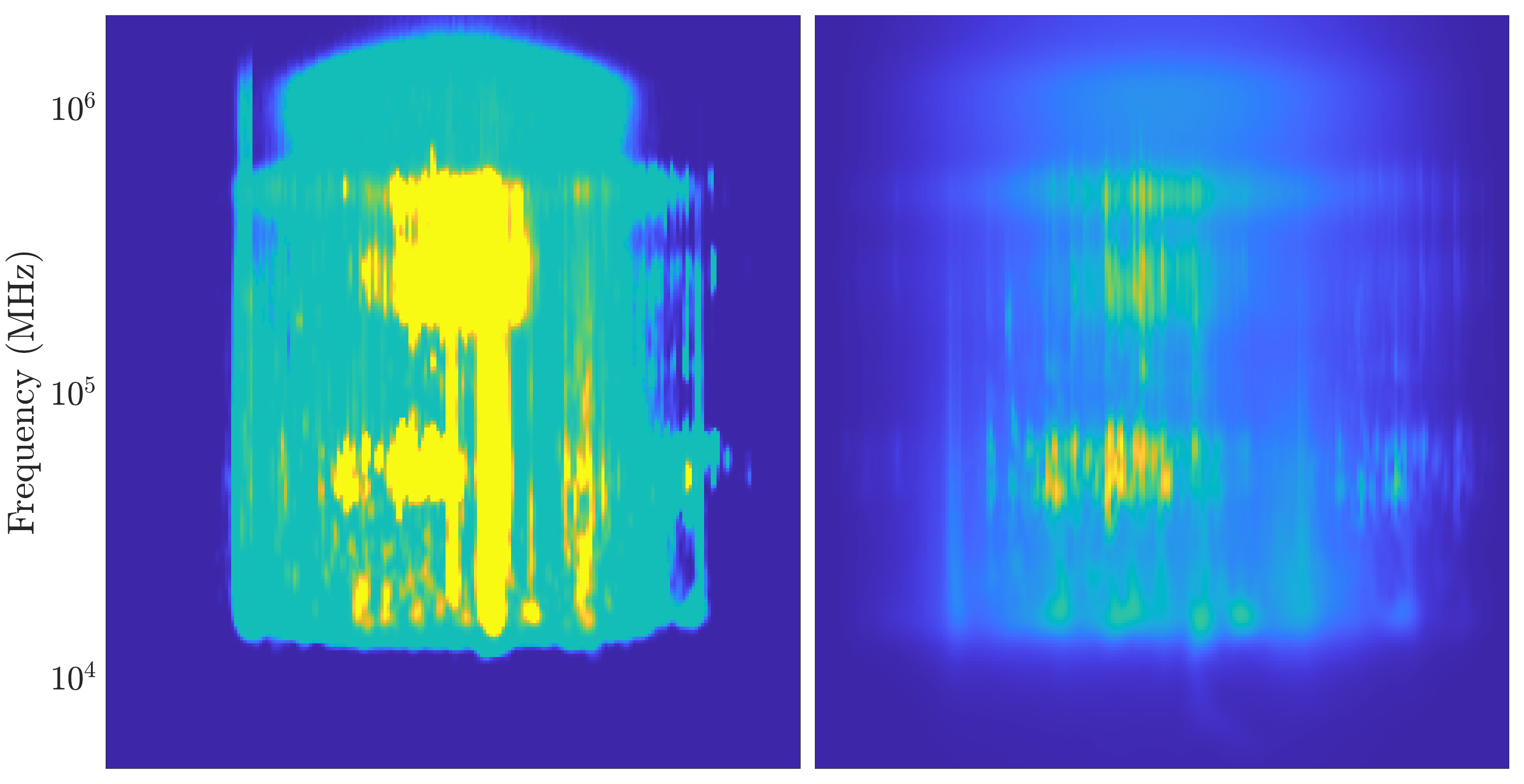}
        \caption{Sensor $\operatorname{F50A}$, $50$cNm}
    \end{subfigure}
    \begin{subfigure}[b]{0.33\textwidth}
        \includegraphics[width=\textwidth]{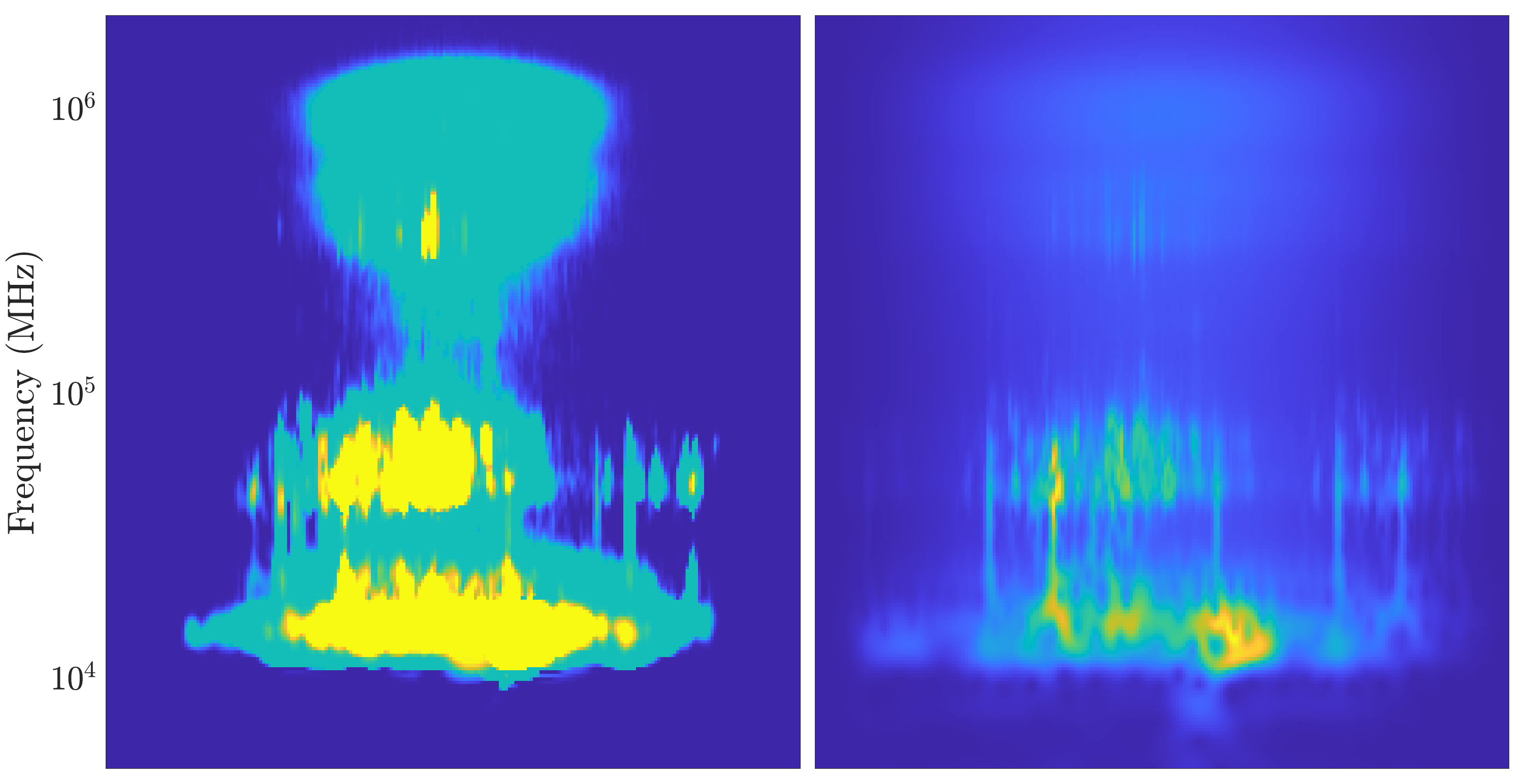}
        \caption{Sensor $\operatorname{mu200HF}$, $50$cNm}
    \end{subfigure}
    \begin{subfigure}[b]{0.33\textwidth}
        \includegraphics[width=\textwidth]{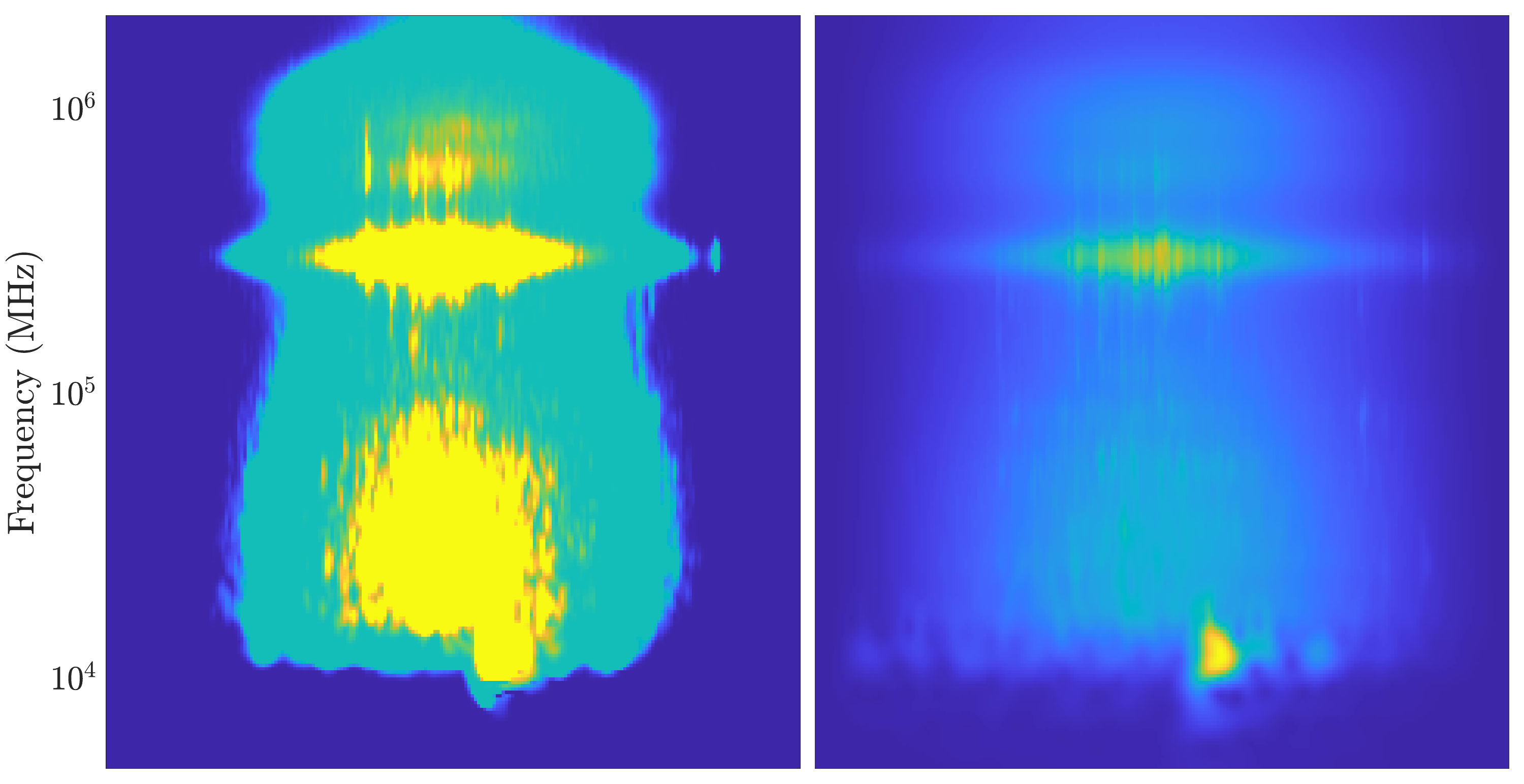}
        \caption{Sensor $\operatorname{mu80}$, $60$cNm}
    \end{subfigure}
    \begin{subfigure}[b]{0.33\textwidth}
        \includegraphics[width=\textwidth]{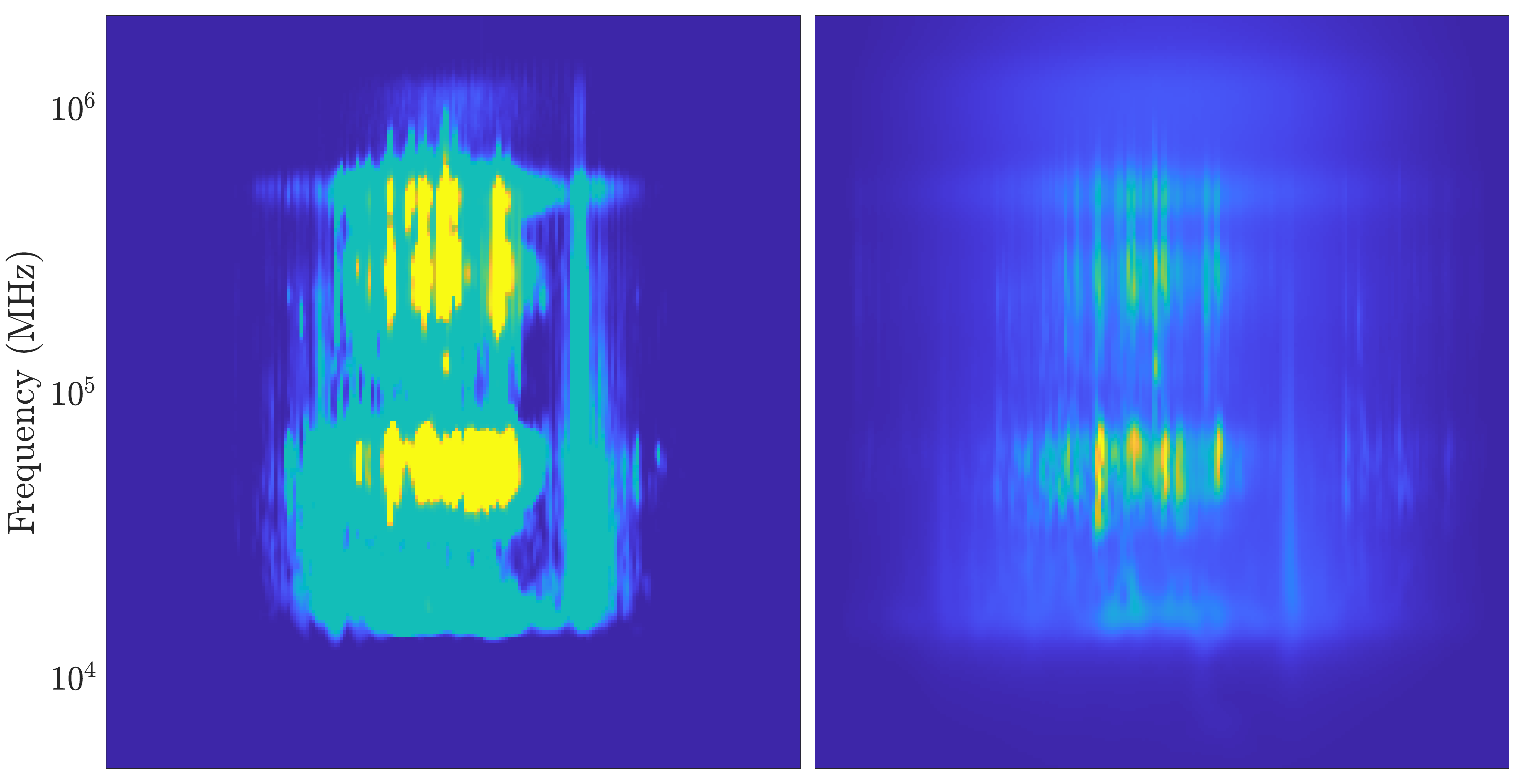}
        \caption{Sensor $\operatorname{F50A}$, $60$cNm}
    \end{subfigure}
    \begin{subfigure}[b]{0.33\textwidth} 
        \includegraphics[width=\textwidth]{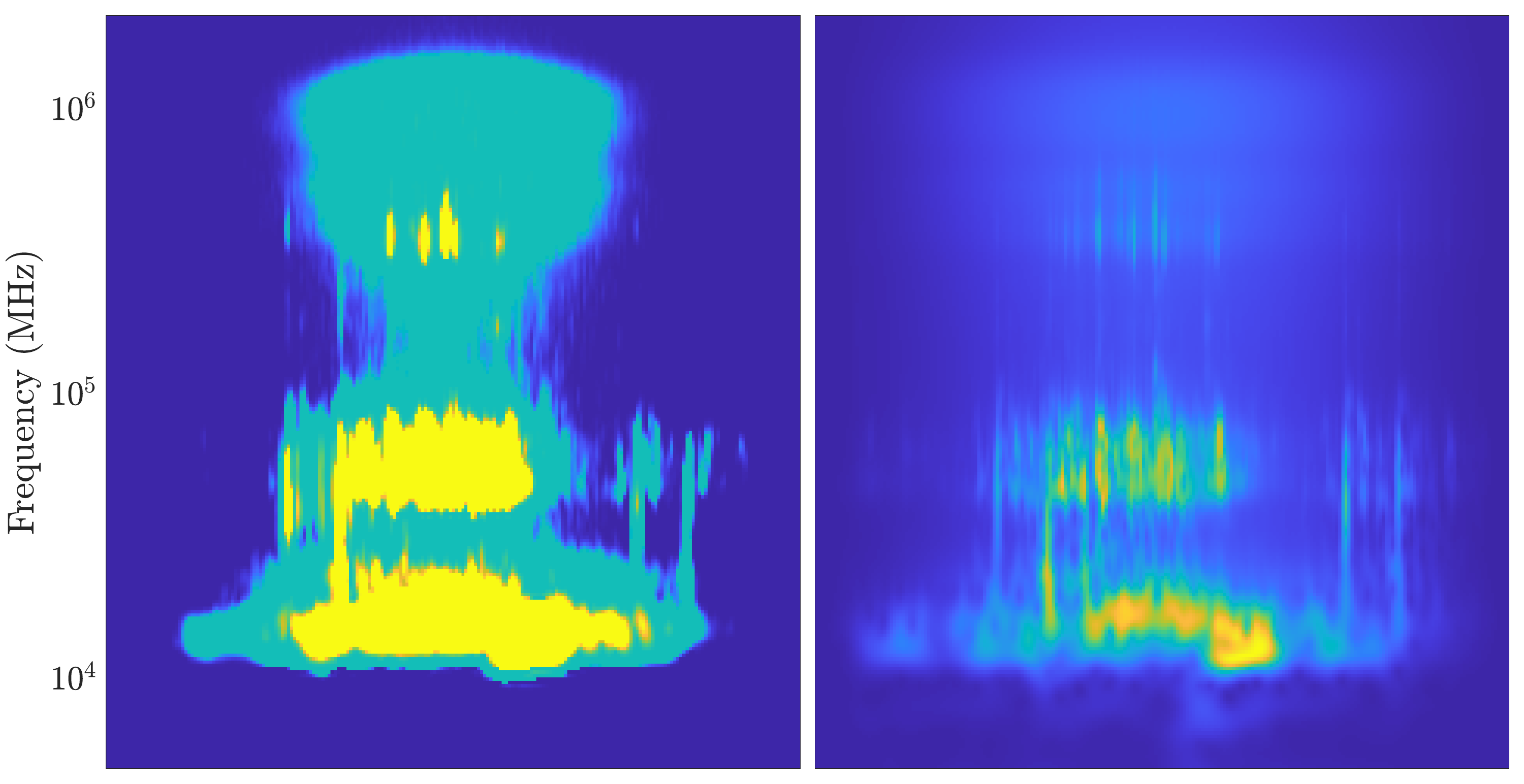}
        \caption{Sensor $\operatorname{mu200HF}$, $60$cNm}
    \end{subfigure}
    \caption{Average CWT images over about 1200 cycles for each tightening levels. Dark blue is for low values, green for median values and yellow for high values. Lines: tightening levels, columns: Sensors ($\operatorname{mu80, F50A, mu200HF}$). For each couple tightening-sensor, there are two images: One encoded in unsigned integer (first image) as used in networks, one in float (second image). The unit of y-axis is MHz and in log-scale, and the x-axis is normalized time (224 pixels).}
    \label{im:cumcwt}
\end{figure*}

{{Figure \ref{im:cumcwt} depicts the average of CWT images in each tightening level, for each sensor. There are about 1200 pairs of images per case. In each sub-figure, the left-hand side image of the pair represents the unsigned integer encoding of the CWT images as used in neural networks, whereas the right-hand side image is encoded using float encoding. These images (where dark blue is for low values, green for median values and yellow for high values) show different aspects of the dataset:
\begin{itemize}
    \item The amount of noise is important and the data are not trivial. Indeed the whole frequency range is concerned, characterizing a significant noise level, and a large variability can be observed with no clear trend across levels.
    
    \item The average responses are very different across sensors but the mu80 seems the most relevant:
    \begin{itemize}
        \item The mu80 sensor (column 1) shows three different frequency bands (below 100kHz, between 100kHz and 200kHz, and above 200kHz). Of particular interest is the area above 200kHz, which differs clearly from the two other sensors. These patterns are well reproducible across tightening levels.
        \item The F50a sensor depicts vertical patterns, indicating that large frequency bands are affected. However, these patterns are not as reproducible as those observed with the mu80 sensor. Furthermore, evaluating the evolution of these patterns across different levels proves to be complex.
        \item The mu200HF sensor exhibits horizontal patterns similar to those of the mu80 sensor; however, the highest frequency components (above 100 kHz) are less pronounced, as indicated by the absence of yellow coloration above this threshold.   
    \end{itemize}
    
    \item Encoding in 8 bits significantly alters the content but it is a necessary evil for the good with reduced memory requirements, lower power consumption, enhanced computational efficiencies, and better quantization and hardware compatibility, particularly with GPUs. Floating-point representations has a very different interpretation as shown in CWT representation. 
    
\end{itemize}
}}

\subsection{Tightening level identification module}

The identification of the tightening torque level relies on a training phase of a CNN, using as inputs the CWT images, arranged according to each tightening level. The seven levels correspond to the classes that the CNN has to predict from images. Pretrained neural networks and transfer learning was shown to be efficient in many SHM applications \cite{bao2019computer,azimi2020structural,liu2022efficient,postorino2023cross,yu2023corrosion,yu2022vision}.
Figure \ref{fig:schema} presents a schematic representation of the main neural network learning configurations discussed in this work, such as batch size, fine-tuned or frozen layers, optimizer, CNN architecture, learning rate scheduler and loss functions. Specifically, these last three items are discussed in detail below: 

\subsubsection{Network architecture}
\label{sec:netarchi}


An indication of the validation accuracy and prediction time of the 20 most widely used deep neural networks on ImageNet dataset, made of more than 14M natural images representing 1000 classes, is available on Mathworks\footnote{\url{https://fr.mathworks.com/help/deeplearning/ug/pretrained-convolutional-neural-networks.html}}. Although this comparison does not prejudge about their performance on a SHM task, \textit{in particular using scalograms that are not natural images as those considered in ImageNet}, it does provides a good indication on the difference between the various networks in terms of complexity (number of parameters), accuracy and inference time. Among these networks, the literature on SHM shows that $\operatorname{Resnet18}$ or VGG16 are often used. In many transfer learning applications, all layers in these networks except the top ones were frozen and additional task-specific layers (generally fully connected and dropout ones) with trainable parameters were incorporated at the head of the network. For example, the method proposed by Fu et al. \cite{fu2023automatic} using ORION-AE dataset relies on this strategy. Generally, for a given application, the trade-off behind using frozen layers is to decrease the computation time. However, it also implicitly assumes that the features extracted from ImageNet dataset share common characteristics with scalograms - although the last type of images are not natural ones. 
An experimental study \cite{ozgenel2018performance} showed that, in the context of SHM using very small datasets, it can be better to fine-tune all layers instead of freezing them, which is confirmed in our experiments. As presented by Wang et al.~\cite{wang2022egeria}, while freezing layers can reduce training cost, prematurely freezing under-trained layers in a static manner will impair the final accuracy. Generalization of frozen \textit{vs} non frozen networks on unseen SHM AE-data are the core of our experiments using the following CNN architectures:
\begin{description}
 
\item[GOOGLENET] Proposed in 2015 \cite{szegedy2015going}, $\operatorname{GoogLeNet}$ architecture is made of ``inception modules'' which allow sparsely connected blocks, where some layers are arranged in parallel instead of series, to limit overfitting and improve generalization. Those layers select which filter size is pertinent to learn the relevant information. The top-1 accuracy (proportion of images for which the model's single most confident prediction is correct) was 69.778\% on ImageNet1K dataset. In the following tests, the two auxiliary classifiers (represented as two additional network's heads equipped with a loss function) are not used since results were almost the same on CIFAR-10 dataset (10 classes) with or without them. The pytorch implementation of $\operatorname{GoogLeNet}$ was also made of 1x1 and 1x3 convolution filters in the inception modules. The model considered is composed of 5.6M training parameters. 

\item [RESNET] Proposed in 2016 \cite{he2016deep}, the $\operatorname{ResNet}$ architecture is made of skip connections which merges the output layers with input ones by addition. The skip connections act as shortcuts between blocks of layers. Bottleneck layers are used to significantly reduce the number of features or channels passing through it, making the network more compact and computationally efficient. In the residual units within ResNet architectures, information can flow more effectively in both the forward and backward pathways. This enhancement mitigates the vanishing gradient issue, facilitating the efficient propagation of gradients across multiple layers, thereby reducing the training process. The main motivation of the ResNet architecture was to increase the number of layers, therefore going deeper for tackling complex problems. For our tasks, we limit the tests to $\operatorname{ResNet18}$, one of the smallest ResNet network used on CIFAR and ImageNet datasets. $\operatorname{ResNet18}$ has yet 11M parameters and led to a top-1 accuracy of 69.758\% on ImageNet1K. 

\item[MOBILENETV2] Proposed in 2018 \cite{sandler2018mobilenetv2}, the $\operatorname{MobileNetV2}$ architecture was motivated by finding an optimal balance between accuracy and performance during tuning of deep neural networks for mobile applications. For that, they used lightweight Depthwise Separable Convolutions blocks, which replace a full convolutional operator with a simplified version that splits convolution into two separate layers. Compared to ResNet, the MobileNet architecture is based on an inverted residual structure, where the input and output of the residual blocks are thin bottleneck layers, unlike traditional residual models that use expanded representations in the input, requiring less computational resources. MobileNetV2 is made of 3.5M parameters and led to a top-1 accuracy of 72.154\% on ImageNet1K. 

\item[EFFICIENTNETB5] Proposed in 2019 \cite{tan2019efficientnet}, the main motivation of $\operatorname{EfficientNet}$ architecture was to create a set of networks that carefully balance network depth, width, and resolution for enhanced performance. The most common way of increasing the capacity of a network was to scale up ConvNets by their depth or width, more or less randomly. Another one was to scale up models by image resolution. Based on these observations, they proposed a new scaling method that uniformly scales all dimensions of depth/width/resolution using a ``compound'' coefficient. They demonstrated the effectiveness of this method on scaling up MobileNets and ResNets. Additionally, their research on the neural architecture design resulted in the creation of a novel baseline network, which was further scaled up to produce a family of models known as EfficientNets, ranging from B0 to B7. These models achieved improved accuracy and efficiency than previous ConvNets, with smaller number of parameters. EfficientNets rely on inverted bottleneck convolution blocks developed in MobileNets as well as squeeze-and-excitation blocks \cite{hu2018squeeze} for increased efficiency in computing relevant feature maps. In this work, we considered the $\operatorname{EfficientNetB5}$ architecture, which leds to a remarkable increase in performance on ImageNet1K with $83.444\%$ using 30M parameters.

\end{description}

\subsubsection{Learning rate scheduler}

When all parameters are set to be trainable, the training stage becomes complex for large networks and the choice of the initial learning rate as well as the scheduler becomes crucial. The task of the scheduler is to \textit{implement a policy that dynamically adjusts the learning rate over epochs or iterations}. Both learning rate (LR) and choice of policy stand out as important parameters for transfer learning due to their impact on training dynamics.

The $\operatorname{1cycle}$ scheduler\cite{smith2018disciplined,smith2019super} outperformed many standard scheduling approaches and is used in our methodology. This learning rate scheduler, a member of the cyclical learning rates (CLR) family, operates in cycles consisting of two ``steps'': the first step involves a gradual increase in the learning rate from a minimum value to a maximum value, followed by a second step in which the learning rate is reduced. The stepsize is chosen by the end-user and can represent the number of iterations, or epochs. Several functions are used to gradually increase and decrease the LR, as the Pytorch implementation with cosine evolution. The $\operatorname{1cycle}$ LR is thus made of a single cycle, and there is a variant that includes a third step, gradually decreasing the LR. To determine the minimum and maximum LR values, we trained the models through some iterations with different LR values. The maximum LR is identified as the value just before overfitting occurs during the validation process. When implementing the 1cycle LR policy, the minimum LR is typically set at a factor of 10-25 less than the established maximum bound. As underlined by Smith \cite{smith2019super}, if the speed of the learning rate evolution is too large, then the training becomes unstable, requiring to vary the number of epochs.

This scheduler has emerged as a promising approach due to a phenomenon known as ``super-convergence'', where training time takes orders of magnitude less than usual. According to Smith \cite{smith2019super}, a key element contributing to this phenomenon is the utilization of the $\operatorname{1cycle}$ policy with a relatively high maximum learning rate. In our tests, a warm-up step is performed, where the learning rate is set to 1/25 of the maximum allowed value. Then, it is gradually increased until its maximum. This first step takes 30\% of the number of iterations. In the second step, the learning is decreased until the maximum number of iterations is reached. In this scheduler, the learning rate is adapted at each mini-batch. 


\subsubsection{Loss functions}
\label{sec:lossf}

ORION-AE dataset is made of $K=7$ classes, therefore the cross-entropy loss (CRE) is a natural choice:
\begin{equation}
  \operatorname{CRE}(T, P) = -\sum_{i=1}^N \sum_{k=1}^K T_i(k) \log(P_i(k)) ,
\end{equation}
where $T_i(k)$ is the target ($0$ or $1$ for categorical target) for $i$-th input and $k$-th class, and $P_i(k)$ the prediction made by the network for this input for each of the $K$ classes. 

In the present work, additional loss functions are incorporated to account for the specific characteristics of the application. Indeed, the seven classes in ORION-AE dataset can be considered as ordered. Therefore, considering a loss that promotes ordinality in classification can be relevant for SHM purposes. In this type of loss, predictions that are close to the true class, but incorrect, are penalized less severely than predictions that are far from the truth, thereby \textit{encouraging misclassification errors to be in adjacent classes}. The different loss functions considered in this study are illustrated on a simple example in Figure~\ref{fig:schema}.

One natural loss, referred to as $\operatorname{CDW1}$ (Class-Distance Weighted \#1) assigns a weight proportional to the difference between predicted and true labels:
\begin{equation}
\operatorname{CDW1} (T,P) = \sum_{i,k} \left( \frac{w_{ik}}{K - 1} + 1 \right) \operatorname{CRE}(T_i(k),P_i(k)) ,
\end{equation}
with 
\begin{equation} 
w_{ik} =  \left| \arg \max_{k} {T}_i(k) - \arg \max_{k} {{P}}_i(k) \right| .
\end{equation}
This ordinal categorical cross-entropy loss is implemented in Keras\footnote{\url{https://www.tensorflow.org/guide/keras?hl=fr}} as an adaptation of the standard CRE for ordinal classification problems. An alternative loss tested in experiments is:
\begin{equation}
\operatorname{CDW2} (T,P) = \sum_{i,k} \exp \left( w_{ik} \right)  \operatorname{CRE}(T_i(k),P_i(k)) ,
\end{equation}
which amplifies the difference between the prediction and the true class. 

In Hou et al.\cite{hou2016squared}, a loss was proposed for ordinal classification problems using cumulative density functions ($cdf$) of targets and of predictions:
\begin{equation}
\operatorname{CDF} (T,P) = \sum_{i,k} \left( cdf(T_i(k)) - cdf(P_i(k)) \right)^2.
\end{equation}
Finally, we propose to consider two losses called $\operatorname{POM1a}$ and $\operatorname{POM1b}$ (``Plus Or Minus 1''), which considers the probability on adjacent classes:
\begin{equation}
  \operatorname{POM1a}(T, P) = -\sum_{i,k}  T_i(k)   \cdot \log \sum_{\substack{l\in\{-1,0,1\} \\ s.t. 0<k-l<=K}} P_i(k-l).
\end{equation}
In this loss, if the network predicts a high probability to an adjacent class of the true one, then the loss remains low, encouraging mis-classification errors to be in adjacent classes. Similarly, the sum can be taken on the logarithm:
\begin{equation}
  \operatorname{POM1b}(T, P) = -\sum_{i,k}  T_i(k)   \cdot \sum_{\substack{l\in\{-1,0,1\} \\ s.t. 0<k-l<=K}} \log  P_i(k-l).
\end{equation}


\section{NO-SHM Case: Single Campaign Classification}
\label{sec:NO-SHM case}
This section presents the results of CNN architectures for the classification of the tightening levels in a given campaign. In the following tests, we studied the effect of denoising the data before creating images, as well as the fusion of sensors. 

These tests are qualified as ``NO-SHM'', since training and testing phases share data from the same campaign, even if different data are used for each phase. The fact that training and test data originate from the same campaign \textit{makes the task simpler than isolating data from a different campaign for testing}. When mixing images from all campaigns and applying a CNN (as explained in Introduction), the problem boils down to a classification task with a pre-normalization of the data, which is different from a ``SHM'' task that should aim at generalizing to unseen test conditions \cite{FarrarBook,bull2021foundations}. The difficulty of a CNN to generalize from a set of training campaigns to a distinct testing one will be studied in the next section. This distinction is important to position our approach regarding the work of Fu et al. \cite{fu2023automatic}, which was a NO-SHM case. In this section, as in  Fu et al., accuracy on testing data is estimated based on the number of correct predictions divided by the total number of data, and denoted as $\operatorname{ACC}$ - the standard accuracy.


\subsection{Effect of denoising}

Wavelets denoising (WD) is widely used in applications related to {{AE}} \cite{bianchi2015wavelet,xin2020fracture,Pomponi2015110}. The basic idea of WD is to use a filter bank based on two quadrature mirror filters (low and high pass) derived from a mother wavelet ($mw$). It works in two stages. The first stage, so-called \textit{decomposition}, is performed from level $0$ (raw signal) until level $l$ by applying both filters, with decimation by a factor $2$ between each level. A thresholding method ($th$) is applied at each level to set to zero the wavelets coefficients corresponding to noise \cite{donoho1995noising}. The second stage, so-called \textit{reconstruction}, allows to generate the denoised signal, from level $l$ back to its original size, from the coefficients obtained at each level.  Wavelets are used to denoise raw AE data stream before creating images to study whether it improves or not the accuracy of the classification. The three sensors available in ORION-AE dataset are considered ($mu 80$, $F50A$ and $mu 200HF$) individually. Blocks of $1 s$ ($5$M sampling points) were used. Similar parameters to Kharrat et al. \cite{kharrat2016signal} are used for denoising: $mw=db45$, $th$ set to the universal threshold method with level dependent rescaling except the level of decomposition $l$. 

The level of decomposition $l$ was set to $0,1,2\dots,9$ for campaign $\#B$ (identified as the most difficult one \cite{ramasso2022clustering}) and to $0,1,4,9$ for the other campaigns. For each level, the pretrained network $\operatorname{GoogLeNet}$ is applied using images generated during the ``\nameref{sec:step1}" and ``\nameref{sec:step2}". All parameters are set as trainable, i.e. without freezing any layer. For training, a Pytorch implementation is put forward considering ADAMW optimizer\cite{loshchilov2019Adamw}, momentum equal to $0.9$, weight decay of $0.0005$, constant learning rate equal to $0.001$ (no scheduler policy), minibatch size of $16$, and $15$ epochs. The dataset is divided into training ($80\%$), validation ($10\%$) and testing ($10\%$), as in Fu et al. \cite{fu2023automatic}. This configuration is summarized in Table \ref{tab:comp1zdzdz}. 

\begin{table*}
\small
    \centering
    \begin{tabular}{|c|c|c|c|c|c|c|c|c|c|}
        \hline
        Approach & Net. & Frozen  & Loss & Optim. & LR    & Mom. & Sched. & MB  & Epochs  \\
        
                 &         &  layers  &    &      &     &        &     &   & \\   
        \hline
        Fu et al. \cite{fu2023automatic} & $\operatorname{Resnet18}$ & Yes & CRE & SGDM & N.C. & N.C. & No & N.C. & 150 \\
        \hline
        This work & $\operatorname{GoogleNet}$ & \textbf{No} & CRE & ADAMW & 0.001 & 0.9 & No & 64 & \textbf{15}  \\
        \hline
    \end{tabular}
    \caption{Configuration for studying the noise level and sensor fusion. ``N.C.'' stands for ``not communicated''.  }
    \label{tab:comp1zdzdz}
\end{table*}

Results from campaigns $\#B$ to $\#F$ are presented in Tables \ref{campB1} to \ref{campF1}, respectively. These tables show that there is \textit{no improvement in performance when using a pre-filtering} of data, for all campaigns. Conversely, a decrease of performance can be observed. The results indicate that denoising removes relevant information from signals even at low level of decomposition. The tables additionally demonstrate high performances without pre-filtering, with an average accuracy $\operatorname{ACC} \sim 99\%$. This performance is attributed to the pretrained CNN's capacity of managing noisy input through regularization, dropout and batch normalization. The input signals are highly noisy, but the decomposition of the signals into frequency bands in the scalograms facilitates the training process. 

\begin{table}[hbtp]
\centering
\begin{tabular}{|c|c|c|c|c|}
\hline
	& $mu 80$	& $F50A$  & 	$mu 200HF$	& All sensors \\
\hline
0	&99.61 \%	&100 \%	&100 \%	&99.81 \% \\
\hline
1	&99.57 \%	&100 \%	&100 \%	&99.94 \% \\
\hline
2	&99.38 \%	&99.94 \%	&99.94 \%	&99.86 \%\\
\hline
3	&99.44 \%	&99.94 \%	&100 \%	&99.86 \%\\
\hline
4	&97.96 \%	&100 \%	&100 \%	&99.50 \%\\
\hline
5	&99.07 \%	&100 \%	&99.94 \%	&99.65 \%\\
\hline
6	&98.70 \%	&99.88 \%	&100 \%	&99.55 \%\\
\hline
7	&98.88 \%	&99.88 \%	&99.88 \%	&99.23 \%\\
\hline
8	&95.11 \%	&99.94 \%	&99.81 \%	&97.50 \%\\
\hline
9	&66.05 \%	&99.94 \%	&99.76 \%	&88.71 \%\\
\hline
\end{tabular}
\caption{\#B: Performance according to the level of decomposition in denoising and to the sensors. \label{campB1}}
\end{table}

\begin{table}[hbtp]
\centering
\begin{tabular}{|c|c|c|c|c|}
\hline
	& $mu 80$	& $F50A$  & $mu 200HF$	& All sensors \\
\hline
0	&99.21 \%	&99.88 \%	&100 \%	&99.51 \% \\
\hline
1	&98.07 \%	&99.86 \%	&100 \%	&99.31 \% \\
\hline
4	&97.72 \%	&100 \%	&100 \%	&99.33 \%\\
\hline
9	&67.76 \%	&99.94 \%	&99.65 \%	&88.87 \%\\
\hline
\end{tabular}
\caption{\#C: Performance according to the level of decomposition in denoising and to the sensors. \label{campC1}}
\end{table}

\begin{table}[hbtp]\centering
\begin{tabular}{|c|c|c|c|c|}
\hline
	& $mu 80$ 	& $F50A$  & $mu 200HF$	& All sensors \\
\hline
0	&98.5 \%	&99.25 \%	&99.83 \%	&99.46 \% \\
\hline
1	&97.86 \%	&98.72 \%	&99.11 \%	&98.39 \% \\
\hline
4	&97.68 \%	&99 \%	&99 \%	&97.96 \%\\
\hline
9	&74.18 \%	&98.72 \%	&97.25 \%	&90.69 \%\\
\hline
\end{tabular}
\caption{\#D: Performance according to the level of decomposition in denoising and to the sensors. \label{campD1}}
\end{table}

\begin{table}[hbtp]\centering
\begin{tabular}{|c|c|c|c|c|}
\hline
	& $mu 80	$& $F50A$  & 	$mu 200HF$	& All sensors \\
\hline
0	&99.51 \%	&100 \%	&100 \%	&99.56 \% \\
\hline
1	&98.84 \%	&100 \%	&100 \%	&99.45 \% \\
\hline
4	&98.47 \%	&100 \%	&100 \%	&99.55 \%\\
\hline
9	&72.39 \%	&100 \%	&99.57 \%	&91.24 \%\\
\hline
\end{tabular}
\caption{\#E: Performance according to the level of decomposition in denoising and to the sensors. \label{campE1}}
\end{table}

\begin{table}[hbtp]\centering
\begin{tabular}{|c|c|c|c|c|}
\hline
	& $mu 80$	& $F50A$  & $mu 200HF$	& All sensors \\
\hline
0	&99.56 \%	&100 \%	&100 \%	&99.95 \% \\
\hline
1	&99.51 \%	&100 \%	&100 \%	&99.90 \% \\
\hline
4	&99.33 \%	&100 \%	&100 \%	&99.74 \%\\
\hline
9	&82.65 \%	&99.57 \%	&99.33 \%	&93.46 \%\\
\hline
\end{tabular}
\caption{\#F: Performance according to the level of decomposition in denoising and to the sensors. \label{campF1}}
\end{table}

Table \ref{tab:my_label} presents the confusion matrix obtained by the $\operatorname{GoogLeNet}$ on campaign $\#B$, considering the sensor $mu80$, and without denoising. Note that the matrix is diagonal, showing the high performance of the classification task in the ``NO-SHM" case, considering the sensor that provided slightly lower performances than the other sensors (on average, less than 1\% difference). 

\begin{table}
\centering
\begin{tabular}{c|c|c|c|c|c|c|c|}
\cline{2-8}
\begin{tabular}[c]{@{}c@{}}Torque\\ Levels\end{tabular} & 1   & 2   & 3   & 4   & 5   & 6   & 7   \\ \hline
\multicolumn{1}{|c|}{1}                                 & 247 & 0   & 1   & 0   & 0   & 2   & 0   \\ \hline
\multicolumn{1}{|c|}{2}                                 & 1   & 216 & 0   & 0   & 0   & 0   & 0   \\ \hline
\multicolumn{1}{|c|}{3}                                 & 0   & 0   & 236 & 0   & 0   & 0   & 0   \\ \hline
\multicolumn{1}{|c|}{4}                                 & 0   & 1   & 0   & 199 & 0   & 1   & 0   \\ \hline
\multicolumn{1}{|c|}{5}                                 & 0   & 0   & 0   & 0   & 218 & 0   & 0   \\ \hline
\multicolumn{1}{|c|}{6}                                 & 0   & 0   & 1   & 0   & 0   & 233 & 0   \\ \hline
\multicolumn{1}{|c|}{7}                                 & 0   & 0   & 0   & 0   & 0   & 0   & 256 \\ \hline
\end{tabular}
\caption{Confusion matrix obtained with the configuration presented in Table \ref{tab:comp1zdzdz} using $\operatorname{GoogLeNet}$ on campaign $\#B$. Accuracy of $98.5\%$ was reached after $9$ epochs, and $99.5\%$ after $12$ epochs.}
    \label{tab:my_label}
\end{table}


\subsection{Performance using sensor fusion}

{Tables \ref{campB1} to \ref{campF1} show the classification accuracy using sensor fusion. In this work, the fusion process consists of considering all images from all sensors during training, validation and testing.}


Note that the fusion slightly decreases the accuracy when considering the single best sensor for each line. This result is in accordance with the study made by Ai et al. \cite{ai2021evaluation}, in the context of AE-based monitoring of reinforced concrete block, who observe that single sensor is more efficient. Other fusion strategies could be considered, e.g. combining the outputs of different networks \cite{ai2023localizing}, but the marginal improvement (compared to $99\%$ with the $\operatorname{mu80}$ sensor) does not justify the increase in computation time. 

These results on the NO-SHM case \textit{extends the study} of Fu et al. \cite{fu2023automatic} to all campaigns, to every sensor\footnote{Fu et al. \cite{fu2023automatic} do not specify which sensor was used.} as well as to sensor fusion. 

\section{SHM Case: Classification on Unseen Campaign}
\label{sec:SHM case}
In the previous section, data from different levels of tightening torque were mixed, regardless of the campaign. This section extends the analysis to the so-called SHM case, in which one campaign of measurements is isolated for testing, while the CNN model is trained on the remaining four campaigns. Therefore, each campaign is interpreted as being generated by a distinct structure (heterogeneous characteristics discussed in subsection ``\nameref{subsec: Challenges}"). This section begins by establishing a baseline, which will be used to compare the performance of different configurations when training CNNs for the SHM case, as well as to assess the influence of including \textit{a priori} information from the structure (campaign) not initially used for training. 

Moreover, accuracy on testing data is calculated in two ways: (1) the standard accuracy $\operatorname{ACC}$ introduced previously; and (2) the accuracy on $\pm 1$ class, denoted as $\operatorname{ACC_{\pm1}}$ (see  
\hyperref[sec:metrics]{Appendix B}). Note that in the ORION-AE dataset, there may be overlapping classes \cite{ramasso2022clustering} due to uncertainties on the actual value of the tightening levels, as torque measurements were conducted using an analog torque wrench \cite{ORIONdata}. Therefore, if a network predicts a tightening level of $\hat{c}$ while the ground truth is $c$, it is considered correct if $\lvert \hat{c}-c \rvert \leqslant  1$. The dataset from $mu80$ sensor is considered in the following tests. 

\subsection{Baseline}

Three network architectures are considered: $\operatorname{GoogLeNet}$, $\operatorname{Resnet18}$ and $\operatorname{EfficientNetB5}$, trained with the same parameters presented in Table \ref{tab:comp1zdzdz}. To establish the baseline, as the campaign $\#B$ is the one that presents considerable challenges for the classification \cite{ramasso2022clustering}, it is used for testing. Therefore, the following configurations are explored: SHM vs. NO-SHM (training, validation and test datasets formed from only campaign $\#B$) situations, and freezing vs. non freezing layers, leading to $4$~(configurations)~$\times$~$3$~(architectures)~$=12$ cases. Results are summarized in Table \ref{tab:comp1zdzdz2} with the standard accuracy ($\operatorname{ACC}$) and the accuracy at plus or minus one class ($\operatorname{ACC_{\pm1}}$). In this table, the \textbf{NO}-SHM case with $\operatorname{Resnet18}$ corresponds to the same configuration used by Fu et al. \cite{fu2023automatic}. In that paper, the authors froze the layers, but we were not able to reproduce the same results ($99\%$), except when all the parameters were considered as trainable (mentioned as ``NO-Freeze'' in the table). In this case, the networks converged after $15$ epochs, i.e. in $10$ times less epochs than Fu et al.\cite{fu2023automatic}. 

Table \ref{tab:comp1zdzdz2} also indicates that layer freezing is not relevant compared to training all parameters, neither for the SHM case nor for NO-SHM, in these datasets. For example, if the simpler task of classifying images from the same campaign is considered (NO-SHM case), freezing layers leads to an accuracy $\operatorname{ACC_{\pm1}}~\sim ~68.8\%$, whereas $100\%$ is obtained when all parameters become trainable. Therefore, if the task is more complex, as in the SHM case, the accuracy drops dramatically by freezing layers - accuracy $\operatorname{ACC_{\pm1}}~\sim~42 - 45\%$ using $\operatorname{Resnet18}$ or $\operatorname{EfficientNetB5}$ architectures against $\operatorname{ACC_{\pm1}}~\sim 65\%$ when both CNNs have trainable parameters. 


\begin{table*}
\small
\centering
\begin{tabular}{c|c|cc|cc|cc|}
\cline{2-8}
                                 & \multirow{2}{*}{Configurations} & \multicolumn{2}{c|}{GoogleNet}   & \multicolumn{2}{c|}{Resnet18}    & \multicolumn{2}{c|}{EfficientNetB5} \\ 
                                 &                                 & \multicolumn{1}{c|}{ACC}  & $\operatorname{ACC_{\pm1}}$  & \multicolumn{1}{c|}{ACC}  & $\operatorname{ACC_{\pm1}}$  & \multicolumn{1}{c|}{ACC}    & $\operatorname{ACC_{\pm1}}$   \\ \hline
\multicolumn{1}{|l|}{Baseline 1} & \textbf{NO}-SHM + \textbf{NO}-Freeze              & \multicolumn{1}{c|}{99.8} & 99.8 & \multicolumn{1}{c|}{99.9} & 100  & \multicolumn{1}{c|}{99.9}   & 99.9  \\ \hline
\multicolumn{1}{|l|}{Baseline 2 \cite{fu2023automatic}} & \textbf{NO}-SHM + Freeze                 & \multicolumn{1}{c|}{\cellcolor[rgb]{0.9,0.5,0.5}{49.0}} & \cellcolor[rgb]{0.9,0.5,0.5}{62.2} & \multicolumn{1}{c|}{\cellcolor[rgb]{0.9,0.5,0.5}{57.1}} & \cellcolor[rgb]{0.9,0.5,0.5}{68.8} & \multicolumn{1}{c|}{\cellcolor[rgb]{0.9,0.5,0.5}{55.0}}   & \cellcolor[rgb]{0.9,0.5,0.5}{66.5}  \\ \hline \hline
\multicolumn{1}{|l|}{Baseline 3} & SHM + \textbf{NO}-Freeze                 & \multicolumn{1}{c|}{26.4} & 52.2 & \multicolumn{1}{c|}{31.4} & \cellcolor[rgb]{0.5,0.9,0.5}{65.4} & \multicolumn{1}{c|}{35.4}   & \cellcolor[rgb]{0.5,0.9,0.5}{65.9}  \\ \hline
\multicolumn{1}{|l|}{Baseline 4} & SHM + Freeze                    & \multicolumn{1}{c|}{ \cellcolor[rgb]{0.9,0.5,0.5}{17.4}} & \cellcolor[rgb]{0.9,0.5,0.5}{43.7} & \multicolumn{1}{c|}{\cellcolor[rgb]{0.9,0.5,0.5}{18.6}} & \cellcolor[rgb]{0.9,0.5,0.5}{42.9} & \multicolumn{1}{c|}{\cellcolor[rgb]{0.9,0.5,0.5}{17.0}}   & \cellcolor[rgb]{0.9,0.5,0.5}{44.9}  \\ \hline
\end{tabular}
\caption{Accuracy for four baseline approaches; baselines 1 $\&$ 2 are variants of the Fu et al. \cite{fu2023automatic} work.}
    \label{tab:comp1zdzdz2}
\end{table*}

\subsection{Influence of the amount of prior using the baseline}

The influence of the amount of prior data from the tested structure is studied by gradually adding knowledge on tightening levels, from none (fully unsupervised) to the case where all classes, except class $5$ cNm, are used in training (almost NO-SHM). The tightening levels are added from $60$ cNm to $10$ cNm \textit{in this order, to mimic the situation where regular inspections on the structure allow to label the previously collected data}. 
The pretrained $\operatorname{Resnet18}$ is used \textit{without freezing any layer}, i.e., all the parameters of the model (11M) are considered trainable, as this is a necessary condition for successful generalization in this dataset. Training procedure is performed using the SGDM (Stochastic Gradient Descent with Momentum) optimizer, and a piecewise learning rate schedule starting with an initial learning rate of 0.01 and a drop factor of 0.1. The model leading to the best validation loss was retained after 15 epochs. A 5-fold cross validation is performed in each test, leading to a total $5 \times 7 \times 5 = 175$ runs.

\begin{figure*}
\centering
\includegraphics[width=1.0\columnwidth]{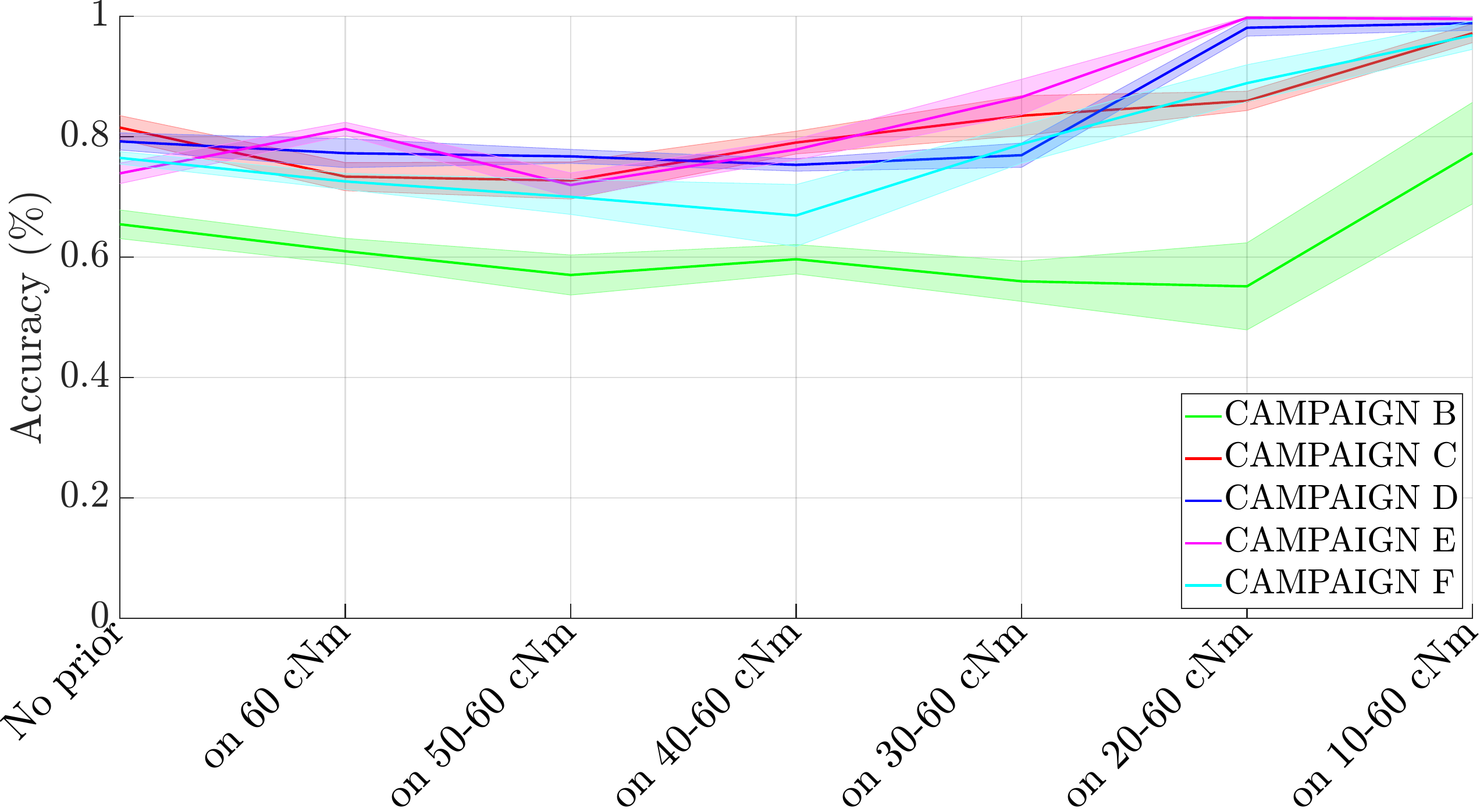}
\caption{$\operatorname{ResNet18}$, sensor $mu 80$: Performance on generalization on all campaigns with gradual amount of prior.
\label{general1}}
\end{figure*}

Figure \ref{general1} depicts the results. On the left-hand side (fully unsupervised), no images from the unseen campaign are used during training. From left to right, more prior knowledge is added progressively with respect to the state of the structure. Note the natural trend of the model to better classify the data, as expected. Moreover, in most campaigns, an accuracy $\operatorname{ACC_{\pm1}}$ between $98$ and $100\%$ is obtained in the most supervised case (almost NO-SHM case with only one level, 5 cNm, remaining). A difference around $15\%$ can be observed between SHM and NO-SHM situations for each campaign. 

Campaign $\#B$ is the most difficult one to guarantee satisfactory results, with an average classification accuracy $\operatorname{ACC_{\pm1}}~\sim~65\%$. In the NO-SHM case, the results of this campaign depicts the highest variability (see confidence intervals in Figure \ref{general1}), demonstrating the sensitivity of the model to the training data during cross-validation. The average of $\operatorname{ACC_{\pm1}}$ values on all campaigns, except for $\#B$, is about $78\% \pm 2$ with maximum \textit{a priori}. 


\section{SHM Case: Super-Convergence}
\label{sec: Super convergence}
In this section, the tests aim at:
\begin{itemize}
\item Identifying configurations of the networks able to quickly train with the best performance in generalization, without freezing any layers; 
\item Demonstrating the phenomenon of super-convergence with the $\operatorname{1cycle}$ scheduler;
\item Demonstrating the relevance of the ordinal losses;
\item Comparing these configurations with the baseline.
\end{itemize} 

Due to the number of tests required, only campaign~$\#B$ is considered in the tests. The other campaigns will be studied subsequently, testing the best configurations found for $\#B$. 
Four types of CNN architectures are considered, as presented in subsection ``\nameref{sec:netarchi}". These architectures comprise one ``small", two ``medium" and one ``large" network, respectively with $\operatorname{MobileNetV2}$ (3M parameters), $\operatorname{GoogLeNet}$ (6M), $\operatorname{Resnet18}$ (11M) and $\operatorname{EfficientNetB5}$ (33M). 

\subsection{Identifying a configuration without freezing any layer}

The training uses ADAMW optimizer, considering a weight decay of $5\times10^{-4}$ coupled with the $\operatorname{1cycle}$ scheduler \cite{smith2018disciplined,smith2019super}, and a maximum learning rate equal to 0.01. The batch size is set to $8$. The five loss functions described in Section ``\nameref{sec:lossf}" are tested. To identify configurations that ensure rapid convergence, the number of epochs is considered from $2$ to $5$. The tests were repeated five times leading to a total more than $4\times 5\times 5 = 100$ tests. 

\begin{figure*}
    \centering    
    \begin{subfigure}{0.49\textwidth}
        \centering
        \includegraphics[width=1\linewidth]{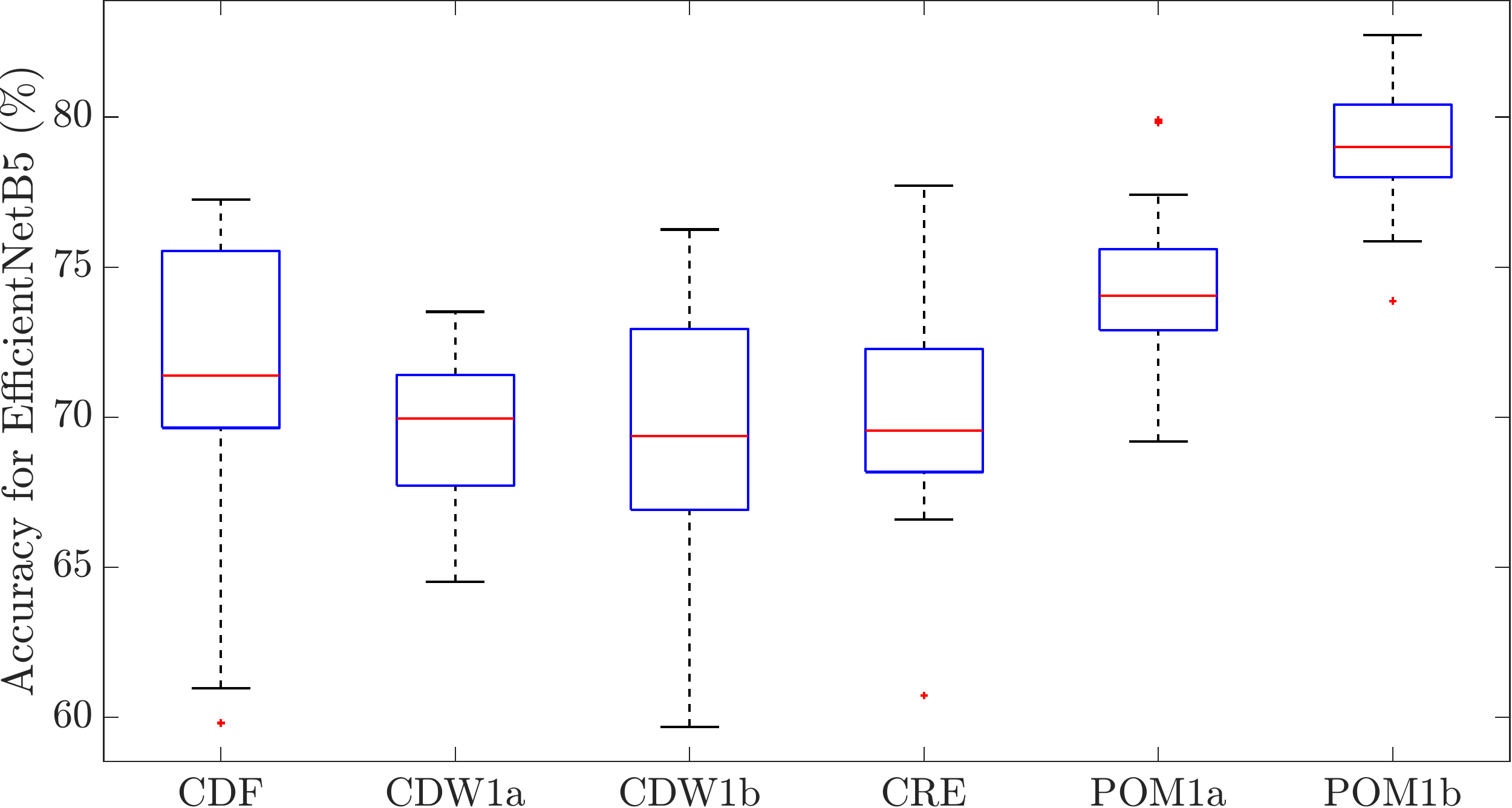}
        \caption{$\operatorname{EfficientNetB5}$}
        \label{fig:effn111}
    \end{subfigure}
    \hfill
    \begin{subfigure}{0.49\textwidth}
        \centering
        \includegraphics[width=1\linewidth]{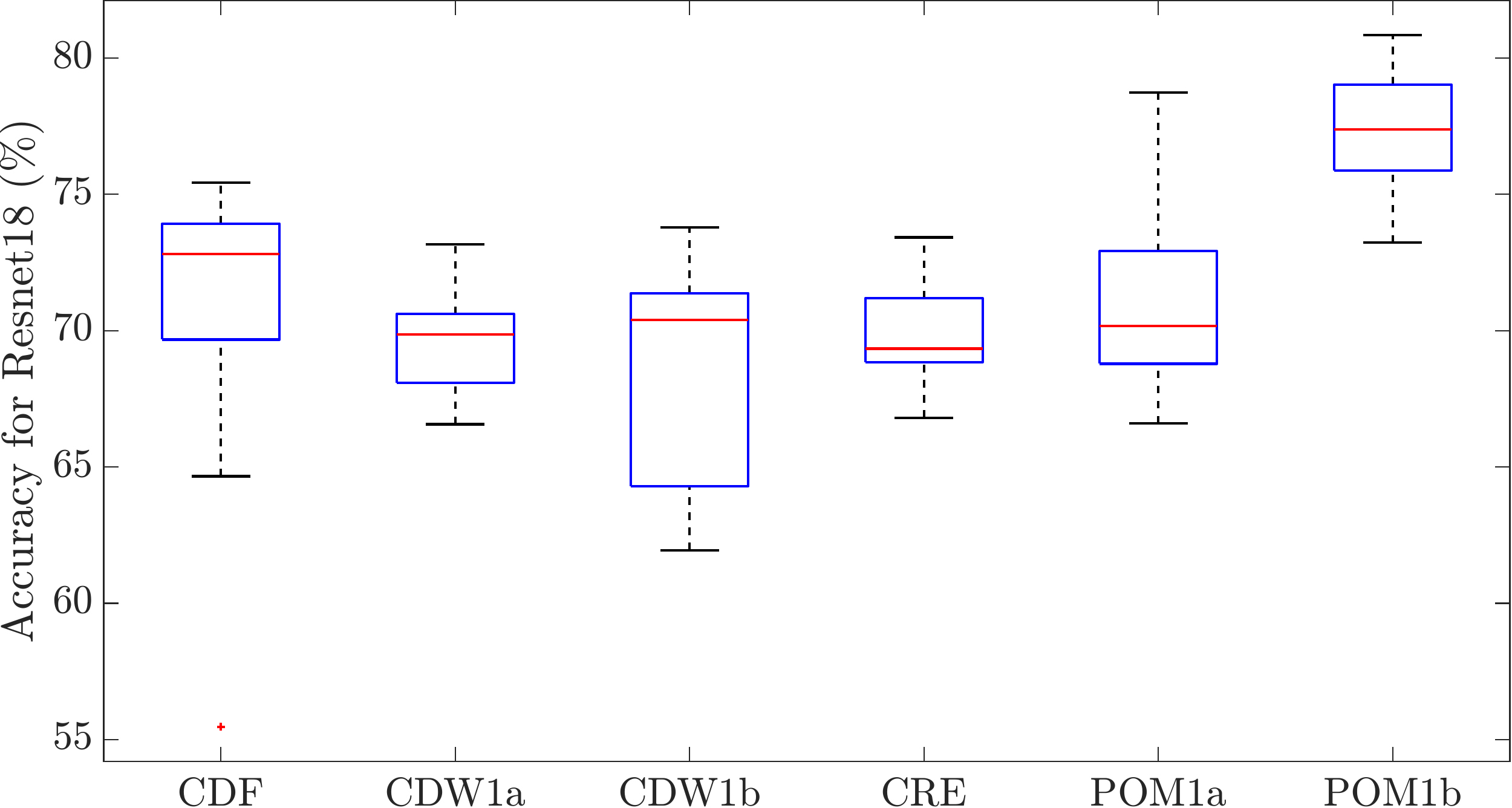}
        \caption{$\operatorname{Resnet18}$}
        \label{fig:resn111}
    \end{subfigure}
    \hfill
    \begin{subfigure}{0.49\textwidth}
        \centering
        \includegraphics[width=1\linewidth]{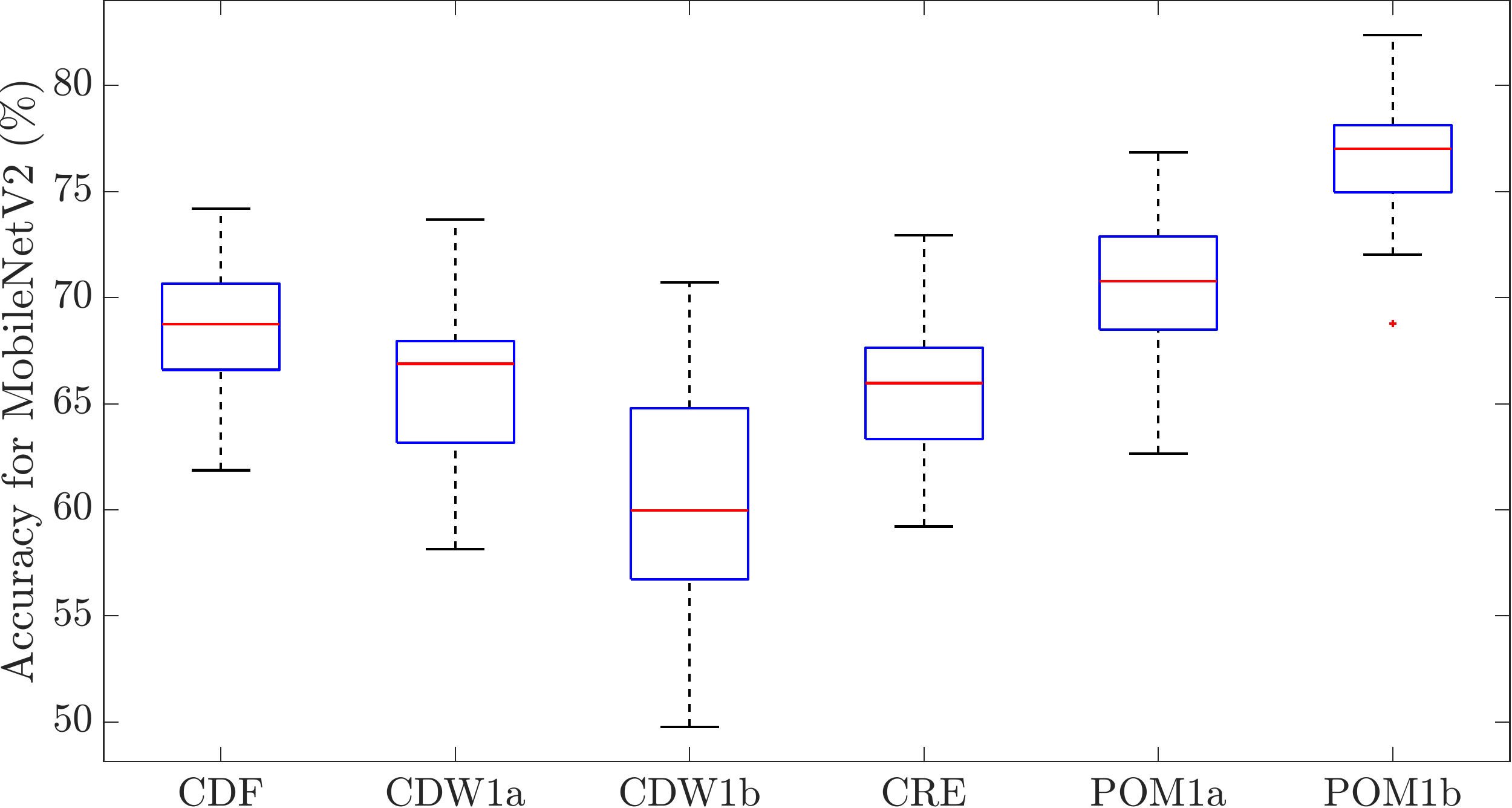}
        \caption{$\operatorname{MobileNetV2}$}
        \label{fig:mbn111}
    \end{subfigure}
    \hfill
    \begin{subfigure}{0.49\textwidth}
        \centering
        \includegraphics[width=1\linewidth]{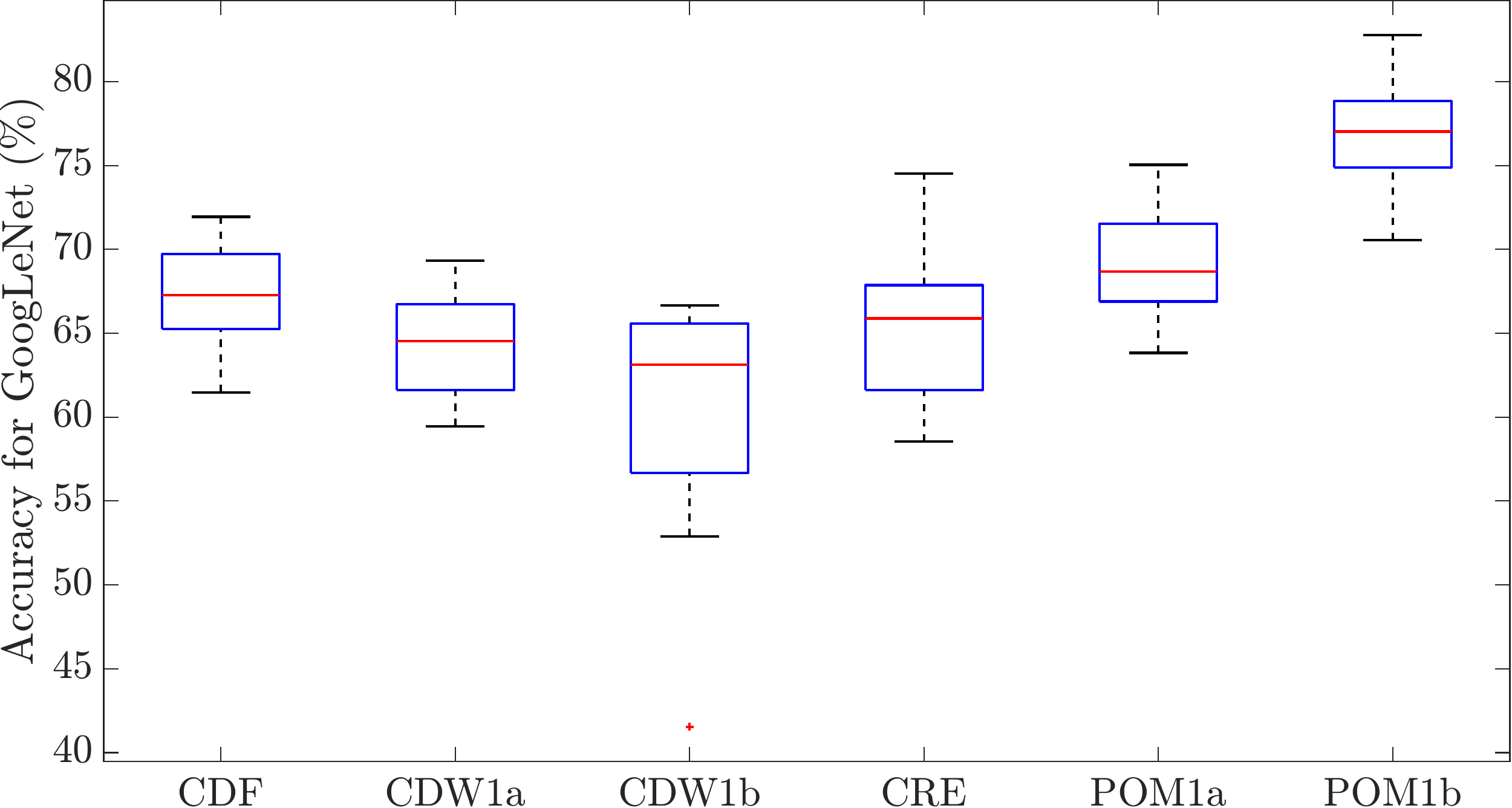}
        \caption{$\operatorname{GoogLeNet}$}
        \label{fig:ggn111}
    \end{subfigure}
\caption{Performance of networks according to the loss. {$\bullet$} corresponds to outliers, and a red line is the median.}
    \label{fig:networkLoss}
\end{figure*}

\begin{figure*}
    \centering    
    \begin{subfigure}{0.49\textwidth}
        \centering  
        \includegraphics[width=1\linewidth]{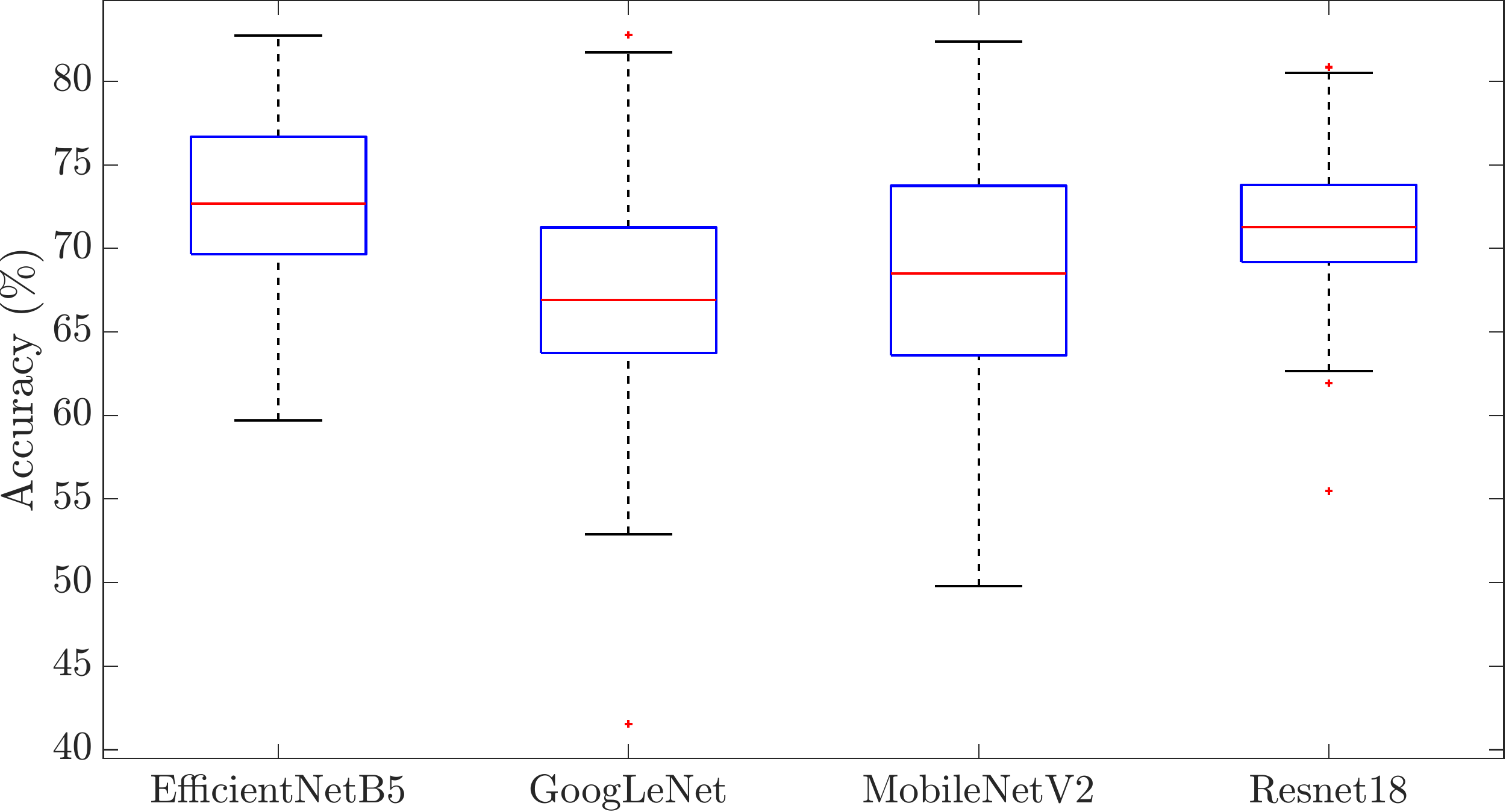}
        \caption{Performance of networks independently of the loss.}
        \label{fig:netindeploss}
    \end{subfigure}
    \hfill
    \begin{subfigure}{0.49\textwidth}
        \centering
        \includegraphics[width=1\linewidth]{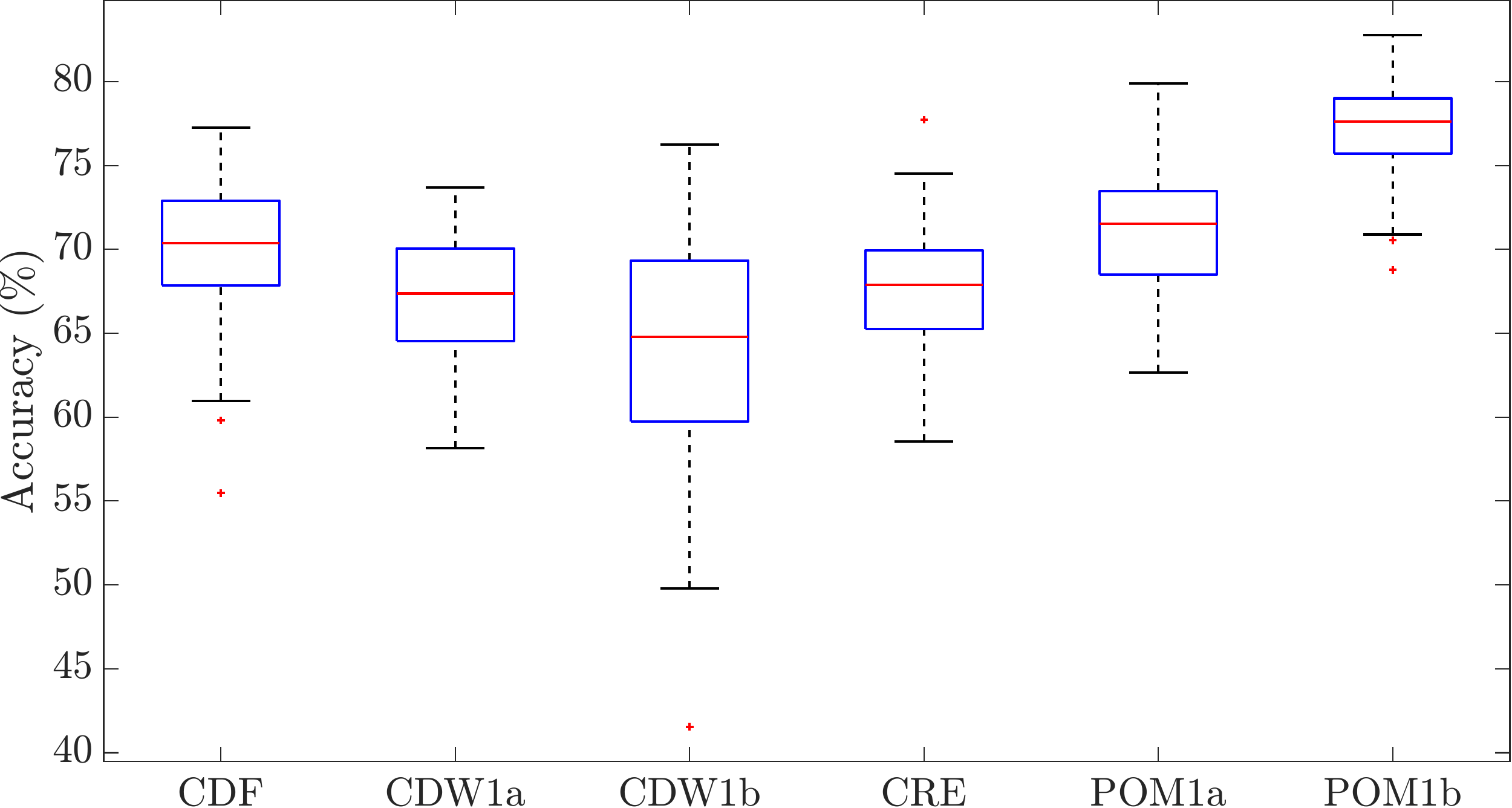}
        \caption{Performance of loss independently of the network.}
        \label{fig:lossindepnet}
    \end{subfigure}
\caption{a) Robustness of the networks according to the loss; b) Robustness of the loss according to the networks.}
    \label{fig:netlossrobus}
\end{figure*}

Figures \ref{fig:networkLoss} and \ref{fig:netlossrobus} present the results through box-plots (median, minimum and maximum values, as well as outliers) of accuracy values $\operatorname{ACC_{\pm1}}$ for the four models. Note the important result provided by $\operatorname{POM1b}$, which outperforms all other losses, for all models. Averaging the $140$ tests over all networks used to compute the box-plot of this specific loss (right-hand boxes in Figure \ref{fig:lossindepnet}, a value of $\operatorname{ACC_{\pm1}} \sim 77.6\%$ is obtained - this is $7\%$ higher than the second best, which is $\operatorname{CDF}$. Throughout tests, $\operatorname{POM1b}$ also depicts the smallest variability, demonstrating the robustness compared to the other losses. More importantly, \textit{all ordinal losses 
provided better performances than the $\operatorname{CRE}$, the loss commonly used in most publications on AE classification through CNNs}. The $\operatorname{CRE}$ loss also depicts the largest variability, inferring a lack of robustness compared to the other losses. This result highlights the importance of considering ordinal losses in SHM applications. 

Figure \ref{fig:netlossrobus} summarizes the performance of networks, independently of the loss (\ref{fig:netindeploss}), and the performance of the losses, independently of the network (Fig~\ref{fig:lossindepnet}). Based on these figures, note that the architecture $\operatorname{EfficientNetB5}$ has the best potential, with a median value of $72.7\%$ when considering the tests with all losses ($146$ cases). The second best network is $\operatorname{Resnet18}$, with $71.26\%$ - this network also has the smallest variability. All models were able to exceed $80-82\%$ for some initialization (according to random draws of the batches). $\operatorname{MobileNetV2}$ has the largest variability. The 30 best models are given in Table \ref{30best} (appendix), confirming the performance of $\operatorname{EfficientNetB5}$ and $\operatorname{Resnet18}$ with the loss $\operatorname{POM1b}$. 

It is important to stress that the results presented in this section are more promising than those presented by the baseline configurations (accuracy around $65\%$), allowing us to assess that better generalization can be achieved (without prior information), without the need for many training epochs - which is desirable in a SHM context. Three epochs seem enough to get good generalization. 



\subsection{Robustness and super-convergence}

\begin{figure*}
    \centering    
    \begin{subfigure}{0.49\textwidth}
        \centering
        \includegraphics[width=1\linewidth]{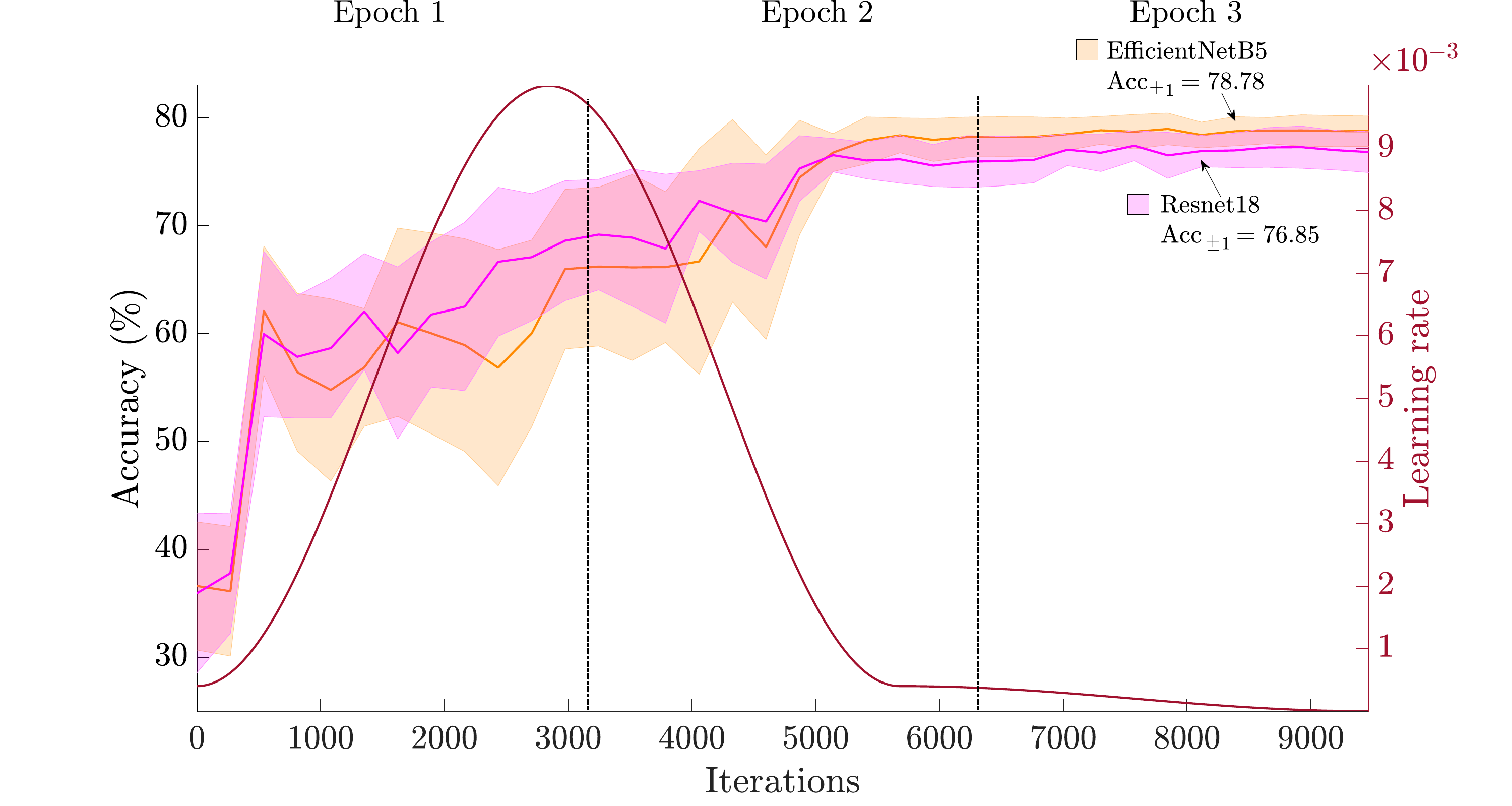}
        \caption{$\operatorname{Resnet18}$ and $\operatorname{EfficientNetB5}$}
        \label{fig:effresn}
    \end{subfigure}
    \hfill
    \begin{subfigure}{0.49\textwidth}
        \centering
        \includegraphics[width=1\linewidth]{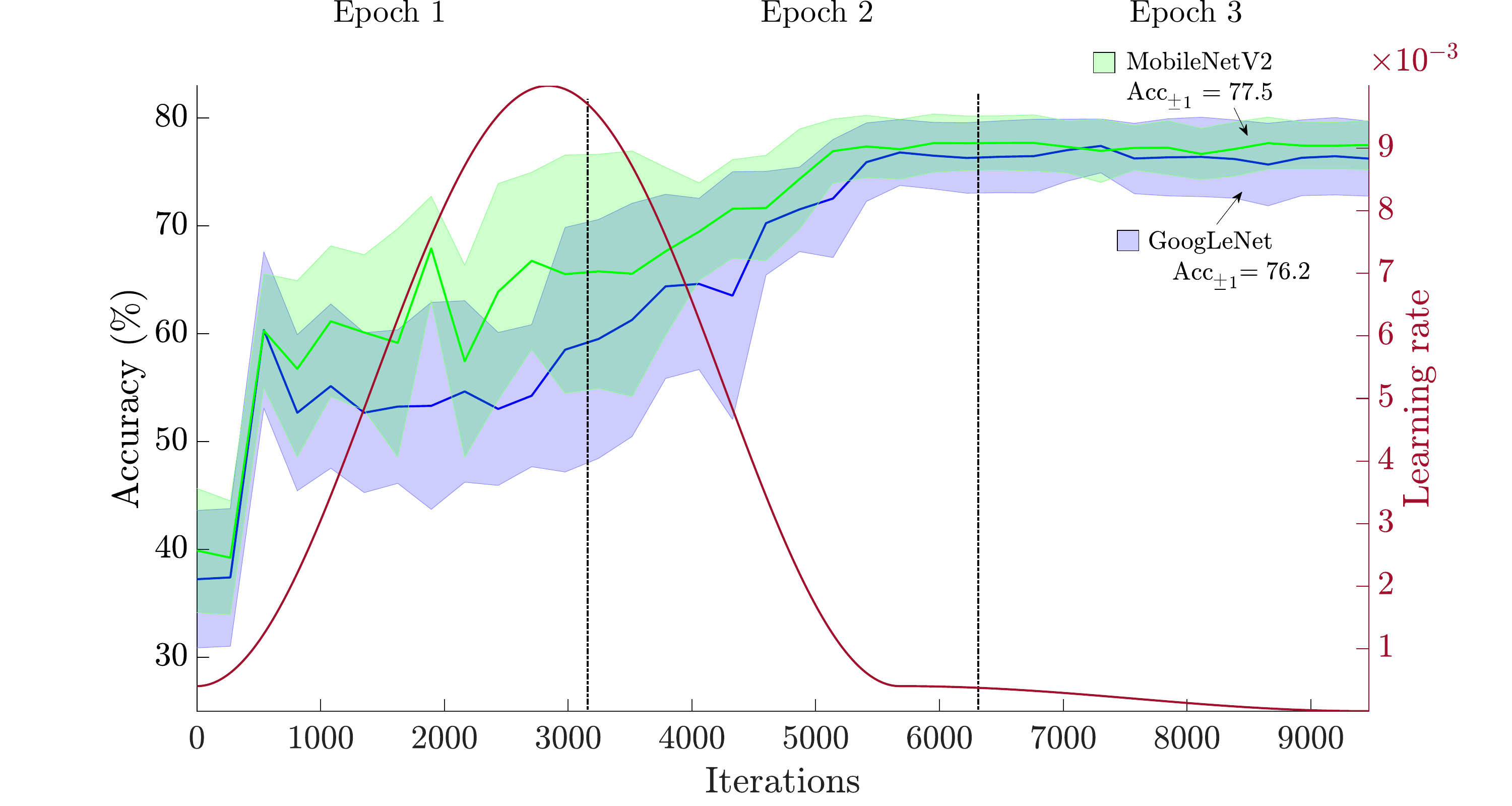}
        \caption{$\operatorname{GoogLeNet}$ and $\operatorname{MobileNetV2}$}
        \label{fig:ggnmbn}
    \end{subfigure}
\caption{Campaign $\#B$: Demonstration of super-convergence. Evolution of the learning rate and of the accuracy in testing for the two best models, using only 3 epochs. }
    \label{fig:ERGMLR}
\end{figure*}

\begin{figure*}
\centering
\includegraphics[width=1.0\columnwidth]{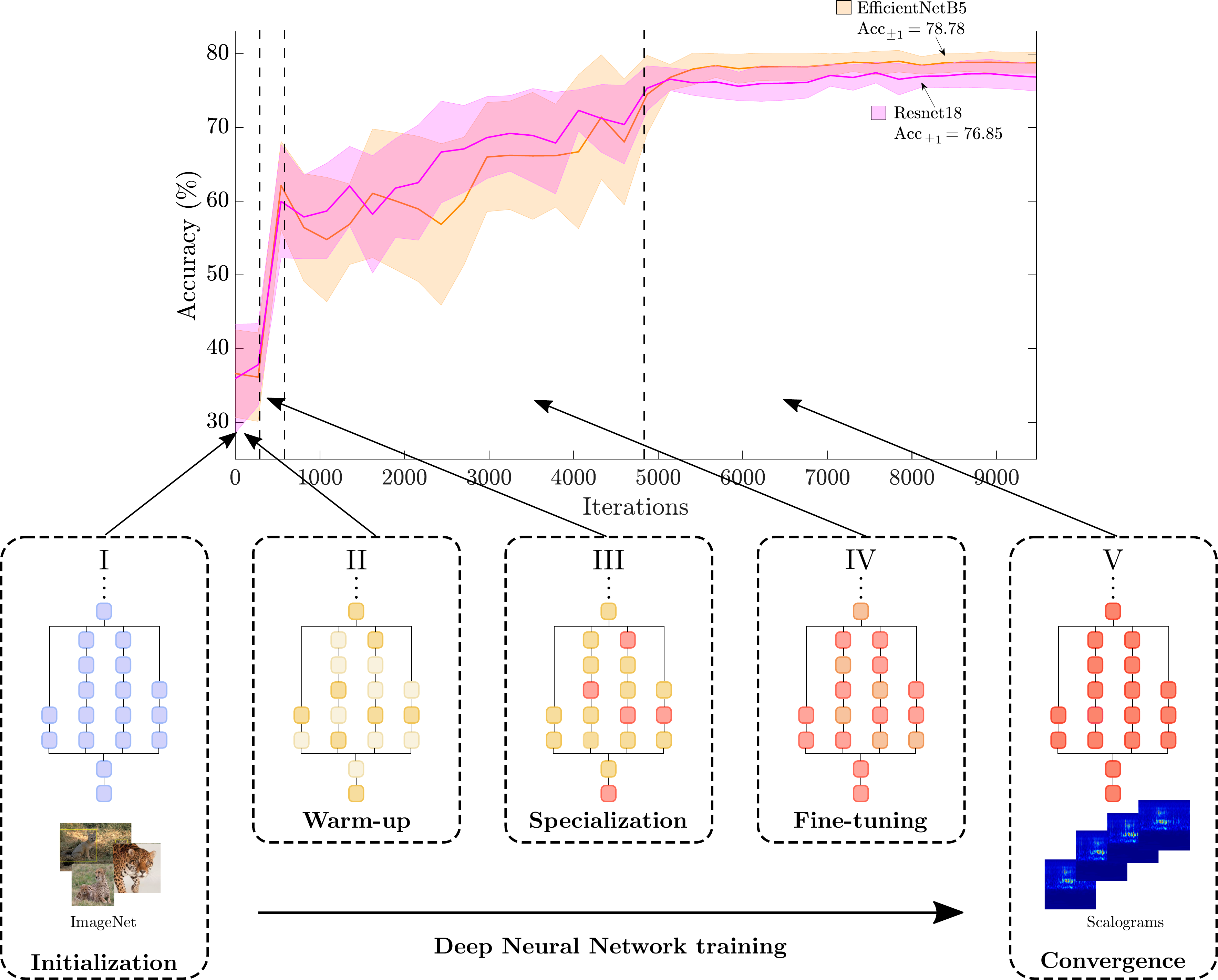}
\caption{Schematic representation of the training dynamics established during the super-convergence study. Training steps illustrated by this figure are the following: (I) Initialization; (II) Warm-up; (III) Specialization; (IV) Fine-tuning; (V) Convergence.
\label{superconv}}
\end{figure*}

Figure \ref{fig:ERGMLR} depicts the evolution of the accuracy on testing data (in generalization without prior on campaign $\#B$) for the four best models, and using only 3 epochs, as deduced in previous tests. The evolution of the learning rate is also presented, computed by the $\operatorname{1cycle}$ policy. Epochs are represented by dashed lines. The variability on accuracy is due to the mini-batches (size $8$) which are drawn randomly at the beginning of each trial, and this variability decreases along iterations as the network is improving. Note that all networks are quite close in terms of accuracy, with a slight advantage to $\operatorname{EfficientNetB5}$ in the first iterations, followed by $\operatorname{Resnet18}$,  $\operatorname{MobileNetV2}$ and  $\operatorname{GoogLeNet}$. Based on these figures, a \textit{training dynamics} can be stated, with a schematic illustration proposed in Figure \ref{superconv}, by the following steps:
\vspace{-0.5cm}
\begin{enumerate}
    \item \textit{The initial accuracy} around $40\%$ obtained from the pretrained networks with ImageNet;
    \item \textit{The warm-up} stage during the very first iterations (0-250), where the accuracy increases a little in most cases;
    \item A phase of network \textit{specialization}, where the accuracy increases  quickly, with $+40\%$ in some cases. This phase is observed for all models;
    \item {Gradual increase} of the accuracy, which corresponds to a \textit{fine tuning} of the network;
    \item A \textit{plateau} that starts around 4500-5500 iterations (end of the second step of the schedule), showing the convergence.
\end{enumerate}
These figures show that the best configurations found in this study improve the accuracy from $65\%$ (baseline, Figure \ref{general1}) to $79\%$ ($\operatorname{EfficientNetB5}$ in Figure \ref{fig:effresn}). Figure \ref{fig:effresn} depicts a characteristic behavior in which the accuracy of $\operatorname{EfficientNetB5}$ increases slowly in the \textit{fine tuning}, and remains lower than $\operatorname{Resnet18}$ before an improvement in the late iterations, leading to a higher accuracy. This behavior is observed on the five datasets as shown in next subsection. 

Note that Smith et al. \cite{smith2018disciplined,smith2019super} recommended to reduce all forms of regularization to preserve a balance between underfitting and overfitting. In particular, the authors advised to use large batch sizes. However, in our study, the minibatch size equal to $8$ worked well on all datasets and can be fitted on small GPU. 

In terms of computational resources, $\operatorname{EfficientNetB5}$ has $3 \times$ more parameters than $\operatorname{Resnet18}$ requiring $2.5 \times$ much space on a hard disk to store the model. The training stage is about $2-3 \times$ longer with $\operatorname{EfficientNetB5}$. On a laptop equipped with a RTX4000 GPU, training $\operatorname{EfficientNetB5}$ took $39\,s$ on average for about $8000$ images (one campaign of $70\,s$), whereas $\operatorname{Resnet18}$ required $19\,s$. Note that these conclusions can vary with the computer used for the test. Tobiasz et al. \cite{Tobiasz23} compared $\operatorname{Resnet(s)}$ and $\operatorname{EfficientNet(s)}$ in terms of inference time, and showed that they are much quicker than older networks like $\operatorname{VGG16}$ or $\operatorname{InceptionNet(s)}$, but largely slower than $\operatorname{MobileNet(s)}$. According to our tests, $\operatorname{MobileNetV2}$ is indeed fast during training but is clearly outperformed by the three others networks for the generalization task (Figures \ref{fig:netindeploss}). Therefore, a good compromise could be $\operatorname{Resnet18}$ with the loss $\operatorname{POM1b}$ or $\operatorname{EfficientNetB5}$ if the computer has enough resources because its performances were generally better.

\subsection{Test on all campaigns}

Previous tests were made on campaign $\#B$ for illustrative purposes, once it represents the most difficult campaign. Based on the previous tests, two efficient configurations stand out. In this section, these same configurations are applied for evaluating their performance on SHM on the other campaigns, namely $\#C$, $\#D$, $\#E$ and $\#F$. Results are showed in Figure \ref{fig:generalizationlast} and performances (using four metrics, see \hyperref[sec:metrics]{Appendix B}) are summarized in Table \ref{tab:my_label}.
 
\begin{table*}[ht]
{{
    \centering
    \begin{tabular}{|c|c|c||c|c||c|c|c|c|}
    \hline
    \multicolumn{3}{|c||}{Training} & \multicolumn{2}{c||}{Testing} & \multicolumn{4}{c|}{Performance in testing}  \\
    \hline
    Campaigns & $\#$ train.  & $\#$ valid. & Campaigns  & $\#$ test. & $\operatorname{ACC_{\pm 1}}$ & $\operatorname{R_{\pm  1}}$ & $\operatorname{P_{\pm 1}}$  & $\operatorname{F1_{\pm 1}}$ \\
    \hline 
    $\# C,\# D,\# E,\# F$ &  $25244$  & $6311$ & $\#B$  & $8064$ & $78.8$  & $71.4$ & $64.1$  & $66.1$ \\
    $\# B,\# D,\# E,\# F$ &  $22348$ & $5587$ &  $\#C$ &  $7006$ &  $86.4$ & $73.7$ &  $76.7$ &  $75.1$ \\
    $\# B,\# C,\# E,\# F$ &  $25148$ & $6286$ &  $\#D$ &  $8185$ &  $86.3$ & $77.9$ &  $72.8$ &  $75.3$ \\
    $\# B,\# C,\# D,\# F$ &  $25152$ & $6287$ &  $\#E$ &  $8180$ &  $78.8$ & $71.1$ &  $65.8$ &  $68.3$\\
    $\# B,\# C,\# D,\# E$ &  $25148$ & $6287$ &  $\#F$ &  $8184$ &  $86.1$ & $77.1$ &  $70.8$ &  $73.8$\\
    \hline
    \end{tabular}
    \caption{Metrics of performance averaged over 5 runs in $\%$ for all campaigns and in generalization, with details on the number of images in training, validation and testing datasets.}
    \label{tab:my_label}
    }}
\end{table*}

\begin{figure*}
    \centering    
    \begin{subfigure}{0.49\textwidth}
        \centering
        \includegraphics[width=1\linewidth]{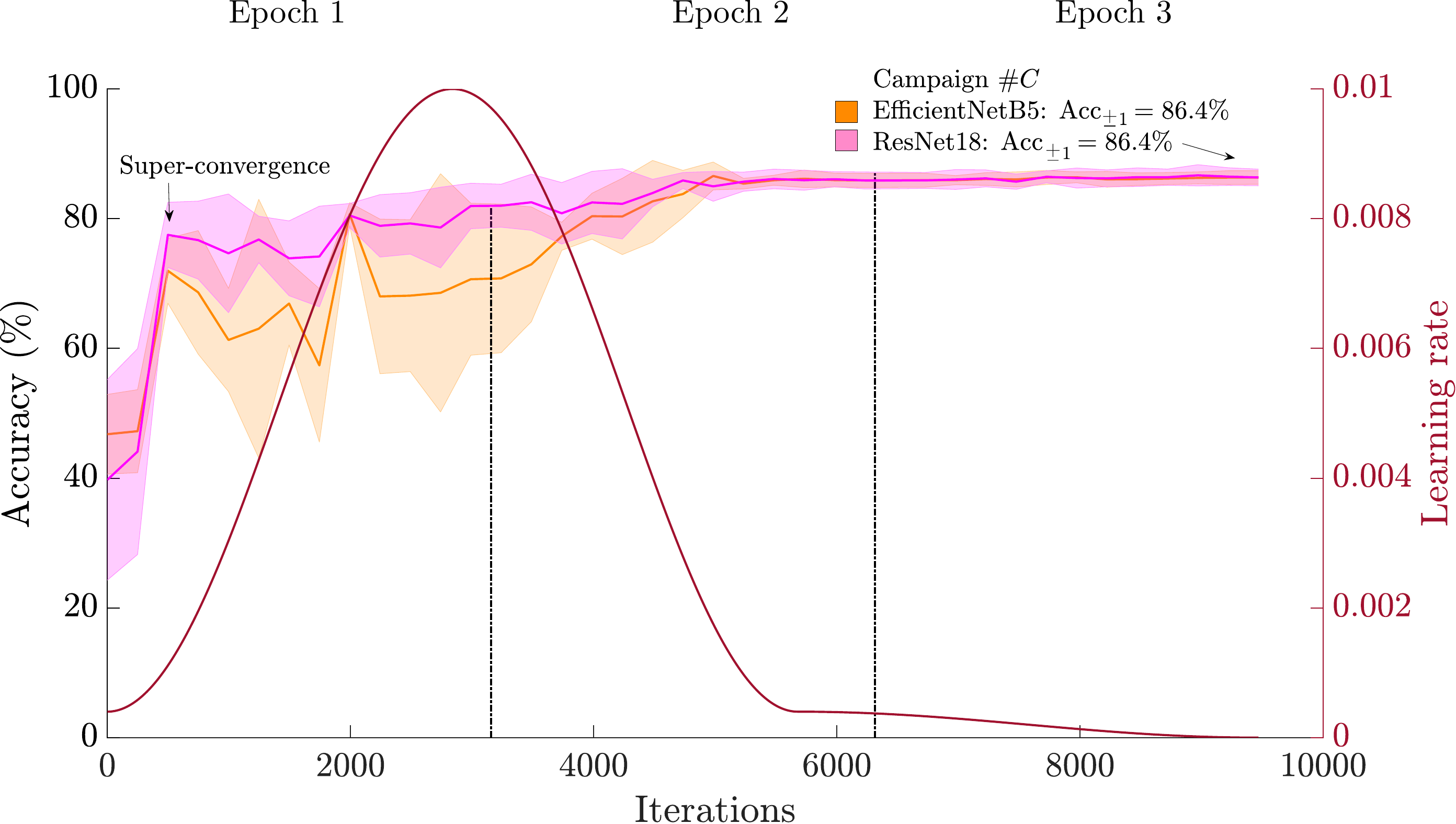}
        \caption{$\#C$}
        \label{fig:RESCDEFn3C}
    \end{subfigure}
    \hfill
    \begin{subfigure}{0.49\textwidth}
        \centering
        \includegraphics[width=1\linewidth]{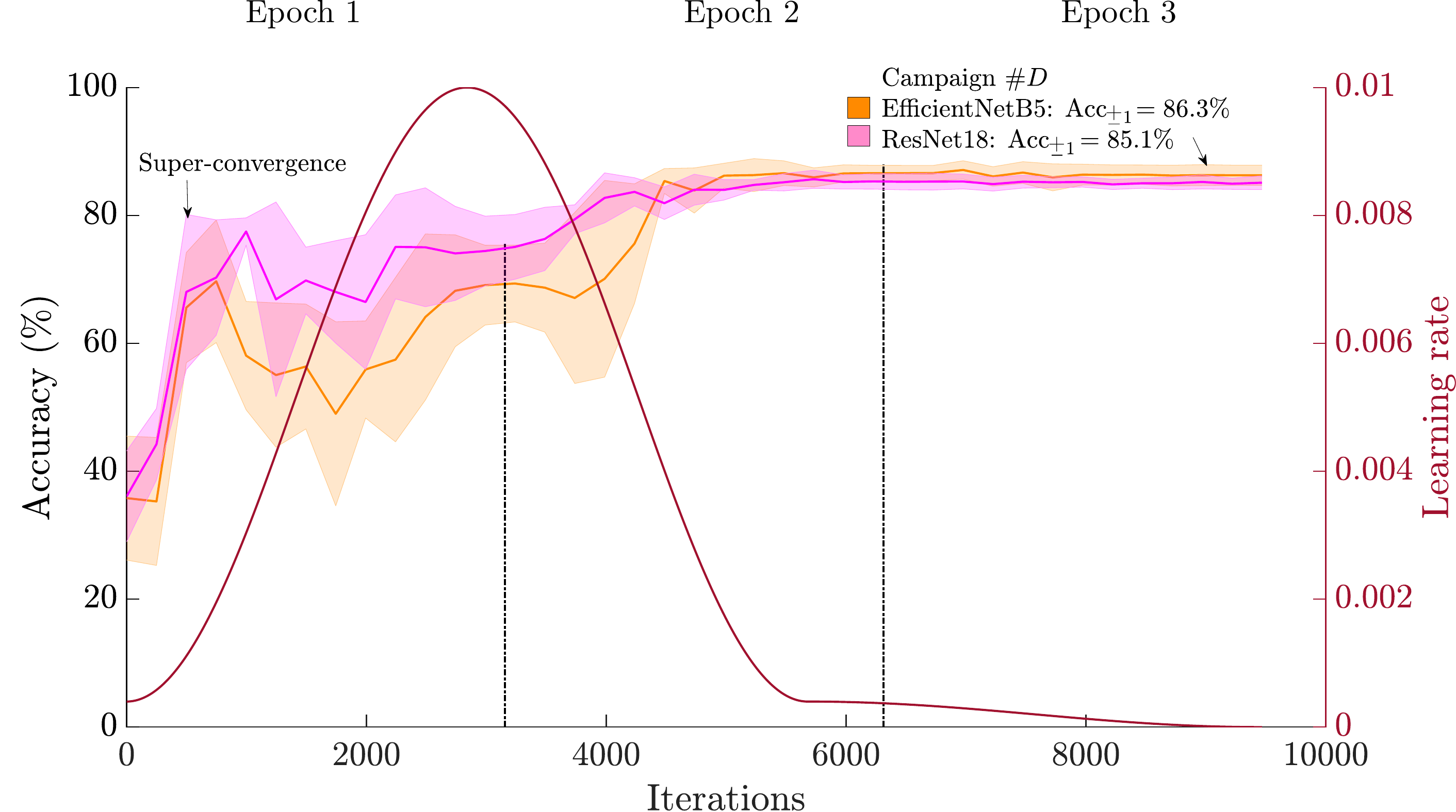}
        \caption{$\#D$}
        \label{RESCDEFn3D}
    \end{subfigure}
    \hfill
    \begin{subfigure}{0.49\textwidth}
        \centering
        \includegraphics[width=1\linewidth]{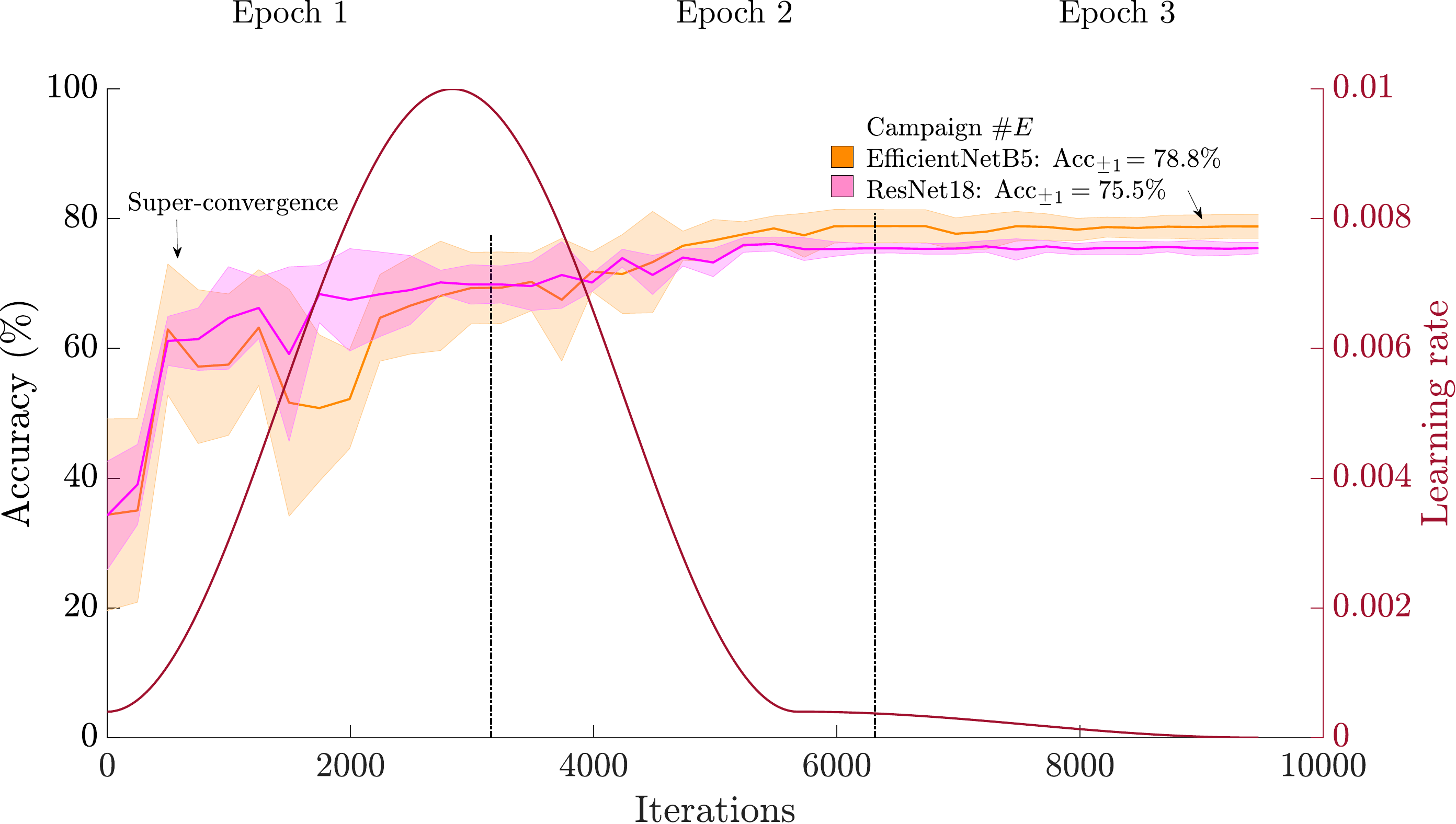}
        \caption{$\#E$}
        \label{RESCDEFn3E}
    \end{subfigure}
    \hfill
    \begin{subfigure}{0.49\textwidth}
        \centering
        \includegraphics[width=1\linewidth]{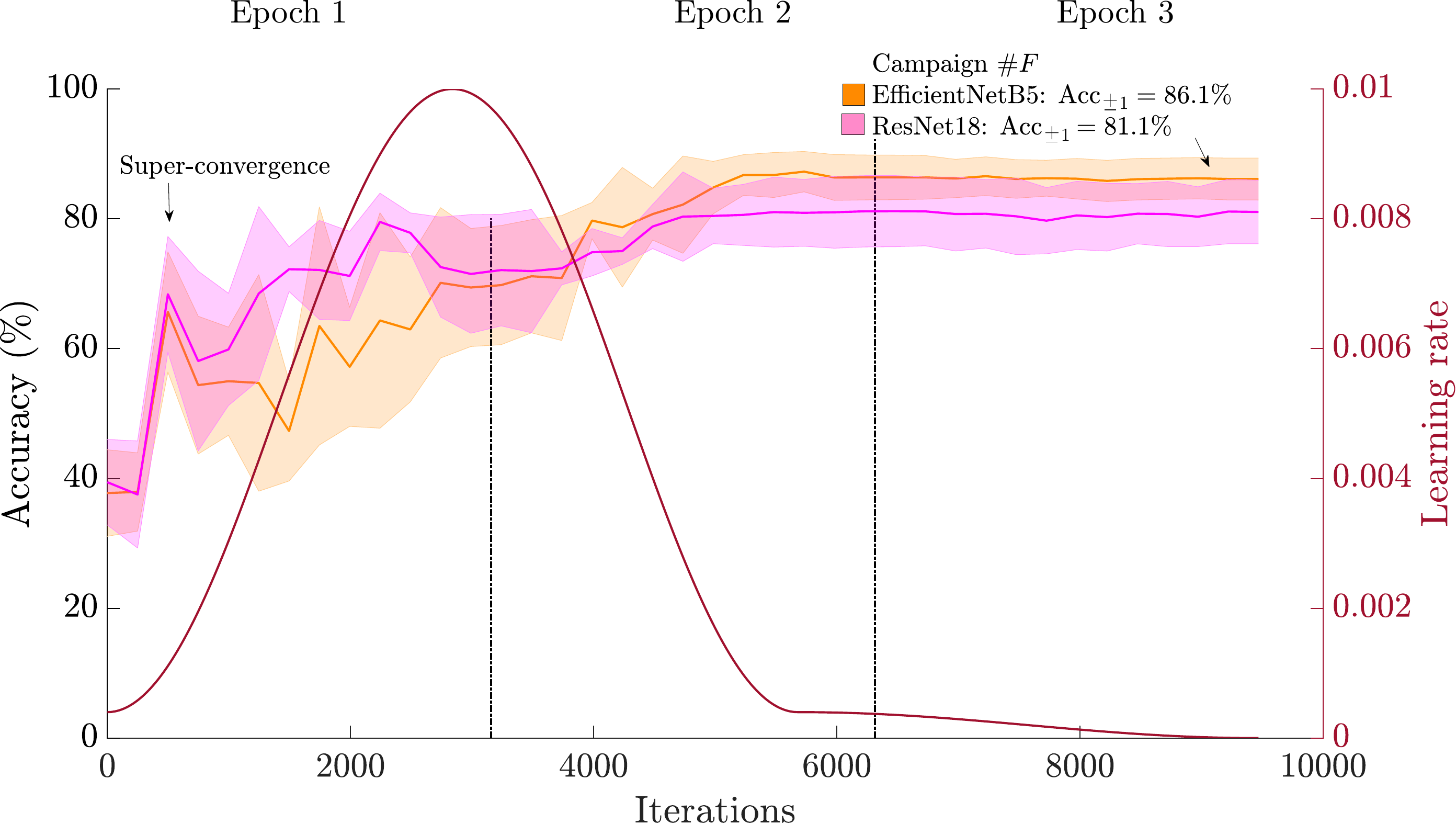}
        \caption{$\#F$}
        \label{RESCDEFn3F}
    \end{subfigure}
\caption{Performance of the best configuration on the other datasets ($\#C$, $\#D$, $\#E$ and $\#F$) by using the same configuration of the two best networks as done for $\#B$.}
    \label{fig:generalizationlast}
\end{figure*}

{{
In these figures, the phenomenon of super-convergence is clearly visible, achieving rapid convergence within just three epochs using the $\operatorname{1cycle}$ policy. Note the characteristic behavior of $\operatorname{EfficientNetB5}$ during the fine-tuning stage. Initially, its generalization accuracy is lower than that of $\operatorname{ResNet18}$, but it subsequently improves and surpasses $\operatorname{ResNet18}$ in the later iterations. The final performance metrics, as shown in Table \ref{tab:my_label}, were achieved using $\operatorname{EfficientNetB5}$ with $\operatorname{POM1b}$ loss across all datasets.
}}

\section{Conclusion}
\label{sec: Conclusions}

Population-based SHM has emerged as an interesting framework to design and validate SHM algorithms. ORION-AE dataset appears to be well-suited to illustrate this framework because the repeated measurements through different campaigns of a bolted joint have modified the contact conditions at the lap-joint, and its microscale topology has changed between experiments. These microscale changes have an important impact on the behavior of the joint and therefore the different measurement campaigns can be considered as a set of heterogeneous data. Moreover, this dataset can be used in two ways: either campaign-wise or for SHM purposes. In the first case, the normalization of the data, i.e. same campaign being used for training, validation and testing, makes the data homogeneous and the classification problem simple, resulting in an accuracy of over $99\%$, using relatively simple models. However, in the latter, the task becomes more difficult because of the modified operational conditions, which make the data heterogeneous and challenge the generalization capabilities of the models.  

The main conclusions of this study are summarized as follows:
\begin{itemize}
    \item Previous approaches based on deep neural networks with frozen layers developed for homogeneous population failed to generalize, making them inefficient for SHM purposes within bolted joints using the {{AE technique}}. Scalograms are not natural images and therefore it is required to update all parameters of the network. 

    \item Denoising AE data before computing the CWT is not necessary. Denoising is embedded in deep neural networks through convolutional layers that process scalograms.
    
    \item Using the $\operatorname{mu80}$ sensor was enough to get high performance, without the need of sensor fusion. 
    
    \item Different loss functions were compared, in particular ordinal losses and the standard cross-entropy ($\operatorname{CRE}$). We identified one loss, called $\operatorname{POM1b}$, that led to the best results overall, independently on the network. Conversely, the $\operatorname{CRE}$ depicted the worst results in terms of accuracy and standard deviation.
    
    \item To the best of our knowledge, the phenomenon of super-convergence is presented for the first time on SHM data. For that, we explored a scheduling strategy of the learning rate called $\operatorname{1cycle}$ scheduler that allows to get high accuracy in a few number of iterations. 

   \item We conducted many tests, with different architectures of CNNs and our study showed that the two best configurations in terms of accuracy and convergence speed were: $\operatorname{EfficientNetB5}$ and $\operatorname{Resnet18}$, both with $\operatorname{POM1b}$-loss, with accuracy of $78.8\%, 86.4\%, 86.3\%, 78.8\%, 86.1\%$ for the five campaigns of measurements $\#B$, $\#C$, $\#D$, $\#E$ and $\#F$, respectively.

   \item {{Based on the results of this study, a promising direction for enhancing generalization in SHM of bolted joints involves exploring the performance of sensors with higher frequency ranges for condition monitoring. This approach could potentially improve the accuracy and reliability of bolt detection and assessment.}}
\end{itemize}

The true labels are not reliable enough because of the uncertainty about the position of the torque screwdriver used to set the tightening levels. Considering a mistake of one class as correct is a possible approximation which can represent a problem in certain applications. As future work, we will focus on how to encode the uncertainty on labels and how to integrate it into deep neural networks and study whether and how it helps at improving the accuracy.

\section*{Acknowledgment}
This work was partly carried out in the framework of the EIPHI Graduate school (contract ANR-17-EURE-0002) and the project RESEM- COALESCENCE funded by the Institut de Recherche Technologique Matériaux Métallurgie Procédés (IRT M2P) and Agence Nationale de
la Recherche (ANR).

\clearpage

\appendix

\section{Appendix A: Identifying a configuration with fast results}

Table \ref{30best} presents the 30 best network configurations found in terms of accuracy as a function of loss, number of epochs, and number of tests carried out for the SHM case. The testing dataset was $\#B$, whereas the training and validation ones are $\{\#C,\#D,\#E,\#F\}$. 

\begin{table}[hbtp]
\centering
\begin{tabular}{|c|c|c|c|c|}
\hline
Network & Loss &  Epoch & Acc & $\sigma$ \\
\hline
    EfficientNetB5  &     POM1b &        3    &      79.268   &    2.1144   \\
    EfficientNetB5  &     POM1b &        4    &      79.264   &    2.1685  \\
    EfficientNetB5  &     POM1b &        2    &      79.045    &   1.5405  \\
    EfficientNetB5  &     POM1b &        5   &       78.683   &    1.7855   \\
    Resnet18     &     POM1b &        5    &      77.741    &   1.6415   \\
    Resnet18     &     POM1b &        3    &      77.545  &      2.019   \\
    GoogLeNet     &    POM1b &        5    &      77.235   &    1.6483  \\
    MobileNetV2   &    POM1b &        3    &      77.169    &   2.3464  \\
    GoogLeNet     &    POM1b &        3      &    76.886    &   3.2553  \\
    Resnet18      &    POM1b &        2     &     76.704   &    2.2842  \\
    MobileNetV2   &    POM1b &        4    &      76.463   &     2.872  \\
    MobileNetV2   &    POM1b &        5    &      76.427  &     1.8958 \\
    Resnet18      &    POM1b &        4    &      76.188  &     2.2991  \\
    GoogLeNet     &    POM1b &        2    &      76.052   &    3.4309  \\
    GoogLeNet     &    POM1b &        4    &      76.034  &     3.6481  \\
    MobileNetV2    &   POM1b &        2       &   75.599    &   3.5895  \\
    EfficientNetB5  &     POM1a   &    5      &    75.164    &   2.7232  \\
    EfficientNetB5  &     POM1a   &    3      &    74.814    &   2.9309   \\
    EfficientNetB5  &     POM1a   &    4      &    74.554   &    3.1466  \\
    MobileNetV2  &     POM1a    &   4        &   73.78      &  1.563     \\
    EfficientNetB5  &     CDF  &       2    &      73.229   &    3.2286  \\
    EfficientNetB5  &     CRE   &      4    &       73.15   &    3.2248  \\
    EfficientNetB5  &     POM1a &      2   &       73.125   &    2.5063  \\
    Resnet18      &    POM1a    &   2      &    72.629    &   4.1272    \\
    MobileNetV2  &     POM1a    &   3     &     72.414     &  2.9934   \\
    Resnet18     &     CDW1b   &    4     &     72.264   &    1.7702   \\
    Resnet18       &   CDF     &    3     &     71.779   &    5.5462   \\
    Resnet18      &    POM1a  &     3      &    71.592    &   1.6987   \\
    EfficientNetB5  &  CDF    &     4      &     71.25   &    1.4033   \\
    MobileNetV2   &    POM1a   &    5      &     70.96    &   1.4625    \\
    \hline
\end{tabular}
\caption{30 best models on average for generalization on campaign \#B, sorted by accuracy at plus or minus one class, and the standard deviation over five tests.}
\label{30best}
\end{table}

\section{Appendix B: Metrics}
\label{sec:metrics}

\begin{figure}[ht]
\centering
\includegraphics[width=0.40\columnwidth]{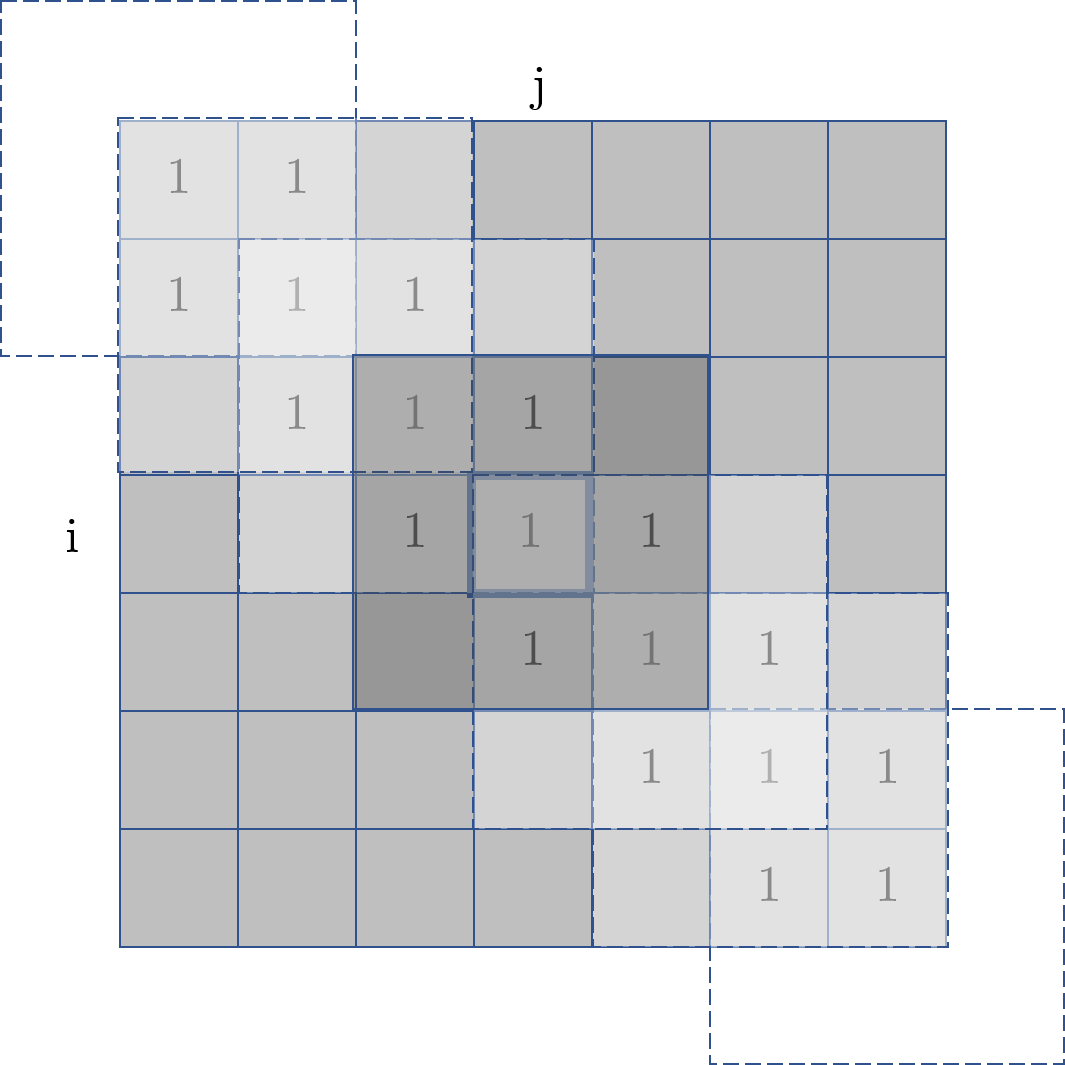}
\caption{Schematic representation of a tridiagonal confusion matrix.}
\label{pm1metrics}
\end{figure}

Figure \ref{pm1metrics} depicts a confusion matrix, say $m$, for $K=7$  classes based on which the following metrics can be obtained. The matrix is tridiagonal which allows to show the specificity of the metrics at plus or minus one class.

The standard accuracy is $$ACC=\sum_{i=1}^K m(i,i),$$, while the accuracy at $\pm 1$ is $$ACC_{\pm 1}=\sum_{i=1}^K \sum_{j\in \{i-1,i,i+1\}} m(i,j),$$ with obvious cases at $i-1>0$, $i+1<K$.

The recall and precision at plus or minus one, $P_{\pm 1}$ and $R_{\pm 1}$ respectively, are obtained as follows. Let $\#correct\;k$ the number of points correctly classified in class $k$, $\#predicted\;k$  the number of points predicted as class $k$, and $N_k$ the number of points in class $k$, then:
$$R_{\pm 1}(k)=\frac{\#correct \; k}{N_k},$$
$$P_{\pm 1}(k)=\frac{\#correct \; k}{\#predicted \; k},$$
with $N_k$ the number of elements in class $k$ in the ground truth (given by the sum of columns for the k-th row) and
$$\#correct \; k = \sum_{(i,j)\in\mathcal{N}(i)\backslash \{(i+1,i+1),(i-1,i-1)\}}m(i,j),$$
where $\mathcal{N}(i)$ is the $3\times 3$ neighborhood of $i$ and 
$$\#predicted \; k = \sum_{i=1}^K \sum_{j\in\{-1,0,1\}} m(i,k-j).$$
After taking the average of recall and precision values over classes we can deduce the F1-score:
$$F_1=\frac{2\overline{R}_{\pm 1}\cdot\overline{P}_{\pm 1}}{\overline{R}_{\pm 1}+\overline{R}_{\pm 1}}.$$

For the example depicted on the figure, we have $ACC=0.36$, $ACC_{\pm 1}=1.0$, $P_{\pm 1}(k)=[   0.8
    0.875,     0.778,      0.778,     0.778,     0.875,     0.8]$, $\overline{P}_{\pm 1}= 0.81$ and $R_{pm_1}(k) = [    0.8,    0.875,    0.778,    0.778,    0.778,     0.875,    0.8]$, $\overline{R}_{\pm 1}= 0.81$, yielding $F_1=0.81$.

\clearpage

\end{document}